\documentclass[11pt]{article}              

\usepackage[margin=25mm]{geometry}
\usepackage{graphicx}
\usepackage{colortbl}
\usepackage{subfigure}
 
\usepackage{fancyhdr}
\usepackage{amssymb,amsfonts,amsmath,amsthm}
\usepackage{natbib}
\usepackage[english]{babel} 
\usepackage{graphicx}
\usepackage{float}
\usepackage{color}
\usepackage{soul}
\usepackage{geometry}
\usepackage{natbib}
\usepackage{setspace}
\usepackage{caption}
\usepackage{multirow}

\usepackage{authblk}
	
\graphicspath{{./},{./Figures/}}

\usepackage{amsopn}

\graphicspath{{Figures/}}

\usepackage[utf8]{inputenc}
\newcommand{\eg}{{e.g. }}
\newcommand{\ie}{{i.e. }}
\newcommand{\ca}{{\mathcal A}}
\newcommand{\cm}{{\mathcal M}}
\newcommand{\cd}{{\mathcal D}}
\newcommand{\cb}{{\mathcal B}}
\newcommand{\ch}{{\mathcal H}}
\newcommand{\cg}{{\mathcal G}}
\newcommand{\ct}{{\mathcal T}}
\newcommand{\cx}{{\mathcal X}}
\newcommand{\cy}{{\mathcal Y}}
\newcommand{\cu}{{\mathcal U}}

\newcommand{\Esp}[1]{{\mathbb E}\left[ #1 \right]}
\newcommand{\Var}[1]{{\rm Var}\left[ #1 \right]}
\newcommand{\Nn}{{\mathbb N}}
\newcommand{\Rr}{{\mathbb R}}
\newcommand{\Pp}{{\mathbb P}}
\newcommand{\prt}[1]{\left(#1\right)}			% entre parenthesese ()
\newcommand{\acc}[1]{\left\{#1\right\}}			% entre accolades {}
\newcommand{\abs}[1]{\left| #1 \right|}			% abs() |x|
\newcommand{\enum}{ , \, \dots \,,}
\newcommand{\eqrefe}[1]{Eq.~(\ref{#1})}
\newcommand{\figref}[1]{Figure~\ref{#1}}
\newcommand{\card}{\text{card}}
\newcommand{\vexi}{\ve{\xi}}
\newcommand{\vepsi}{\ve{\psi}}
\newcommand{\veX}{\ve{X}}

\newcommand{\vex}{\ve{x}}
\newcommand{\vey}{\ve{y}}
\newcommand{\vev}{\ve{v}}
\newcommand{\vea}{\ve{\alpha}}
\newcommand{\veb}{\ve{\beta}}
\newcommand{\matSigma}{\mat{\Sigma}}

\newcommand{\vegamma}{\ve{\gamma}}
\newcommand{\vetheta}{\ve{\theta}}
\newcommand{\mat}[1]{\boldsymbol{#1}}		
\newcommand{\matA}{\mat{A}}

\newcommand{\ve}[1]{\boldsymbol{#1}}		
\newcommand{\matcalf}{\mathcal{F}}
\newcommand{\matcalA}{\mathcal{A}}
\newcommand{\di}[1]{{\rm d}#1} 				
\newcommand{\eqdef}{\stackrel{\text{def}}{=}}
\newcommand{\norme}[2]{\parallel #1\parallel_{#2}}
\newcommand{\innprod}[3]{< #1\,,\,#2 >_{#3}} 
\newcommand{\tr}{^{\textsf T}}			

\newcommand{\vealetter}{\ve{a}}

\begin{document}
\title{Surrogate models for oscillatory systems using
  sparse polynomial chaos expansions and stochastic time warping}

\author[1]{C. V. Mai} \author[1]{B. Sudret} 

\affil[1]{Chair of Risk, Safety and Uncertainty Quantification,
  
  ETH Zurich, Stefano-Franscini-Platz 5, 8093 Zurich, Switzerland}

\date{}
\maketitle

\abstract{   Polynomial chaos expansions (PCE) have proven efficiency in a number
  of fields for propagating parametric uncertainties through
  computational models of complex systems, namely structural and fluid
  mechanics, chemical reactions and electromagnetism, etc. For problems
  involving oscillatory, time-dependent output quantities of interest,
  it is well-known that reasonable accuracy of PCE-based approaches is
  difficult to reach in the long term. In this paper, we propose a fully
  non-intrusive approach based on stochastic time warping to address
  this issue: each realization (trajectory) of the model response is
  first rescaled to its own time scale so as to put all sampled
  trajectories in phase in a common virtual time line. Principal
  component analysis is introduced to compress the information contained
  in these transformed trajectories and sparse PCE representations using
  least angle regression are finally used to approximate the components.
  The approach shows remarkably small prediction error for particular
  trajectories as well as for second-order statistics of the latter. It
  is illustrated on different benchmark problems well known in the
  literature on time-dependent PCE problems, ranging from rigid body
  dynamics, chemical reactions to forced oscillations of a non linear
  system.
  \\[1em]

   }

\maketitle

%%%%%%%%%%%%%%%%%%%%%%%%%%%%%%%%%%%%%%%%%%%%%%%%%%%%%%%%%%%%%%%%%%%%%%%%%%%%%%%%%%%%%%%%%%%%%%%%
\section{Introduction}
%%%%%%%%%%%%%%%%%%%%%%%%%%%%%%%%%%%%%%%%%%%%%%%%%%%%%%%%%%%%%%%%%%%%%%%%%%%%%%%%%%%%%%%%%%%%%%%%

\section{Introduction}
In modern engineering, it is of utmost importance to investigate the
significant effects of uncertainties when considering the behavior of
complex systems. These uncertainties may arise from environmental
factors (\eg excitations, boundary conditions) or inherent sources (\eg
natural variability of the materials) and are usually represented by
random variables.  In this context, the framework of uncertainty
quantification was introduced, of which a major component is the
propagation of uncertainty from the input parameters through the system
to the output quantities of interest.  The outcomes of uncertainty
propagation (\eg statistical, reliability and sensitivity measures)
allow a better understanding of the system and are critical in decision
making.

So far Monte Carlo simulation (MCS) is universally used for solving
uncertainty propagation problems. The idea behind MCS is to perform the
simulation a sufficiently large number of times by varying input
parameters such that the average of the response quantity of interest
converges to the expected value according to the law of large numbers.
However, the use of MCS is hindered by the fact that a large number of
simulations is not affordable in many practical problems (\eg when each
evaluation of the computational model is time- and memory-consuming).

To overcome this issue, spectral methods have been used in the last two
decades as an alternative approach to traditional MCS. The spectral
approach consists in representing the response quantity of interest in a
space spanned by well-defined basis functions. Among a wide variety of
basis functions that have been investigated, polynomial functions have
shown particular effectiveness \cite{Ghanembook2003, LemaitreBook,
  Soize2004}. The spectral approach that uses polynomial chaos functions
as a basis is simply named polynomial chaos expansions (PCEs).

In practice, PCEs are widely used as an approximate model to substitute
a computationally expensive model for uncertainty propagation. They can
be used in an either intrusive or non-intrusive setup. The former
requires knowledge of the mathematical equations describing the
considered system. One has to interfere with the original set of
equations, reformulate it and then solve the reformulated system to
compute the PCE coefficients.  In contrast, the latter does not
necessitate any prior knowledge of the governing equations. It considers
the deterministic computational model as a black box and only requires to
define an experimental design, \ie a set of input and corresponding
output values.  In several studies, PCEs have shown great efficiency
compared to the traditional uncertainty propagation approach with MCS,
see \eg \cite{Dossantos2008,Rajabi2014}.

PCEs, however, face challenges when used for dynamical systems that are
encountered in the fields of structural and fluid dynamics or in
chemical engineering \cite{Beran:2006,Ghosh2007, LeMaitre2009, Wan2006}.
In these cases, the governing equations are a system of ordinary
differential equations with random parameters. First, the response as a
function of time is no longer a scalar quantity but may be cast as a
vector after proper time discretization. Applying PCEs at each time
instant might require large computational resources. To reduce the
computational cost, Blatman and Sudret \cite{BlatmanIcossar2013} used principal component
analysis to capture the main stochastic features of the vector-valued
response quantities by means of a small number of variables which can be
represented by PCEs.  The greatest challenge is the decrease in time of
the accuracy of the PCE model as reported in numerous publications
\cite{Beran:2006,Gerritsma2010,Ghosh2007,LemaitreBook,LeMaitre2009,Wan2006}
though. The features of the accuracy degeneration, \ie its onset (the
instant at which PCEs start being insufficiently accurate) or its rate
(how fast the accuracy decreases), depend on the considered problem.

The cause of the decaying accuracy of PCEs in dynamics can be classified
into an approach-related cause and an inherent cause.  The
approach-related cause refers uniquely to intrusive techniques. In fact,
the latter solves a system of reformulated ordinary differential
equations which are derived from the original system of equations by
substituting PCE for the quantity of interest. At any given instant, the
PCE is truncated after $P$ terms, thus introducing a truncation error.
The latter is accumulated in time, therefore the results deteriorate
\cite{Ghosh2007}.
By means of the non-intrusive approach, one can avoid this source of
error since the responses at different instants can be examined
``independently'', which prevents the accumulation of error at later
instants provided the deterministic solver is equally accurate whatever
the realization of the input parameters. The inherent cause refers to
the fact that the problem itself demonstrates increasing complexity as
time evolves, as shown through examples in \cite{Ghosh2007, Pettit2006}.
The growing complexity makes it increasingly hard for PCEs to capture the
behavior of the system.

The growth in time of the inherent complexity of the problem is
characterized by the increasingly complicated relationship between the
output quantity and the input parameters, exhibiting important
non-linearity, abrupt changes and possibly discontinuities (see \eg
\cite{Desai2013, Witteveen2013}). It may be related to the difference in
terms of frequency and phase content of the various response time-series
obtained with distinct values of the uncertain input parameters
\cite{LeMaitre2009,Witteveen2008}. These discrepancies tend to be more
and more severe when time passes. In other words, trajectories tend to
be similar at early instants and less and less in phase in the long term
\cite{Wan2006,Wan2005}.

To alleviate this issue, Blatman and Sudret \cite{Blatman2010b} introduced adaptive sparse
PCEs that allow one to take advantage of the sparsity in the structure
of the model (if this sparse structure exists), thus extending the time
range where the computation of PCEs is tractable and the result is
sufficiently accurate. In other words, adaptive sparse PCEs may delay
the onset of the accuracy degeneration. 
{Le Ma\^{i}tre et al. \cite{LeMaitre2004} developed adaptive methods for multi-resolution
analysis, which relies on a multi-wavelet basis of compact piecewise-smooth polynomial functions.}
Lucor and Karniadakis \cite{Lucor2004} used adaptive
generalized PCEs, which consists in detecting the first-order terms with
the most important effects on the fluctuation of the response and then
building the higher-order terms that only include the selected
first-order terms.  From the same perspective, Mai and Sudret \cite{MaiUncecomp2015}
developed the hierarchical PCEs which aims at updating the set of
candidate polynomials adaptively by adding selected interaction terms
while selecting only the regressors with the most importance.
In most papers, the proposed high-order PCE approaches consist in using
assumptions to reduce the size of the high-order PCE basis or using
advanced computational techniques for computing them.

Wan and Karniadakis \cite{Wan2006, Wan2006a, Wan2005} proposed multi-element PCEs, in which
the random space is divided into multiple subspaces in such a way that
the complexity of the model in each subspace is reduced, thus requiring
only low-order PCEs. Jakeman et al. \cite{Jakeman2013} also used multi-element PCEs
with a discontinuity detector in order to minimize the number of
subspaces. Nouy \cite{Nouy2010} and Soize \cite{Soize2015} approximated a
multimodal random variable (\ie the output quantity of interest) by a
mixture of unimodal random variables, each modeled by PCEs. This
approach might help to improve the effectiveness of PCEs in the context
of dynamical systems, when the responses at late instants usually
exhibit multi-modal distributions as will be shown in the current paper
through numerical applications.  In the above approaches, the input
space is divided into subspaces according to the detected
discontinuities or dissimilarities.  One then builds a local PCE in each
subspace and combines those PCE models to obtain a global metamodel.
Therefore these approaches can be classified as local PCEs.  The use of
polynomial functions in local domains, however, requires an accurate
decomposition of the input space and will not be straightforward in
high-dimensional problems.

From a different perspective, Gerritsma et al. \cite{Gerritsma2010} proposed to compute
time-dependent PCEs by updating the polynomial chaos basis on-the-fly.
If the approximation error is excessive at a considered time instant,
the authors add to the existing set of random variables a new variable,
which is the response quantity at the previous instant. This is based on
the idea that a fixed set of random variables at the beginning of the
process is not sufficient to model the system in the long term and thus,
the set of random variables the PCEs depend on needs to be updated.
This approach can be viewed as a nested PCE model, \ie a PCE model of
another PCE model. Luchtenburg et al. \cite{Luchtenburg2014} used flow map composition,
which is in principle similar to time-dependent PCEs.  The time-history
response is composed of short-term flow maps, each modeled by PCE.  The
idea of constructing the basis on-the-fly was also applied by
Cheng et al. \cite{Cheng2013} and Choi et al. \cite{Choi2014}, who derived intrusively a system
of equations governing the evolution of the time-dependent spatial and
stochastic basis.  In the context of structural dynamics,
Spiridonakos and Chatzi \cite{Spiridonakos2015, Spiridonakos2015a} proposed the combination of
PCEs and autoregressive models which consists in representing the
response as a function of its past values. This approach is currently
investigated with the use of sparse adaptive PCEs by
Mai et al. \cite{Mai2016IJ4UQ2}. Recently, Ozen and Bal \cite{Ozen2016} introduced the
dynamical PCEs, which is also based on the idea that the future
evolution of the response depends on the present solution.

As explained earlier, the accuracy of PCEs may degenerate in time due to
the time-increasing dissimilarity between the response trajectories when
considering distinct values of the uncertain input parameters.  To
alleviate the accuracy decay, one may naturally think of increasing the
similarity between the response trajectories. For this purpose, an
attractive approach is to pre-process the response trajectories in order
to increase the similarity between them.
To this end, Witteveen et al. \cite{Desai2013, Witteveen2008} represented
the dynamic response trajectories as functions of the phase $\phi$
instead of time $t$ in order to obtain in-phase vibrations. The phases
are extracted from the observations, based on the local extrema of the
time series. The response trajectories are then transformed from
time-histories to phase-histories. PCEs are eventually applied in the
phase space.
Le Ma\^{i}tre et al. \cite{LeMaitre2009} represented the responses in a rescaled
time $\tau$ such that the dynamic responses vary in a small
neighborhood of a reference trajectory.  The time scale $\tau$ is
intrusively adjusted at each time step so that the distance between the
dynamic response and the reference solution is minimized, thus in-phase
vibrations are achieved.
{From the same perspective, Alexanderian et al. \cite{Alexanderian2012, Alexanderian2014} introduced a   
multiscale  stretching of the responses which allows an efficient PC representations
of the stochastic dynamics with non-intrusive spectral projections.}

As a summary, PCEs fail to represent long-term time-dependent system
responses because of their inherent increasing complexity. To the
authors' knowledge there is no versatile tool that helps overcome the
problem in a \emph{non-intrusive} setup. This paper aims at filling this
gap by introducing a fully non-intrusive approach that allows efficient
use of PCEs for time-dependent problems showing oscillatory behaviors.
The proposed approach relies on a \emph{stochastic time warping} and the
subsequent rescaling of the response trajectories.

The paper is organized as follows: in Section 2, the fundamentals of
PCEs for time-independent problems are recalled. We introduce so-called
\emph{time-frozen} PCEs that will be used for comparison. In Section 3,
we propose an original non-intrusive PCE approach for uncertain
dynamical systems based on \emph{stochastic time-warping}. Five
applications are finally considered to show the efficiency of the
proposed approach.
%%%%%%%%%%%%%%%%%%%%%%%%%%%%%%%%%%%%%%%%%%%%%%%%%%%%%%%%%%%%%%%%%%%%%%%%%%%%%%%%%%%%%

%%%%%%%%%%%%%%%%%%%%%%%%%%%%%%%%%%%%%%%%%%%%%%%%%%%%%%%%%%%%%%%%%%%%%%%%%%%%%%%%%%%%%
\section{Polynomial chaos expansions}
%%%%%%%%%%%%%%%%%%%%%%%%%%%%%%%%%%%%%%%%%%%%%%%%%%%%%%%%%%%%%%%%%%%%%%%%%%%%%%%%%%%%%
\subsection{Spectral representation}
Let us consider the model $Y=\cm(\veX)$ where $\veX=(X_1 \enum X_M)$ is a $M$-dimensional input vector of random variables with given joint probability density function $f_{\veX}$ defined over an underlying probability space $(\varOmega, \matcalf, \mathbb{P})$ and $\cm:\, \vex  \in \cd_X \subset \Rr^M \mapsto \Rr$ is the computational model of interest, where $\cd_{\veX}$ is the support of the distribution of $\veX$. Herein, we assume that the input random variables are independent, \ie the joint probability density function (PDF) is the product of the marginal PDFs:
\begin{equation}
 f_{\veX}(\vex)= f_{X_1}(x_1) \ldots f_{X_M}(x_M).
 \label{eq2.1}
\end{equation}
Assuming that the scalar output $Y$ is a second order random variable, \ie $\Esp{Y^2} < +\infty$, is equivalent to require that the computational model $\cm$ belongs to the Hilbert space $\ch$ of square-integrable functions with respect to the inner product:
\begin{equation}
 \innprod{u}{v}{\ch} = \int\limits_{\cd_{\veX}} u(\vex) v(\vex) f_{\veX}(\vex) \di \vex .
 \label{eq2.2}
\end{equation}
Denote by $\ch_i$ the Hilbert space of square-integrable functions with respect to the marginal probability measure $\mathbb{P}_{X_i}(\di x_i)= f_{X_i}(x_i) \di x_i$. Let us equip $\ch_i$ with an inner product:
\begin{equation}
 \innprod{u}{v}{\ch_i} = \int\limits_{\cd_{X_i}} u(x_i) v(x_i) f_{X_i}(x_i) \di x_i ,
 \label{eq2.3}
\end{equation}
where $\cd_{X_i}$ is the support of the distribution of $X_i$
and denote by $\{ \phi_k^i, k \in \Nn \}$ an orthonormal basis of $\ch_i$ which satisfies:
\begin{equation}
 \innprod{\phi_k^i}{\phi_l^i}{\ch_i} = \delta_{kl} ,
 \label{eq2.4}
\end{equation}
in which $\delta_{kl}$ is the Kronecker symbol, which is equal to 1 if $k = l$ and equal to 0 otherwise.

As shown by Soize and Ghanem \cite{Soize2004}, the Hilbert space $\ch$ is isomorphic to
the tensor product $\operatorname*{ \otimes}_{i=1}^M \ch_i $. Thus a
basis of $\ch$ may be obtained by the tensor product of the univariate
bases $\acc{\phi_k^i, k \in \Nn}, \, i = 1 \enum M$. As a consequence,
the random variable $Y = \cm(\veX)$ that results of the propagation of
the uncertainties modeled by $\veX$ through the computational model
$\cm$ may be cast as:
\begin{equation}
 Y = \sum\limits_{\alpha_1 \in \Nn} \ldots \sum\limits_{\alpha_M \in \Nn} y_{\alpha_1 \ldots \alpha_M} \phi_{\alpha_1}^1(X_1) \ldots \phi_{\alpha_M}^M(X_M) .
 \label{eq2.8}
\end{equation} 
For the sake of simplicity, introducing multi-indices $\vea = \acc{\alpha_1 \enum \alpha_M}$, $Y$ may be rewritten as:
\begin{equation}
 Y= \sum\limits_{\vea \in \Nn^{M}} y_{\vea} \ve{\phi}_{\vea}(\veX) .
 \label{eq2.9}
\end{equation}
where $\ve{\phi}_{\vea}(\veX)= \prod\limits_{i=1}^{M} \phi_{\alpha_i}^i(X_i) $ are the multivariate basis functions and $y_{\vea}$ are the associated deterministic coefficients.
%%%%%%%%%%%%%%%%%%%%%%%%%%%%%%%%%%%%%%%%%%%%%%%%%%%%%%%%%%%%%%%%%%%%%%%%%%%%%%%%%%%%%
\subsection{Polynomial chaos expansions}
The univariate basis functions $\phi_k^i, k \in \Nn, \, i = 1 \enum M$ may
be constructed using orthonormal polynomials \cite{Abramowitz} leading
to the so-called generalized polynomial chaos expansion \cite{Xiu2002,
  Soize2004}.  For instance when $X_i$ is a uniform (resp. standard
normal) random variable, the corresponding polynomial basis comprises
orthonormal Legendre (resp. Hermite) polynomials. Then Eq.~\eqref{eq2.9}
becomes:
\begin{equation}
 Y= \sum\limits_{\vea \in \Nn^{M}} y_{\vea} \ve{\psi}_{\vea}(\veX),
 \label{eq2.2.1}
\end{equation}
in which $\vea = (\alpha_1 \enum \alpha_M)$ are the multi-indices with $\alpha_i, i=1\enum M$ denoting the degree of the univariate polynomial in $X_i$ and $\ve{\psi}_{\vea}(\veX) = \prod\limits_{i=1}^{M} \psi_{\alpha_i}^i(X_i)$ are multivariate \emph{orthonormal} polynomials obtained by the tensor product of univariate polynomials. 

In practice, the use of infinite-dimensional PCEs is not tractable. One always truncates the expansion to obtain an approximate representation:
\begin{equation}
 Y = \sum\limits_{\vea \in \matcalA} y_{\vea} \ve{\psi}_{\vea}(\veX) + \epsilon ,
 \label{eq2.2.2}
\end{equation}
in which $\matcalA$ is a truncation set and $\epsilon$ is the truncation-induced error. A classical truncation scheme consists in selecting all polynomials of total degree less than or equal to $p$, when the truncation set reads:
\begin{equation}
 \matcalA^{M,p} =\{ \vea \in \Nn^M: \quad \norme{\vea}{1} \eqdef  \alpha_1 + \ldots + \alpha_M  \leqslant p \} .
\end{equation}
%%%%%%%%%%%%%%%%%%%%%%%%%%%%%%%%%%%%%%%%%%%%%%%%%%%%%%%%%%%%%%%%%%%%%%%%%%%%%%%%%%%%%
\subsection{Computation of PC coefficients and error estimation}
\label{section2.3}
The computation of the coefficients $\acc{y_{\vea}, \, \vea \in
  \matcalA}$ in Eq.~\eqref{eq2.2.2} can be conducted using intrusive
(\ie Galerkin scheme) or non-intrusive approaches (\eg projection,
regression and quadrature methods). In the following, we will compute
the coefficients of the expansions using the adaptive sparse PCE
technique proposed by Blatman and Sudret \cite{Blatman2011b} which is a non-intrusive
least-square minimization technique based on the least angle regression
algorithm \cite{Efron2004}. The reader is referred to
\cite{Blatman2011b} for more details on this approach.

The accuracy of the representation is estimated by means of the
leave-one-out (LOO) cross-validation, which allows a fair error
estimation at an affordable computational cost
\cite{Blatman2010b,BlatmanThesis}. The principle of cross validation is
to use different sets of points to build PCEs, then compute the errors
with the actual model.
Assume that one is given a sample set $\cx = \acc{\vex^{(i)}, \, i = 1 \enum n}$. The computational model $\cm$ is run for each point in $\cx$, resulting in the vector of output quantity values $\cy = \acc{y^{(i)}, \, i = 1 \enum n}$.
Setting one point $\vex^{(i)}$ apart from $\cx$, one can build a PCE model $\cm^{\text{PC}\backslash i}(\cdot)$ from the remaining points $\cx \backslash \vex^{(i)} = \acc{\vex^{(1)} \enum \vex^{(i-1)}, \vex^{(i)} \enum \vex^{(n)} }$.
The predicted residual error at point $\vex^{(i)}$ reads:
\begin{equation}
	\Delta^{(i)} \eqdef \cm(\vex^{(i)}) - \cm^{\text{PC}\backslash i}(\vex^{(i)}) .
\end{equation}
The LOO error is defined as follows:
\begin{equation}
	\widehat{\text{Err}}_{LOO} = \dfrac{1}{n} \sum\limits_{i=1}^{n} \Delta_i^2 .
\end{equation}
At first glance, one could think that evaluating the LOO error is
computationally demanding since it requires $n$~different predicted
residuals, each of them obtained from a different PCE. However, by means
of algebraic derivations, one can compute $\widehat{\text{Err}}_{LOO}$
from a \emph{single} PCE $\cm^{\text{PC}}(\cdot)$ built with the full
experimental design as follows \cite{BlatmanThesis}:
\begin{equation}
	\widehat{\text{Err}}_{LOO} =  \dfrac{1}{n} \sum\limits_{i=1}^{n} \prt{ \dfrac{ \cm(\vex^{(i)}) - \cm^{\text{PC}}(\vex^{(i)}) }{ 1 - h_i} }^2  ,
\end{equation}
where $h_i$ is the $i^{\text{th}}$ diagonal term of the projection matrix $\matA \, \prt{ \matA\tr \matA }^{-1} \matA\tr  $ and the information matrix $\matA$ is defined by
$\acc{A_{ij} = \vepsi_j(\vex^{(i)}), \, i = 1 \enum n, \, j = 1\enum \card \, \matcalA}$, \ie the $i^{\text{th}}$ row of $\matA$ is the evaluation of the polynomial basis functions at the point $\vex^{(i)}$ in the ED.
Note that in practice, a normalized version of the LOO error is used:
\begin{equation}
	\hat{\epsilon}_{LOO} = \dfrac{ \widehat{\text{Err}}_{LOO} }{ \Var{\cy}} ,
\end{equation}
where $\Var{\cy}$ is the empirical variance of the sample of outputs.
%%%%%%%%%%%%%%%%%%%%%%%%%%%%%%%%%%%%%%%%%%%%%%%%%%%%%%%%%%%%%%%%%%%%%%%%%%%%%%%%%%%%%
\subsection{Time-frozen polynomial chaos expansions}
In the context of time-dependent problems, \ie $Y(t) = \cm(\veX,t)$, the polynomial chaos representation of the response quantity reads:
\begin{equation}
 Y(t) = \sum\limits_{\vea \in \matcalA} y_{\vea}(t) \ve{\psi}_{\vea}(\veX) + \epsilon(t)
 \label{eq2.3.1}
\end{equation}
in which the notation $y_{\vea}(t)$ indicates the time-dependent
coefficients of PCEs. The representation of a time-dependent quantity by
means of PCEs as in Eq.~\eqref{eq2.3.1} is widely used in the
literature, see \eg \cite{Pettit2006,LeMaitre2009,Gerritsma2010}.  At a
given time instant~$t$, the coefficients $\acc{y_{\vea}(t),\vea \in
  \matcalA}$ and the accuracy of the PCEs are estimated by means of the
above mentioned techniques (see Section~\ref{section2.3}).  The
metamodel of the response is computed independently at each time
instant, hence the name time-frozen PCEs.

We now introduce the use of time-frozen PCEs for computing the time-dependent statistics of the response. The multivariate polynomial chaos functions are orthonormal, \ie:
\begin{equation}
	\Esp{ \ve{\psi}_{\vea}(\veX) \, \ve{\psi}_{\veb}(\veX) }  \eqdef \int\limits_{\cd_{X}} \ve{\psi}_{\vea}(\vex) \, \ve{\psi}_{\veb}(\vex) \, f_{\veX}(\vex) \, \di \vex = \delta_{\vea \veb} \; \forall \vea, \, \veb \in \Nn^M ,
\end{equation}
in which $\delta_{\vea \veb}$ is the Kronecker symbol that is equal to 1 if $\vea = \veb$ and equal to 0 otherwise. Indeed, each multivariate polynomial is orthogonal to $\ve{\psi}_{\ve{0}}(\veX) = 1$, which means
$\Esp{\ve{\psi}_{\vea}(\veX)}  = 0 \, \forall \vea \neq \ve{0}$ and $\Var{\ve{\psi}_{\vea}(\veX)} = \Esp{ \prt{\ve{\psi}_{\vea}(\veX)}^2 } = 1 \; \forall \vea \neq \ve{0}$.
Thus, the time-dependent mean and standard deviation of the response can be estimated by means of a mere post-processing of the truncated PC coefficients (in Eq.~\eqref{eq2.3.1}) with no additional cost as follows:
\begin{equation}
	\Esp{Y(t)} \approx \Esp{ \sum\limits_{\vea \in \matcalA} y_{\vea}(t) \ve{\psi}_{\vea}(\veX)} = y_{0}(t)  ,
\end{equation}

\begin{equation}
	\sigma_{Y(t)}^2 =  \Var{ Y(t) } \approx \Var{ \sum\limits_{\vea \in \matcalA} y_{\vea}(t) \ve{\psi}_{\vea}(\veX) } = \sum\limits_{\substack{{\vea \in \matcalA} \\ \vea \neq \ve{0} }} y_{\vea}^2(t).
\end{equation}
%%%%%%%%%%%%%%%%%%%%%%%%%%%%%%%%%%%%%%%%%%%%%%%%%%%%%%%%%%%%%%%%%%%%%%%%%%%%%%%%%%%%%
%%%%%%%%%%%%%%%%%%%%%%%%%%%%%%%%%%%%%%%%%%%%%%%%%%%%%%%%%%%%%%%%%%%%%%%%%%%%%%%%%%%%%
\section[Stochastic time-warping PCEs for random oscillations]{Stochastic time-warping polynomial chaos expansions for random oscillations}
\label{sec:tw_theory}
%%%%%%%%%%%%%%%%%%%%%%%%%%%%%%%%%%%%%%%%%%%%%%%%%%%%%%%%%%%%%%%%%%%%%%%%%%%%%%%%%%%%%
\subsection{Introduction}
An interesting problem emerges in nonlinear oscillating systems possessing a limit cycle\footnote{Limit cycle is a closed isolated trajectory in the phase-space of self-oscillated oscillators. The nearby trajectories can either spiral in toward or away from the limit cycle.} which may depend on the uncertain parameters.
Limit cycle oscillations (LCO) represent a class of time-dependent
problems that plays an important role in several fields, see \eg
aerospace engineering \cite{Bunton2000} and mechanical engineering
\cite{Sarrouy2013} among others. Use of PCEs to represent LCO systems
has attracted a large attention and actually almost all novel ideas with
PCEs are applied first to LCO systems or systems involving periodicity.
For instance, Wan and Karniadakis \cite{Wan2006} used multi-element PCEs whereas
Beran et al. \cite{Beran2006} proposed different methods namely use of Haar wavelets
as local bases or use of B-spline functions. These approaches aim at
resolving the highly nonlinear behavior of LCO responses in the
stochastic domain.  There are also techniques that are designed
specifically for LCO.
Le Ma\^{i}tre et al. \cite{LeMaitre2009} proposed an intrusive time transform of the
trajectories which aims at representing the transformed time-histories
in a small neighborhood of a reference trajectory, \ie to reduce their
variability by making them in-phase.  A transformed time line $\tau$ is
introduced, of which the varying clock speed $\dot \tau = \dfrac{\di
  \tau}{\di t}$ is adjusted in an intrusive setup at each time step.
This is achieved by minimizing the Euclidean distance between the
distinct trajectories and the reference counterpart.
{The use of a stochastically stretched time variable in a non-intrusive setting  
has been investigated by Alexanderian et al. \cite{Alexanderian2012, Alexanderian2014}.
The random responses were preconditioned by random scalings of the time horizon and their amplitudes. Consequently, the scaled responses exhibit similar dynamical features, more precisely they become in-phase.}
From a similar perspective, Witteveen and Bijl \cite{Witteveen2008} interpolated the oscillatory
responses on the phase space to obtain in-phase oscillations.  Inspired
by the mentioned approaches, a non-intrusive time transform, which
consists in finding a suitable \emph{stochastic warping} of the time
line to increase the \emph{similarity} between different trajectories in
the transformed (warped) time scale, is introduced in this section. The
proposed approach focuses on increasing the frequency and phase
similarity of the considered trajectories in problems involving
periodicity.

It is worth noting that in the engineering literature, the time-warping
technique has been of interest for decades. In the context of voice
recognition, Sakoe and Chiba \cite{Sakoe1978} first proposed the time-warping to
eliminate the timing differences and obtain maximum coincidences between
two speech patterns. Wang and Gasser \cite{Wang1997} introduced a novel cost function
to determine the time-warping function.  Later, Ramsay and Li \cite{Ramsay1998} used
the technique under the name ``curve registration'' for biological data.
The essential idea consists in the registration (or alignment) of
salient curve features by means of a suitable \emph{smooth monotone
  transformation} of the temporal variable $t$. The actual analyses are
then carried out on the aligned curves.  Note that the same idea can
also be conducted in the spatial domain.  For instance,
Bookstein \cite{Bookstein1997} showed particular applications of registering the
outcomes over surfaces or volumes in medical imaging.

Herein, we are adding one dimension to the time-warping technique by
incorporating the effects of uncertainties in the transformation
function. This results in a stochastic time-transform framework. Indeed,
due to the inherent randomness of the stochastic problem, a time
transformation function with deterministic parameters is not suitable.
Therefore, stochastic transform parameters must be used and will be cast
as functions of the original random parameters. The theoretical
foundation of this work was originally presented by Mai and Sudret \cite{MaiIcasp2015}.
%

%%%%%%%%%%%%%%%%%%%%%%%%%%%%%%%%%%%%%%%%%%%%%%%%%%%%%%%%%%%%%%%%%%%%%%%%%%%%%%%%%%%%%
\subsection{Stochastic time-warping polynomial chaos expansions}
Consider a dynamical system (\eg a structural dynamic or chemical system) whose behavior is modeled by a system of ordinary differential equations (ODEs):
\begin{equation}
 \frac{\di \vey}{\di t} = \ve{f}(\vey,\vexi,t) ,
% \label{eq3.1}
\end{equation}
where the initial condition is $\vey (t=0) = \vey_0$ and the random
vector $\vexi$ comprises independent second-order random variables
defined over a probability space $(\varOmega, \matcalf, \Pp)$. $\vexi$
may include the parameters governing the system behavior, \eg masses,
stiffness, damping ratio, reaction parameters, frequency and amplitude
of excitation. The initial condition can also be uncertain, in which
case it becomes a random variable belonging to $\vexi$. The
time-dependent response of the system is denoted by $\vey(t,\vexi)$.
Without loss of generality, we consider {a generic response of the 
uncertain dynamical system}, \eg $y(t,\vexi)$ with the initial condition $y(t=0) = y_0$. 
At each time instant, $y(t,\vexi)$ is assumed to be a second-order random
variable. As in \cite{LeMaitre2009,Wan2006, Wan2005,Witteveen2008},
herein we focus on the class of problems when $y(t,\vexi)$ is an
oscillatory response with random frequencies and amplitudes.

The time-dependent response $y(t,\vexi)$ is represented by time-frozen PCEs as:
\begin{equation}
 y(t, \vexi) = \sum\limits_{\vea \in \matcalA} y_{\vea}(t) \ve{\psi}_{\vea}(\vexi) + \epsilon(t).
 \label{eq:timefroPCE}
\end{equation}
A virtual time variable $\tau(t,\vexi)$, which is obtained by a stochastic time-warping, is introduced as follows:
\begin{equation}
	\tau(t,\vexi) = \sum\limits_{i=1}^{N_{\tau}} c_i(\vexi) \, f_i(t) = F(t, \vexi),
	\label{eq3.3}
\end{equation}
where $\acc{f_i(t), i = 1 \enum N_{\tau}}$ are functions of time $t$ and $\acc{c_i(\vexi), i = 1 \enum N_{\tau}}$ are coefficients which depend on the input random variables $\vexi$. The coefficients $c_i(\vexi)$ can be represented by PCEs as:
\begin{equation}
	c_i(\vexi) = \sum\limits_{\vea \in \Nn^M} {c_i}_{\vea} \, \ve{\psi}_{\vea}(\vexi),
	\label{eq3.4}
\end{equation}
where $\ve{\psi}_{\vea}(\vexi)$ and ${c_i}_{\vea}$ are respectively the orthonormal polynomial functions and the coefficients of the expansion.
The only constraint on the time-warping is that $\tau$ is a strictly monotonically increasing function of $t$ {given a random set of parameters $\vexi$. This constraint ensures that there is no repeated value on the virtual time line}. Then the inverse transform may be cast as: 
\begin{equation}
	t(\tau,\vexi) = F^{-1}(\tau, \vexi).
\end{equation}
Note that, in the sequel, linear transform of the form:
\begin{equation}
	\tau(t,\vexi) = k(\vexi) \, t + \phi(\vexi)
\end{equation}
is considered.
For each realization $\vexi_0$, \ie each trajectory of the system response, we assume a one-to-one mapping between $t$ and $\tau$. The response trajectory may then be represented in the transformed (warped) time scale by:
\begin{equation}
 y(\tau,\vexi) = \sum\limits_{\veb \in \cb} y_{\veb}(\tau) \ve{\psi}_{\veb}(\vexi) + \epsilon(\tau),
 \label{eq:timewarpPCE}
\end{equation}
in which $\cb$ is the truncation set of the multi-indices $\veb$. The inverse time transform allows one to obtain the PCEs of the response in the physical time scale as follows:
\begin{equation}
	y(t,\vexi) = y(F^{-1}(\tau,\vexi),\vexi).
\end{equation}
{It is worth remarking that for complex problems involving a complex
  time transform, the inversion may be delicate to evaluate, as
  discussed by Alexanderian et al. \cite{Alexanderian2012,
    Alexanderian2014}. In practice, one shall make sure that all
  realizations are sampled over sufficiently long time horizons (in the
  original time scale) so that their counterparts in the transformed
  time scale are properly defined over the time interval of interest.}

\begin{figure}[!ht]
	\centering
	\includegraphics[width=0.5\linewidth]{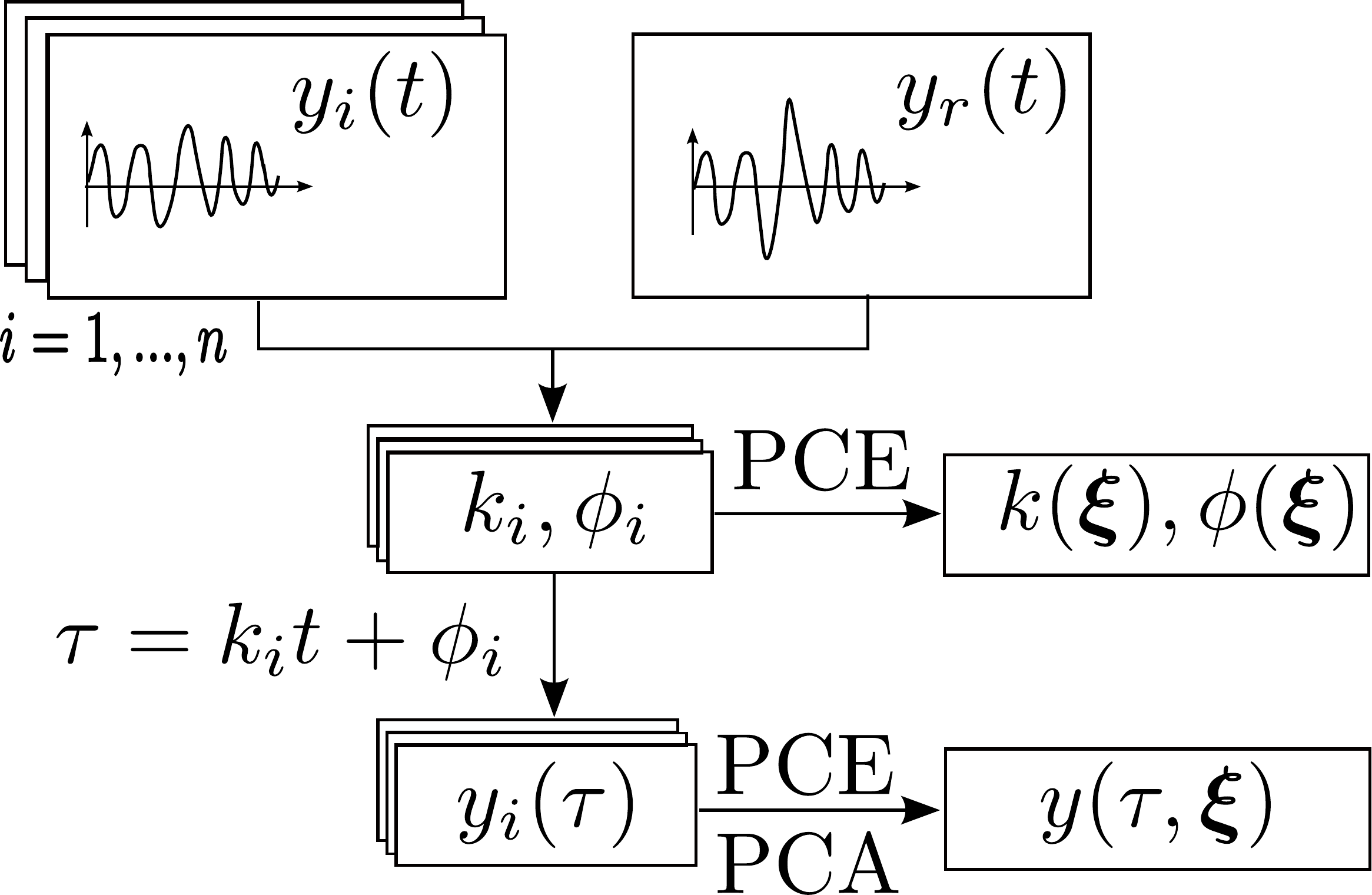}
	\caption{Stochastic time-warping approach: computation of PCEs}
	\label{fig3.1.0}
\end{figure}

The objective is to find a suitable time-warping defined by
Eq.~\eqref{eq3.3} and \eqref{eq3.4} so that the cardinality of $\cb$
remains small (\ie low-degree PCEs can be used) to achieve an acceptable
error $\epsilon(\tau)$ even at late instants. This can be obtained if
the trajectories $y(\tau(t,\vexi))$ become in-phase, as suggested by Le Ma\^{i}tre et al.
\cite{LeMaitre2009} and Witteveen and Bijl \cite{Witteveen2008}.
First, a deterministic reference trajectory $y_r(t)$ is introduced. The
stochastic time-warping (Eq.~\eqref{eq3.3}) is determined by maximizing
the similarity between $y(\tau(t,\vexi))$ and the reference counterpart
$y_r(t)$ for all values of $\vexi$, which makes the responses become
in-phase. This allows the effective computation of
Eq.~\eqref{eq:timewarpPCE}.  Having at hand the time-warping
(Eq.~\eqref{eq3.3}) and the PCEs of the response in the virtual time
line $\tau$ (Eq.~\eqref{eq:timewarpPCE}), one can finally obtain the
PCEs in the physical time line $t$ by conducting the inverse
time-warping.
The proposed non-intrusive time-warping approach is explained in detail in the following. For the sake of clarity, it is graphically summarized in \figref{fig3.1.0}.

% % % % % % % % % % % % % % % % % % % % % % % % % % % % % % % % % % %
\begin{itemize}
\item One first chooses a reference trajectory $y_r(t)$ which is for instance obtained by considering the mean values of the input vector $\vexi$, \ie $y_r(t)=y(t,\Esp{\vexi})$. In general, $y_r(t)$ may be any realization of the response quantity $y(t)$ obtained with a specific sample $\vexi_0$. For the numerical case studies considered in the current chapter, the choice of $y_r(t)$ did not affect the accuracy of the final results.
\item Let us start now the time-warping, which consists in transforming the time line with the purpose of increasing the similarity between different realizations of the output $y(t,\vexi)$. Assume that one is given a set of trajectories $y_i(t) \equiv y(t,\vexi_i), \, i=1\enum n$ for $n$ realizations of $\vexi$ corresponding to an experimental design in the input space $\cd_{\ve{\varXi}}$. 
Then for the realization $\#i$, $i = 1 \enum n$, the following steps are performed:
\begin{itemize}
\item Define a linear time-warping $\tau = k_i \, t + \phi_i$. In
  general, the functions $f_i(t)$ in Eq.~\eqref{eq3.3} might be
  polynomials of $t$. However, when investigating the problem of
  vibration with random frequencies, a linear transform usually
  suffices. This is due to the periodicity of the considered response
  trajectories.  In the intrusive time transform approach
  \cite{LeMaitre2009}, although a linear warping function is not
  specified for the considered examples, the resulting transformed time
  $\tau$ eventually represents a linear relationship when plotted
  against $t$. {Alexanderian et al. \cite{Alexanderian2012, Alexanderian2014}
  investigated linear-based stretching of the time variable.}
  Wang and Gasser \cite{Wang1997} also used a linear
  warping function. In particular, given the complexity of the problems
  under investigation, use of a linear function facilitates the inverse
  transform in the next phase, which is highly convenient. This linear
  warping represents two actions, namely scaling and shifting,
  respectively driven by the parameters $k_i$ and $\phi_i$. The time
  line is stretched (resp. compressed) when $k_i >1$ (resp. $0<k_i < 1$)
  and is shifted to the left (resp. to the right) when $\phi_i<0$ (resp.
  $\phi_i>0$). In fact, the scaling factor $k_i$ (resp. shifting factor
  $\phi_i$) allows to maximize the similarity in frequency (resp. phase)
  between the considered trajectories.
\item Determine the parameters $(k_i, \, \phi_i)$ governing the time-warping as the solution of an optimization problem which aims at maximizing the similarity between the response trajectory $y_i(k_i \, t + \phi_i)$ and the reference counterpart $y_r(t)$. The details of the optimization problem, in which a measure of similarity is introduced, will be described in Section~\ref{sec:determine_k_phi}.
\item Represent $y_i(t)$ on the transformed time line $\tau$. For this purpose, one chooses a grid line of $\tau$ with the desired time interval. In fact, the finer the grid is, the smaller is the error introduced by the \emph{interpolation}. The trajectory $y_i(t)$ is projected onto $\tau_i = k_i \, t + \phi_i$ to obtain $y_i(\tau_i)$. 
In order to assure that all transformed time lines $\tau_i$ start at $0$, when $t \leq t_0$, one uses the following transform $\tau_i = \dfrac{k_i \, t_0 + \phi_i}{t_0} \, t$. The small value $t_0$ is chosen so that $k_i \, t_0 + \phi_i > 0 \quad \forall i = 1 \enum n $. For instance, $t_0 = 0.2$~s is used for the numerical applications that follow. 
Finally the projected trajectory is linearly \emph{interpolated} on the selected time line $\tau$ yielding $y_i(\tau)$.
\end{itemize}
\item One builds PCEs of $k(\vexi)$, $\phi(\vexi)$ and $y(\tau,\vexi)$ using the realizations $\{k_i, \phi_i, y_i(\tau)\}$, $i=1 \enum n$ as the experimental design (or training set):
\begin{equation}
	k(\vexi) = \sum\limits_{\vegamma \in \cg} k_{\vegamma} \, \ve{\psi}_{\vegamma}(\vexi) + \epsilon_k ,
	\label{eq:k_vs_xi}
\end{equation}
\begin{equation}
	\phi(\vexi) = \sum\limits_{\vetheta \in \ct} \phi_{\vetheta} \, \ve{\psi}_{\vetheta}(\vexi) + \epsilon_{\phi},
	\label{eq3.8}
\end{equation}
\begin{equation}
 y(\tau,\vexi) = \sum\limits_{\veb \in \cb} y_{\veb}(\tau) \ve{\psi}_{\veb}(\vexi) + \epsilon_{y}(\tau).
 \label{eq3.9}
\end{equation}
In the above equations, $\vegamma$, $\vetheta$ and $\veb$ are
multi-indices belonging to the truncation sets $\cg$, $\ct$ and $\cb$ of
the expansions.  $k_{\vegamma}$, $\phi_{\vetheta}$ and $y_{\veb}(\tau)$
are coefficients computed by means of sparse adaptive PCEs
\cite{Blatman2011b}.  $k(\vexi)$ and $\phi(\vexi)$ are scalar
quantities, therefore the computation of their PCE models is
straightforward. However, for the vector-valued response
$y(\tau,\vexi)$, it might be computationally expensive when the number
of discretization points of the $\tau$-line is large. This computational
cost can be reduced significantly by coupling PCEs with the principal
component analysis \cite{BlatmanIcossar2013}. The combination of PCA and
PCEs will be described in detail in Section \ref{sec:pcapce}.
\end{itemize}
%%%%%%%%%%%%%%%%%%%%%%%%%%%%%%%%%%%%%%%%%%%%%%%%%%%%%%%%%%%%%%%%%%%%%%%%%%%%%%%%%%%%%
\subsection{Determination of time-warping parameters}
\label{sec:determine_k_phi}
% % Uniqueness of solution
This section describes the optimization problem used for determining the parameters $k$ and $\phi$ of the time-warping process.
We first propose a function to measure the similarity between two trajectories $y_1(t)$ and $y_2(t)$:
\begin{equation}
	g(y_1(t),y_2(t))= \dfrac{\abs{ \int\limits_{0}^T y_1(t) y_2(t) \di t}}{\| y_1(t) \| \| y_2(t) \| },
	\label{eq29}
\end{equation}
in which $\int\limits_{0}^T y_1(t) y_2(t) \di t$ is the inner product of
the two considered time histories and $\| \cdot \|$ is the associated
$L^2$-norm. In practice, the trajectories are discretized and thus, the
inner product (resp. the $L^2$-norm) becomes the classical dot product
between two vectors (resp. the Euclidean norm). By the Cauchy-Schwarz
inequality, this similarity measure always takes values in the interval
$[0,1]$. {For responses of limit cycle oscillation systems which
  feature a dominant frequency, the proposed similarity measure} attains
its maximum when the considered trajectories have the same frequency and
phase content. {In the following, constraints on the parameters will
  be imposed so that the solution of the optimization problem is
  unique.}

The parameters $(k_i, \, \phi_i), i = 1 \enum n$ are determined as the maximizers of the similarity measure between $y_i(\tau)$ and $y_r(t)$. The objective function reads:
\begin{equation}
 g(k_i,\phi_i) = \dfrac{\abs{ \int\limits_{0}^T y_i(k_i\,t+\phi_i) y_r(t) \di t}}{\| y_i(k_i\,t+\phi_i) \| \| y_r(t) \| } .
 \label{eq:tw_objfunc}
\end{equation}
Note that the optimal warping parameters $(k_i, \, \phi_i)$ are different for each trajectory. This results in varying total durations of the trajectories after the warping process. 
This also occurred in the intrusive time transform approach \cite[Figure 4]{LeMaitre2009}.
The objective function is therefore computed on the overlapped duration between the warped trajectory and the reference one.

Let us now examine the solution $(k_i,\phi_i)$ of the proposed optimization problem. 
The constraint that $\tau$ is a strictly monotonically increasing function of $t$ requires that $k_i >0$.
In case $y_r(t)$ and $y_i(t,\vexi_i)$ are both monochromatic signals, the value of $k_i$ that maximizes their similarity in frequency is unique.
However, there are multiple values for the shifting factor $\phi$ that make the considered trajectories in phase. This will be investigated in the next paragraph.

\begin{figure}[!ht]
	\centering
	\includegraphics[width=0.5\linewidth]{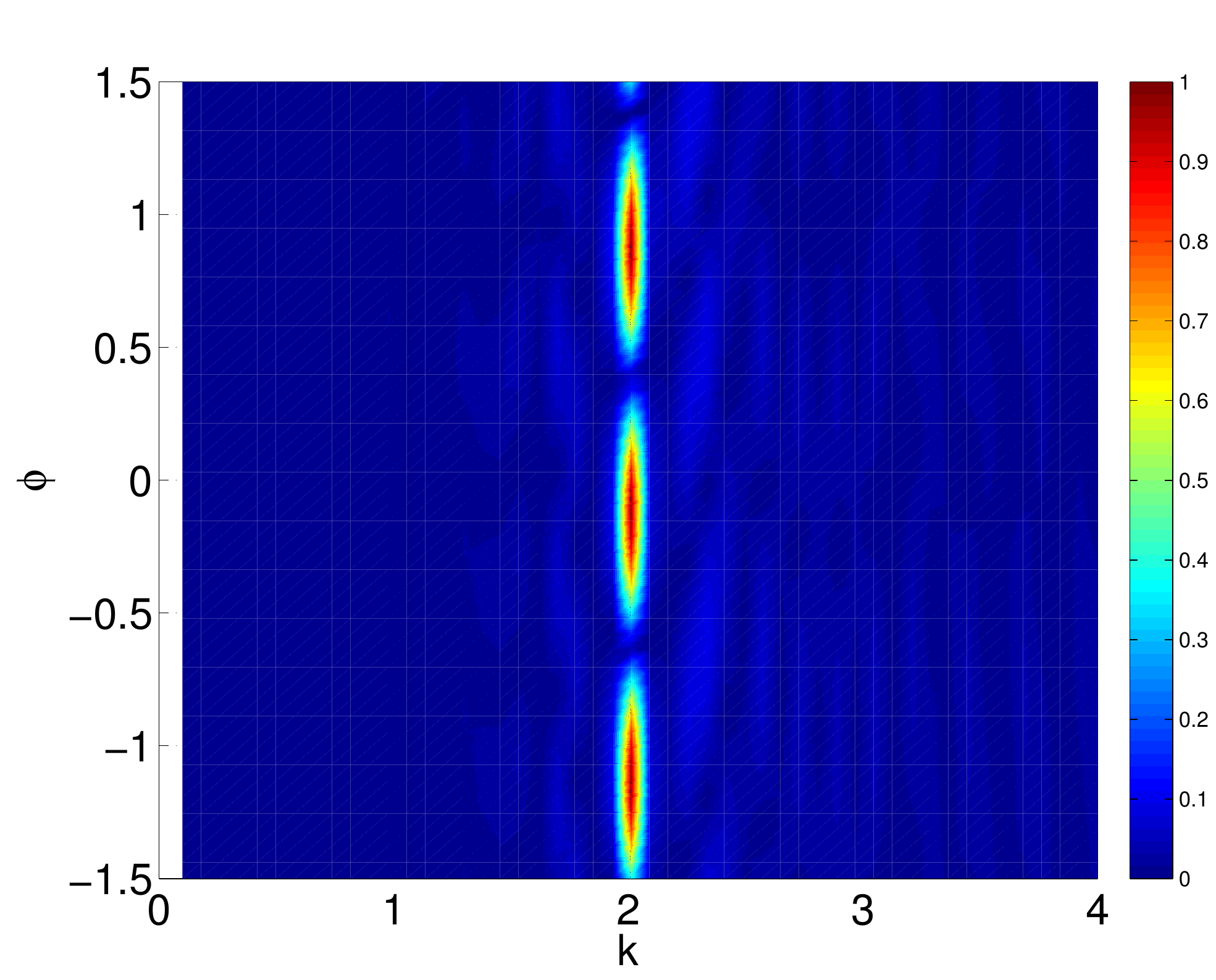}
	\caption{Similarity measure as a function of $k$ and $\phi$}
	\label{fig3.2}
\end{figure}

\figref{fig3.2} depicts the objective function $g(k,\phi)$ as a similarity measure between the reference trajectory $y_r(t)=\sin(\pi \, t)$ and a response $y(t)=\sin(2 \, \pi \,t)$. The two trajectories are chosen in such a way that $(k, \phi)=(2,0)$ is the maximizer of $g(k,\phi)$. However, there are three global maxima in the depicted interval $[-1.5,\,1.5]$ of $\phi$. This is due to the fact that in the virtual time line $\tau$, if the transformed trajectory $y(\tau)$ is shifted (whether to the left or to the right) a distance equal to one half of the period $T_r = 2~s$ of the reference counterpart, the similarity measure reaches another global maximum.
In fact, if $T_r/4 \leq \phi \leq T_r/2$ (resp. $-T_r/2 \leq \phi \leq -T_r/4$) maximizes the similarity measure, then $\phi-T_r/2$ (resp. $\phi+T_r/2$) in the interval $[-T_r/4,\,T_r/4]$ is also a maximizer.
In addition, for the sake of simplicity, it is preferable that $\phi$ is as close to $0$ as possible, \ie the time line of the scaled trajectory is shifted as least as possible.
Therefore, the selected value of $\phi$ needs to satisfy the condition that the shifted distance (in time) is not larger than $1/4$ of the period $T_r$ of the reference trajectory $y_r(t)$, \ie $\abs{\phi}\leq T_r/4$.  This constraint ensures that the solution is unique. By adopting the constraint on $\phi$, one finds the solution $(k, \, \phi) = (2,\,0)$ for the considered example. 

Finally, one can set up the global optimization problem for determining the time-warping parameters as follows:
\begin{equation}
 (k_i,\phi_i) = \arg \max\limits_{\substack{k_i \in \Rr^+ \\ \abs{\phi_i}\leqslant T_r/4}} g(k_i,\phi_i) .
% \label{eq3.12}
\end{equation}
This problem can be solved by means of global optimization methods.

{At this point, it is worth noting that the similarity metric
  (Eq.~\eqref{eq29}) is computed over the time horizon of the simulation
  under consideration. Therefore, the parameters of the time transform
  are optimized with respect to this metric, \ie in a time-average
  sense. It might be of interest to optimize the parameters on shorter
  time horizons. This has been recently investigated by Yaghoubi et al.
  \cite{Yaghoubi2016} in which a {\em multi-linear} transform is
  conducted in the frequency domain. The problem consists in determining
  the break-points between the various intervals. In other cases, one
  might also consider a bilinear time transform, namely for the
  transient (resp. the stationary phase) of the response trajectories.}

%%%%%%%%%%%%%%%%%%%%%%%%%%%%%%%%%%%%%%%%%%%%%%%%%%%%%%%%%%%%%%%%%%%%%%%%%%%%%%%%%%%%%
\subsection{Principal component analysis and time-warping polynomial chaos expansions}
\label{sec:pcapce}
The instant-wise application of PCEs to model the response in the
transformed time line (\eqrefe{eq3.9}) might lead to an important
computational burden when the discretized vector $\tau$ is of large
length. To overcome this issue, Blatman and Sudret \cite{BlatmanIcossar2013} proposed a
two-step approach which combines principal component analysis (PCA) and
PCEs.  The first step consists in conducting PCA to capture the
stochastic features of the random vector-valued response with a small
number of deterministic principal components and the associated
non-physical random variables. The second step relies on representing
the resulting random variables with adaptive sparse PCEs.

Consider a sample set of the response trajectories $\cy = \acc{
  y^{(1)}(\tau) \enum y^{(n)}(\tau) }$ represented at the discretized
points $\acc{\tau_1 \enum \tau_{K}}$ in the transformed time line.
By stacking up the discretized responses, one obtains a matrix of
trajectories of size $n \times K$ denoted by $\mat{Y}$. The response can
be represented by PCA as follows:
\begin{equation}
	{\vey}(\tau,\vexi) = \bar{\vey}(\tau) + \sum\limits_{i=1}^{K} A_i(\vexi) \, \tilde{\vev}_i(\tau), 
\end{equation}
where $\bar{\vey}(\tau)$ is the empirical mean vector, $\tilde{\vev}_i(\tau)$ is an empirical eigenvector determined with $\mat{Y}$ and $A_i(\vexi)$ is a finite variance random variable. 
Only a few eigenvectors are retained in the decomposition, which leads to:
\begin{equation}
	{\vey}(\tau,\vexi) = \bar{\vey}(\tau) + \sum\limits_{i=1}^{K'} A_i(\vexi) \, \tilde{\vev}_i(\tau) + \epsilon_1(\tau).
\end{equation}
The number of principal components is selected so that the relative error $1 - \dfrac{ \sum\limits_{i= 1}^{K'} \lambda_i }{ \sum\limits_{i= 1}^{K} \lambda_i}$ is smaller than a prescribed threshold, \eg $\epsilon = 0.01$.
The samples of the random coefficient $A_i(\vexi)$ can be obtained using $\vealetter_i = (\mat{Y} - \bar{\mat{Y}}) \, \tilde{\vev}_i$ with $\bar{\mat{Y}} = \acc{\bar{\vey}(\tau) \enum \bar{\vey}(\tau) } $ being a $n \times K$ matrix obtained by replicating $n$ times the empirical mean $\bar{\vey}(\tau)$. The computed samples of $A_i(\vexi)$ are then used as the experimental design to compute the PCE of this random coefficient:
\begin{equation}
	A_i(\vexi) = \sum\limits_{\vea \in \ca} c_{i, \vea} \, \vepsi_{\vea}(\vexi) + \epsilon_{2,i}.
\end{equation}
Finally, the response in the transformed time scale is represented by coupling PCA and PCEs as follows:
\begin{equation}
	{\vey}(\tau,\vexi) = \bar{\vey}(\tau) + \sum\limits_{i=1}^{K'} \sum\limits_{\vea \in \ca} c_{i, \vea} \, \vepsi_{\vea}(\vexi) \, \tilde{\vev}_i(\tau) + \epsilon(\tau).
	\label{eq:PCAPCEtrares}
\end{equation}
Note that Blatman and Sudret \cite{BlatmanIcossar2013} introduced a
measure of the \emph{upper bound} of the total error induced by the
truncation of the principal component analysis and the approximation of
the random coefficients $A_i(\vexi)$ by PCEs. The reader is referred to
the mentioned publication for more details. Herein this error measure
can be used as an indicator of the accuracy of the computed surrogate
models.
%%%%%%%%%%%%%%%%%%%%%%%%%%%%%%%%%%%%%%%%%%%%%%%%%%%%%%%%%%%%%%%%%%%%%%%%%%%%%%%%%%%%%
\subsection{Predicting random oscillations with time-warping polynomial \\chaos expansions}

\begin{figure}[!ht]
	\centering
	\includegraphics[width=0.25\linewidth]{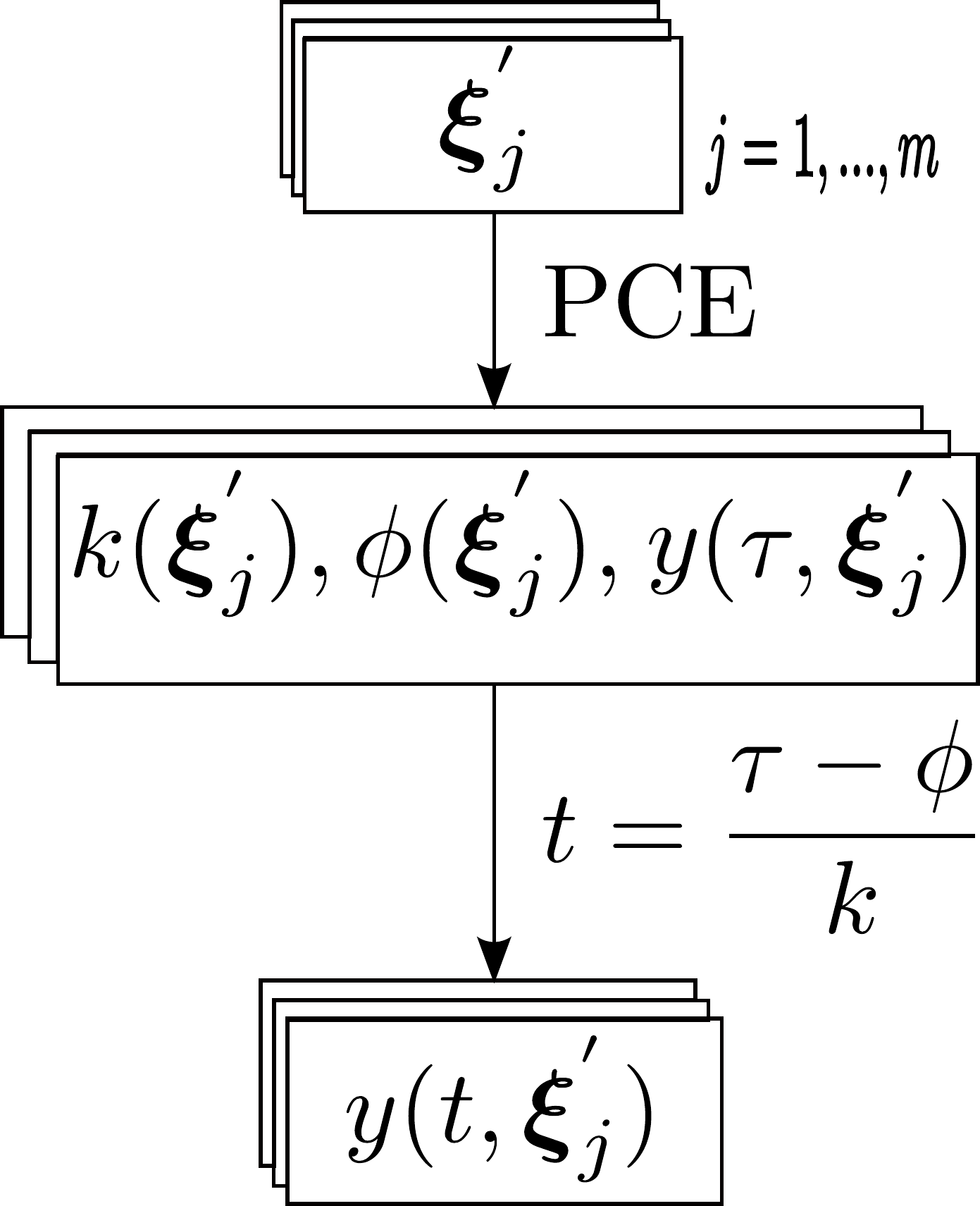}
%	}
	\caption{Stochastic time-warping approach: prediction of the response trajectories using PCEs}
	\label{fig:tw_predict}
\end{figure}

Let us now demonstrate the use of time-warping PCEs to predict responses of the model given a new set of input parameters $\vexi'$. For the sake of clarity, the procedure is depicted in \figref{fig:tw_predict} and explained in two steps as follows:
\begin{itemize}
\item First, one predicts $k(\vexi')$, $\phi(\vexi')$ and $y(\tau, \vexi')$ using the computed PCEs in equations \eqref{eq:k_vs_xi}, \eqref{eq3.8} and \eqref{eq:PCAPCEtrares}.
\item Second, one maps $y(\tau, \vexi')$ into $y(t, \vexi')$ using the inverse time-warping $t=\dfrac{\tau-\phi(\vexi')}{k(\vexi')}$. To this end, the discretized trajectory in the warped time $\acc{y(\tau_1, \vexi') \enum y(\tau_K, \vexi') }$ is attached to the real time instants $t_1 = \dfrac{\tau_1-\phi(\vexi')}{k(\vexi')}$ $\enum t_K = \dfrac{\tau_K-\phi(\vexi')}{k(\vexi')} $
\end{itemize}

%%%%%%%%%%%%%%%%%%%%%%%%%%%%%%%%%%%%%%%%%%%%%%%%%%%%%%%%%%%%%%%%%%%%%%%%%%%%%%%%%%%%%
%%%%%%%%%%%%%%%%%%%%%%%%%%%%%%%%%%%%%%%%%%%%%%%%%%%%%%%%%%%%%%%%%%%%%%%%%%%%%%%%%%%%%
\section{Numerical applications}
\label{sec:tw_applications}
Time-warping-based polynomial chaos expansions (PCEs) developed in Section~\ref{sec:tw_theory}
are now applied to five engineering problems, namely a model of rigid body dynamics, the Kraichnan-Orszag three-mode model, the so-called Oregonator model describing the chemical reaction between three species and a Bouc-Wen oscillator subject to a stochastic sinusoidal excitation. {The vibration of a nonlinear Duffing oscillator is investigated in the Supplement.}
In each case, time-frozen sparse adaptive PCEs\footnote{The term ``time-frozen sparse adaptive PCEs'' refers to the instantaneous computation of sparse adaptive PCEs.} are applied first to show the degradation of the prediction accuracy after a certain time. Time-warping PCEs with simple linear time transforms are then investigated.
%It is then shown how specific realizations of the trajectories as well as their first and second order statistics are accurately computed using the time-warping PCE approach.
%
The PCE surrogate models are computed using a small number of numerical simulations of the original model as experimental design, then validated on a large independent validation set of size $N_{val} = 10,000$. The accuracy of the time-frozen and time-warping PCE models are judged on the basis of predicting the responses to specific values of input parameters and estimating the time histories of first- and second-order statistics of the responses.

The accuracy of the prediction $\# i$ is indicated by the relative error, {\ie the mean of squared  error normalized by the variance of the response time series,} which reads:
\begin{equation}
	\epsilon_{val, i} = \dfrac{ \sum\limits_{t=1}^{K} (y(t, \vexi_i) - \hat{y}(t, \vexi_i))^2 }{\sum\limits_{t=1}^{K} (y(t, \vexi_i) - \bar{y}(t, \vexi_i) )^2} ,
%	\label{eq4.4.2}
\end{equation}
where $\hat{y}(t, \vexi_i)$ is the output trajectory predicted by PCEs
and {$\bar{y}(t, \vexi_i) = \dfrac{1}{K} \sum\limits_{t=1}^{K} y(t,
  \vexi_i) $} is the mean value of the actual response time series
$y(t,\vexi_i)$ which is obtained from the original numerical solver.
The above formula is also used to assess the accuracy of the predicted
time-dependent statistics (\ie mean and standard deviation).

These problems are solved in the UQLab framework \cite{Marelli2014},
more specifically using the least angle regression algorithm implemented
in the polynomial chaos expansion module (Marelli and Sudret \cite{UQdoc_09_104}).

%%%%%%%%%%%%%%%%%%%%%%%%%%%%%%%%%%%%%%%%%%%%%%%%%%%%%%%%%%%%%%%%%%%%%%%%%%%%%%%%%%%%%
\subsection{Rigid body dynamics}
\label{ex1}
We first consider the rotation of a rigid body described by Euler's
equations \cite{Peraire2009}.  The conservation of angular momentum
reads:
\begin{equation}
 \left\{
 \begin{array}{l}
  M_x = I_{xx} \, \dot{x} - (I_{yy}-I_{zz}) \, y \, z , \\
  M_y = I_{yy} \, \dot{y} - (I_{zz}-I_{xx}) \, z \, x , \\
  M_z = I_{zz} \, \dot{z} - (I_{xx}-I_{yy}) \, x \, y ,
  \end{array}
 \right.
 \label{eq4.1.1}
\end{equation}
in which $M_x$, $M_y$, $M_z$ are the external moments, $I_{xx}$, $I_{yy}$, $I_{zz}$ are the moments of inertia and $x$, $y$, $z$ are the angular velocities about the principal axes.
In the case when the rigid body rotates freely under no external excitation, \ie $M_x=M_y=M_z=0$
and $I_{xx}=\dfrac{1-\xi}{2} I_{yy}$, $I_{zz}=\dfrac{1+\xi}{2} I_{yy}$, one obtains the following set of reduced equations:
  \begin{equation}
 \left\{          
 \begin{array}{l }
  \dot{x}(t) = y(t) \, z(t) ,\\
  \dot{y}(t) = \xi \, x(t) \, z(t) ,\\
  \dot{z}(t) = -x(t) \, y(t) .
 \end{array}
 \right.
 \label{eq:rigbody}
\end{equation}
The initial conditions are set equal to $x(0)=0$, $y(0)=1$, $z(0)=1$.
Assume that $\xi$ is modeled by a random variable with uniform distribution: $\xi \sim \cu(-1,1)$. Suppose a solver of the coupled ODEs is available. For any realization of $\xi$, this solver provides discretized trajectories $\acc{ \acc{x(t_i), y(t_i), z(t_i)}, t_i = 0 , \Delta_t \enum K \, \Delta_t \equiv T }$. In this example, the equations are solved using the Matlab ordinary differential equation solver \texttt{ode45} (Runge-Kutta method, total duration $T= 50~s$, time step $\Delta_t = 0.01$). We aim at building PCEs of the angular velocity $x(t)$ as a function of the random variable $\xi$. Note that the corresponding polynomial functions are from the family of orthonormal Legendre polynomials since $\xi$ is uniformly distributed.

\begin{figure}[!ht]
	\centering
	\includegraphics[width=0.45\linewidth]{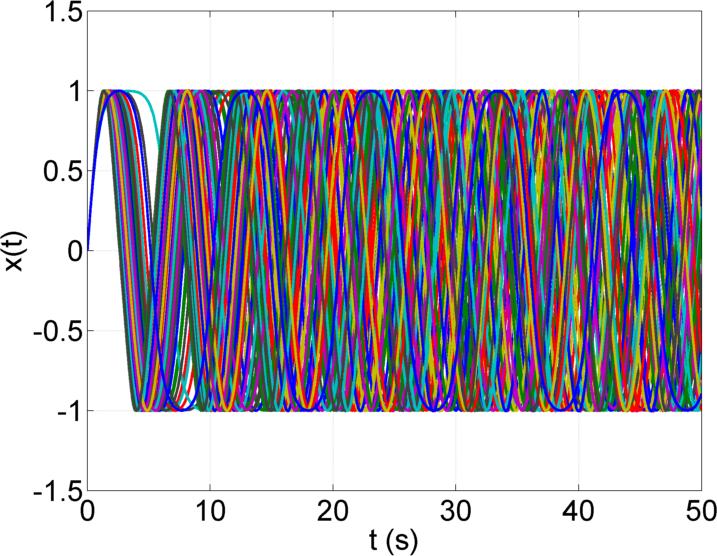}
	\caption[Rigid body dynamics -- Different trajectories $x(t)$ in the original time scale $t$]{Rigid body dynamics -- $N=50$ different trajectories $x(t)$ in the original time scale $t$.}
	\label{fig4.1.2}
\end{figure}

\begin{figure}[!ht]
 	\centering
 	\subfigure[$x(t_o)$ as a function of $\xi$ at instants $t_o = 5,\,10,\,15,\,20$~s]{
 	\includegraphics[width=0.45\linewidth]{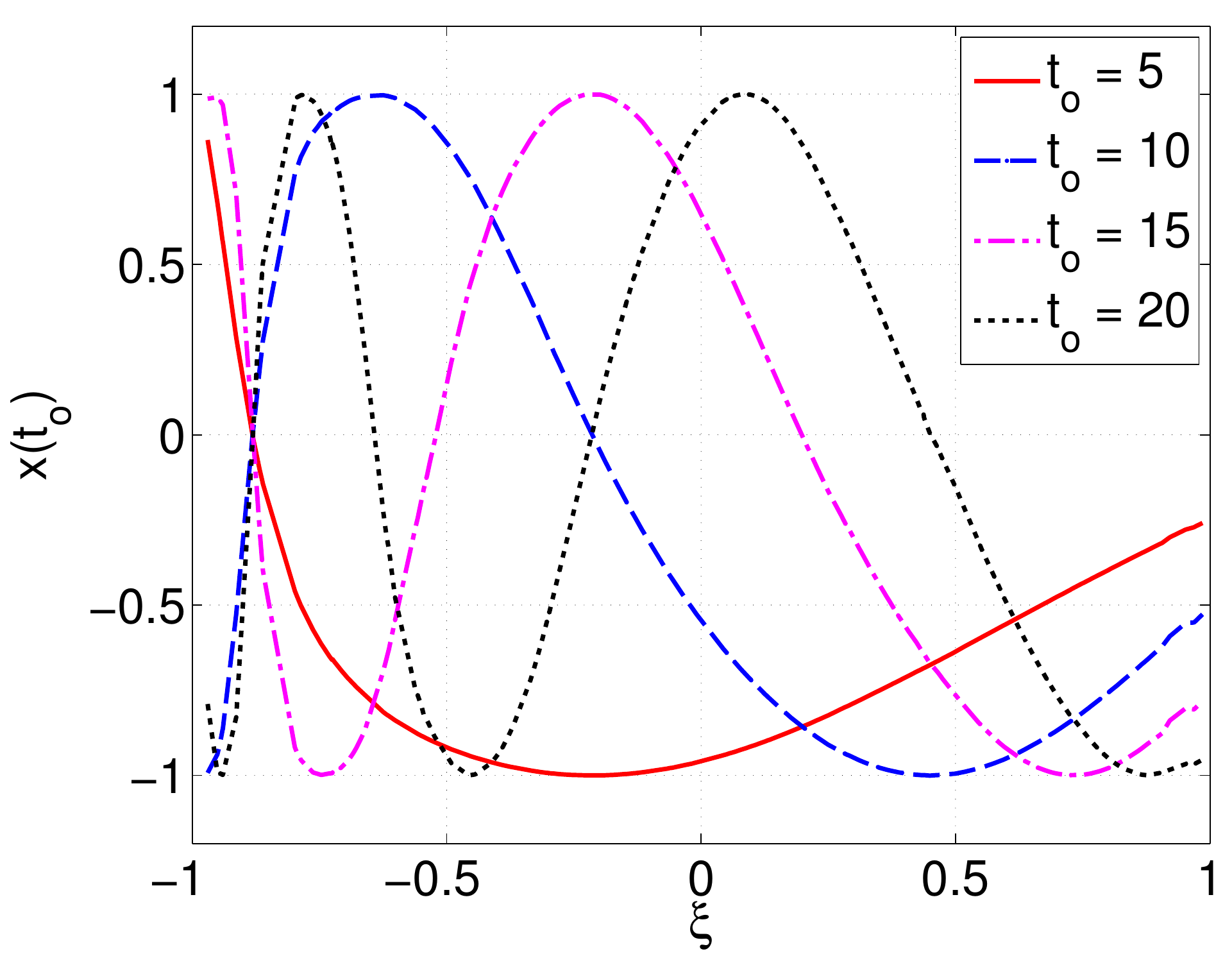}
 	\label{fig4.1.3a}
 	}
 	\subfigure[Probability density function of $x(t_o)$ at $t_o = 5,\,10,\,15,\,20$~s]{
 	\includegraphics[width=0.45\linewidth]{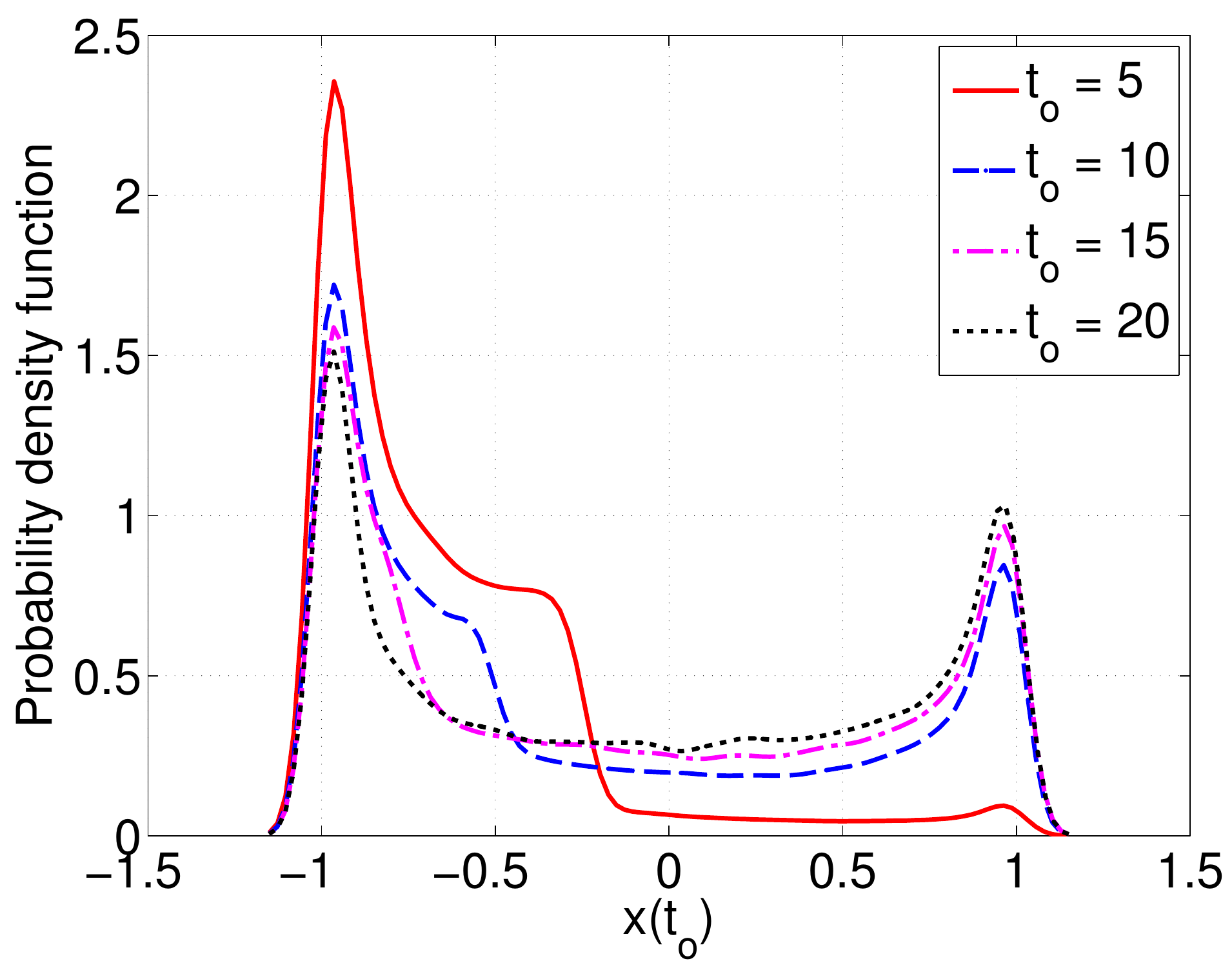}
 	\label{fig4.1.3b}
 	}
 	\caption{Rigid body dynamics -- $x(t,\xi)$ as a function of $\xi$ for particular instants and its probability density function.}
	\label{fig4.1.3}
\end{figure}

\figref{fig4.1.2} depicts a set of 50 trajectories of $x(t)$ obtained
for different realizations of the random variable $\xi$. This set is
used as the experimental design for fitting the time-frozen PCEs. $x(t)$
are oscillatory trajectories which fluctuate around zero at different
frequencies. This is a typical example of the problem of stochastic
oscillation with uncertain frequencies \cite{Wan2005, Wan2006}. At the
early instants ($t<10~s$), one can differentiate between the distinct
trajectories, whereas this is hardly the case at later instants, since
the patterns are mixed up completely. Due to the growing difference in
frequency and phase, $x(t,\xi)$ is more and more non-linear as a
function of $\xi$ for increasing $t$ (\figref{fig4.1.3a}). Subsequently,
the probability density function of $x(t)$ becomes bi-modal at late
instants (\figref{fig4.1.3b}). This explains why increasing-degree
time-frozen PCEs are required in order to represent $x(t)$ properly. As
analyzed previously, this is not a sustainable approach since the
required degree of PCEs will certainly become too high at some point.

\begin{figure}[!ht]
	\centering
	\includegraphics[width=0.45\linewidth]{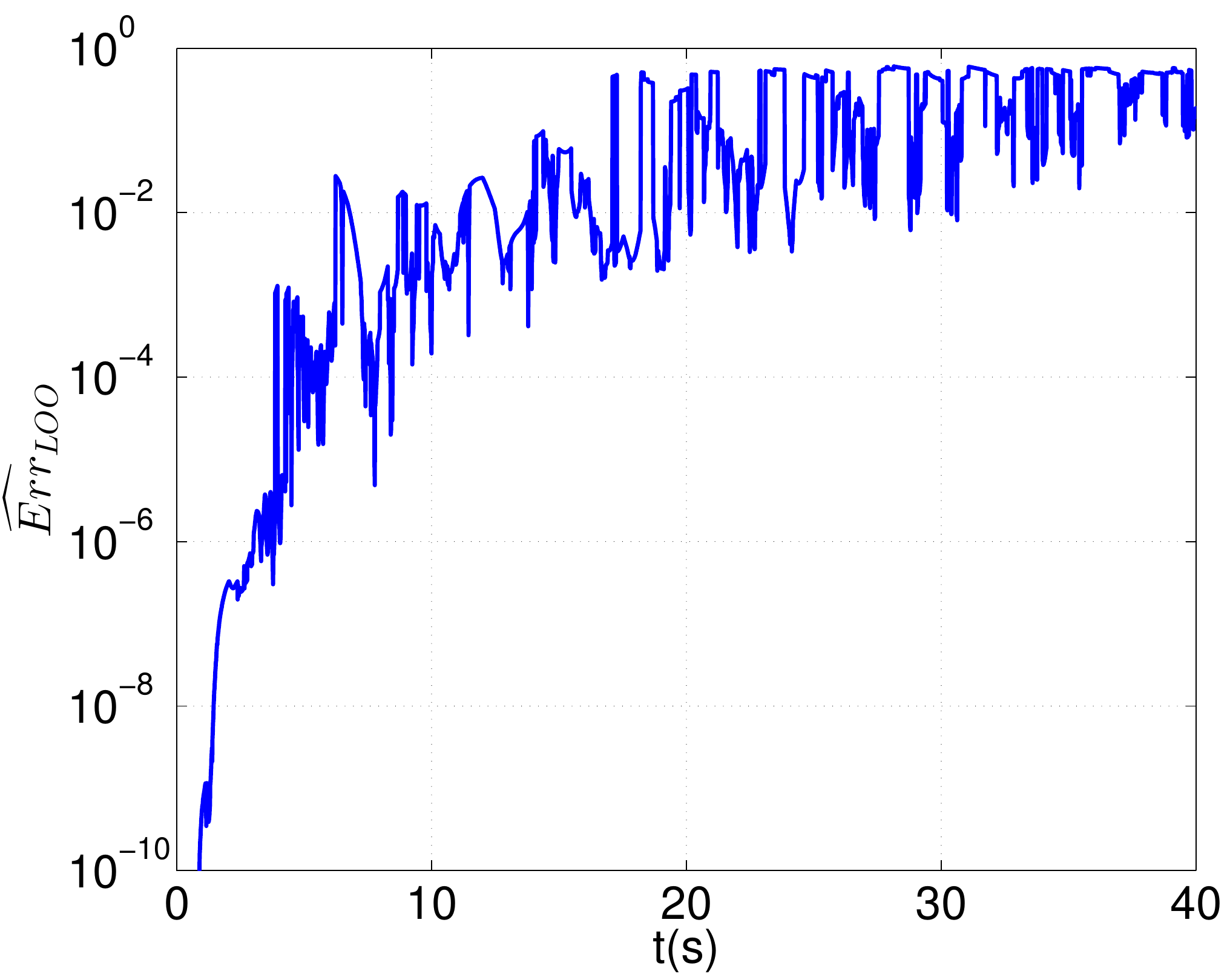}
	\caption{Rigid body dynamics -- Leave-one-out error of time-frozen PCEs.}
	\label{fig4.1.4}
\end{figure}

Time-frozen sparse PCEs are now utilized to model the variability of the response trajectories, and exemplify the deficiency of such an approach. At each instant $t$, an adaptive PCE scheme with candidate polynomials up to total degree 20 is used (Eq.~\eqref{eq:timefroPCE}) based on the available 50 data points from the experimental design made of the 50 trajectories. The PCE model which results in the smallest leave-one-out (LOO) error is retained. \figref{fig4.1.4} depicts the LOO error of these time-frozen PCEs, which is increasing in time, showing that the accuracy of the PCE model degenerates. 

\begin{figure}[!ht]
\centering
\subfigure[$\xi=-0.5385$]
	{
		\includegraphics[width=0.45\linewidth]{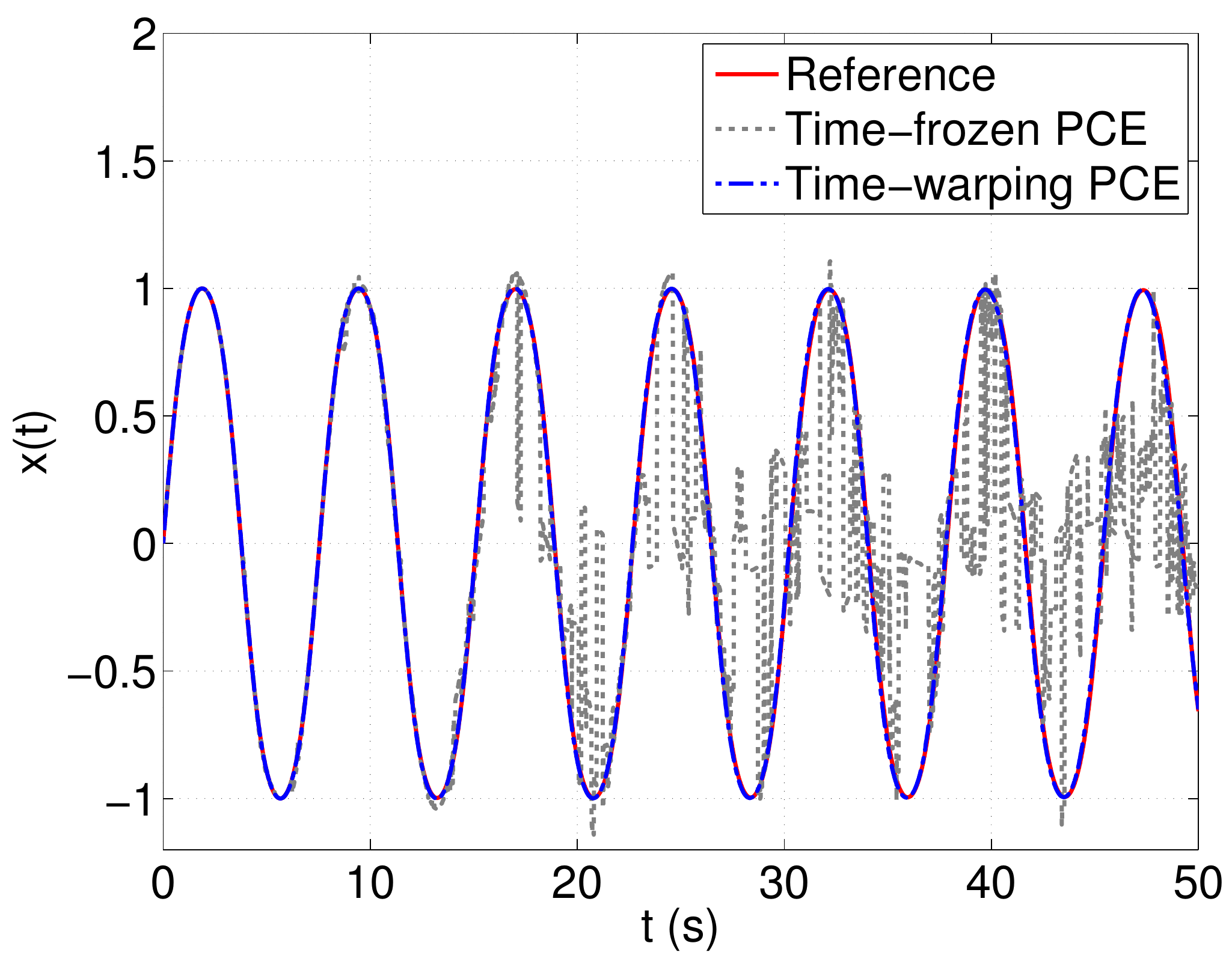}
	}
\subfigure[$\xi=0.6256$]
	{
		\includegraphics[width=0.45\linewidth]{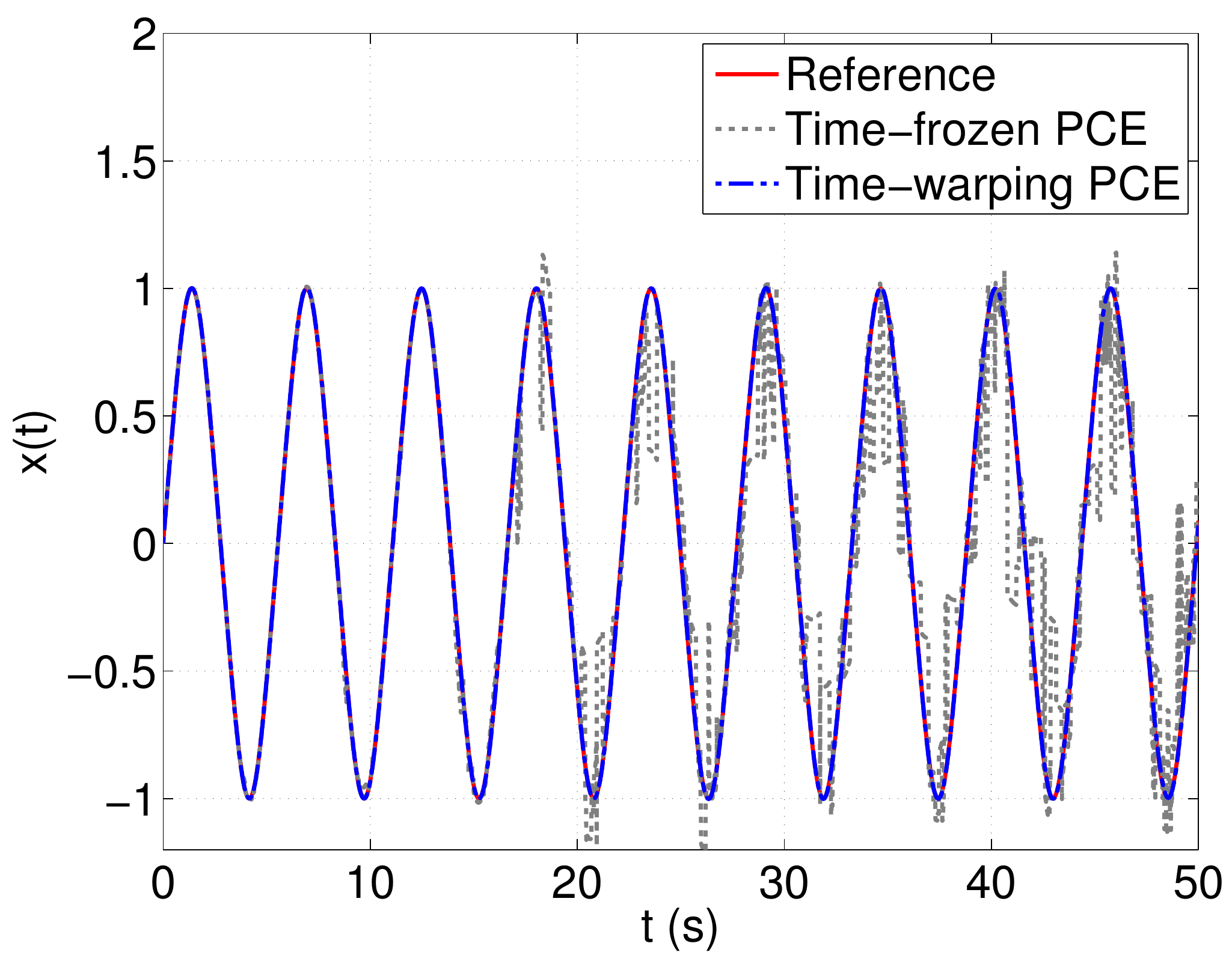}
	}
		\caption{Rigid body dynamics -- Two particular trajectories and their predictions by time-frozen and time-warping PCEs.}
		\label{fig4.1.7}
\end{figure}

For validation purpose, a set of $10,000$ trajectories is computed using the \texttt{ode45} Matlab solver.
\figref{fig4.1.7} depicts two particular response trajectories predicted by time-frozen PCEs versus the actual responses obtained by numerically solving the system of ordinary differential equations \eqref{eq:rigbody}. After $15~s$ (when the LOO error is approximately $10^{-2}$) the PCE prediction deviates significantly from the actual trajectory. In particular, there are signs of instability in the PCE model, \eg the PCE-based prediction for consecutive instants differ noticeably in terms of accuracy.

\begin{figure}[!ht]
	\centering
	\subfigure[Mean trajectory]
	{
	\includegraphics[width=0.45\linewidth]{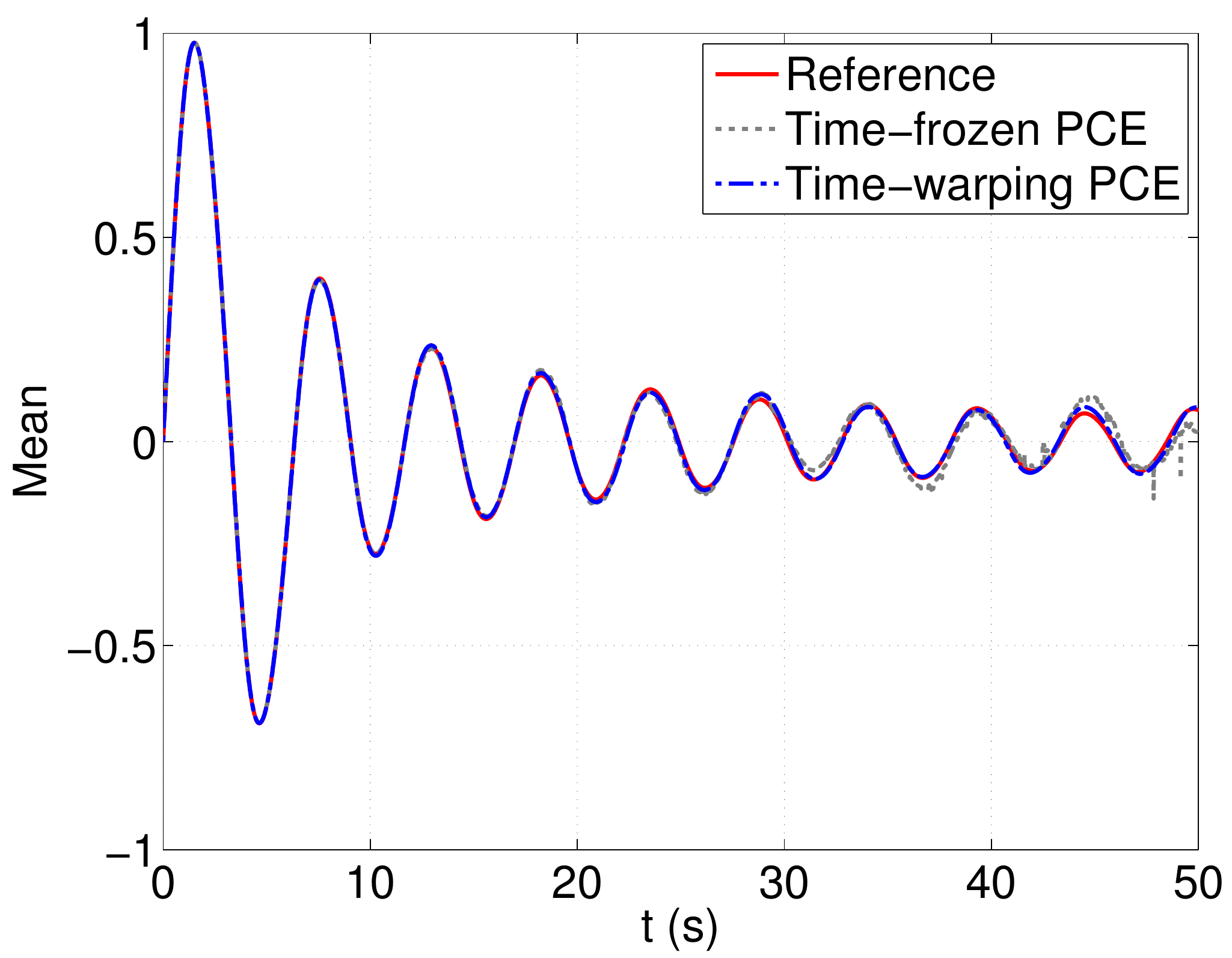}
	}
	\subfigure[Standard deviation trajectory]
	{
	\includegraphics[width=0.45\linewidth]{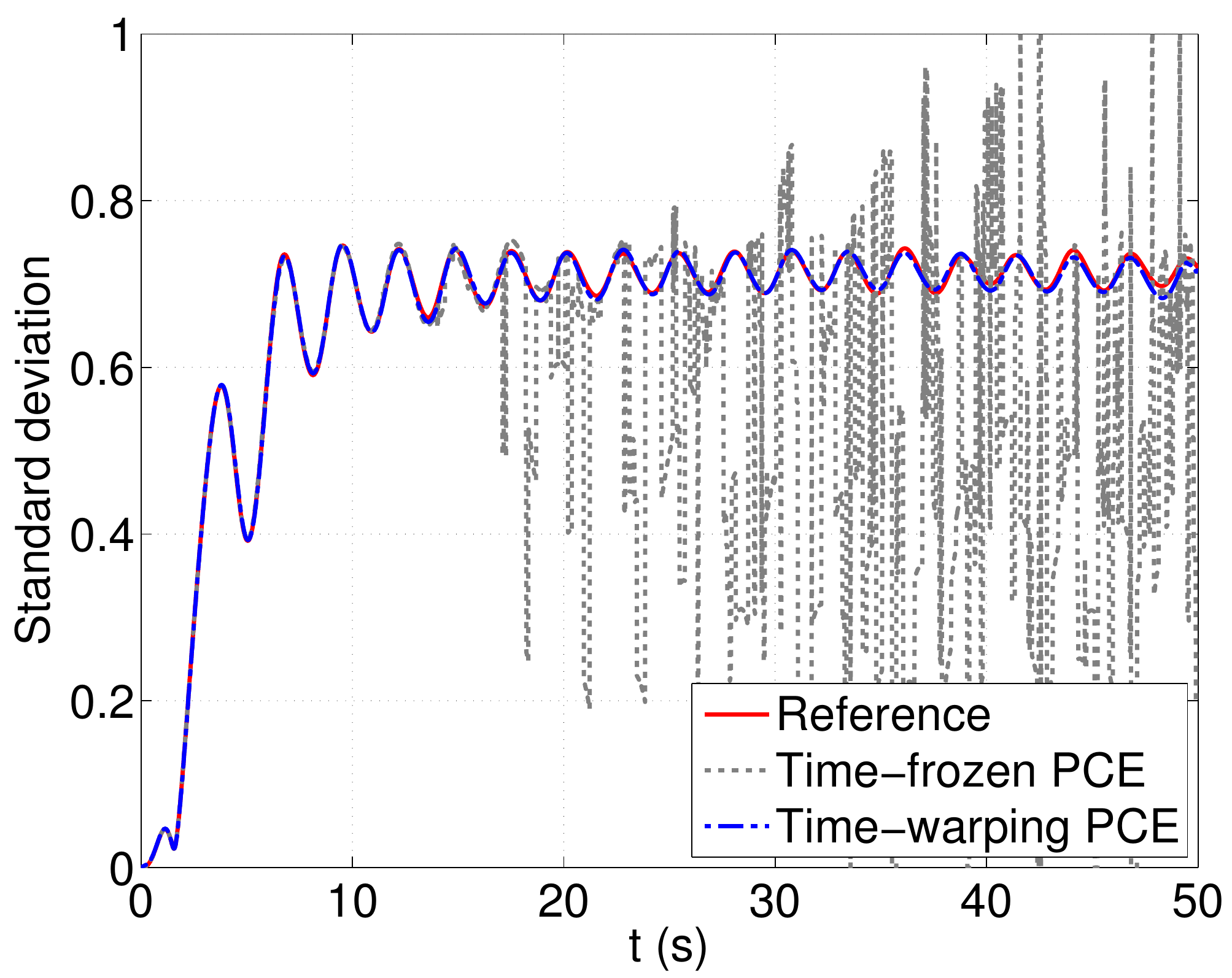}
	}
	\caption{Rigid body dynamics -- Mean and standard deviation of the trajectories: comparison of the two approaches.}
	\label{fig4.1.8}
\end{figure}

We now consider the time-dependent mean and standard deviation of the response $x(t)$ which are depicted in \figref{fig4.1.8}. In the early time instants ($t<15~s$), time-frozen PCEs represent the statistics with relatively small error compared to Monte Carlo simulation (MCS). However, after $15~s$, the accuracy declines quickly. In particular, PCEs cannot mimic the oscillatory behavior of the standard deviation. Another interpretation is that even degree-20 time-frozen PCEs cannot capture the complex distribution of the response at late time instants.

Let us now apply the time-warping approach to pre-process the
trajectories $x(t)$. Provided that the initial condition is equal to 0,
it suffices to use a linear time-warping $\tau = k \, t$. For each
computed realization of the angular velocity $x(t,\xi_i), i = 1 \enum
50$, the parameters $k_i$ is estimated as the maximizer of the
similarity measure described in Eq.~\eqref{eq:tw_objfunc}. Note that the
same 50 trajectories are used as the experimental design for this
approach and the reference trajectory is obtained with the mean value of
the input parameter.
The optimization problem is solved using the global optimization toolbox in Matlab. The function \texttt{fmincon} based upon an interior-point algorithm is used while allowing for a maximum of $2,000$ function evaluations.
Adaptive sparse PCEs for candidate bases up to total degree 20 are used to represent the parameter $k$. The relative LOO error is $3.82 \times 10^{-4}$, which indicates a high accuracy of the PCE model.

\begin{figure}[!ht]
	\centering
	\subfigure[$N=50$ different trajectories $x(\tau)$ in the warped time scale $\tau$]{
	\includegraphics[width=0.45\linewidth]{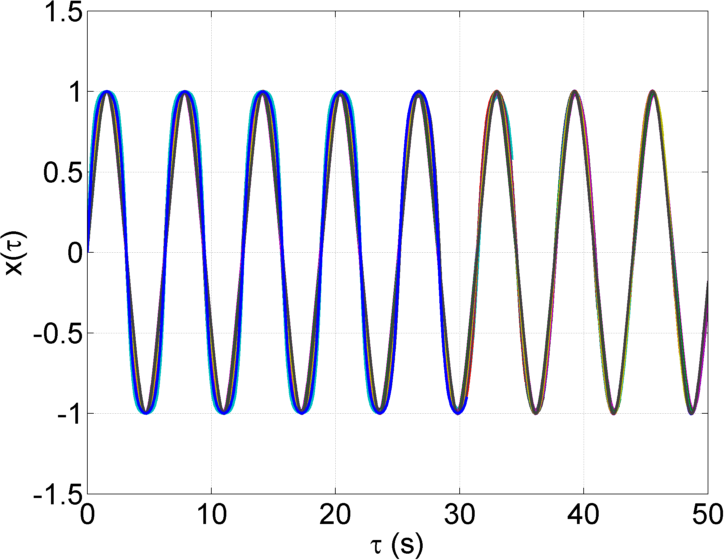}
	\label{fig4.1.5a}
	}
	\subfigure[Relationship between the response $x(\tau)$ and the random variable $\xi$ in the warped time scale $\tau$]{
		\includegraphics[width=0.45\linewidth]{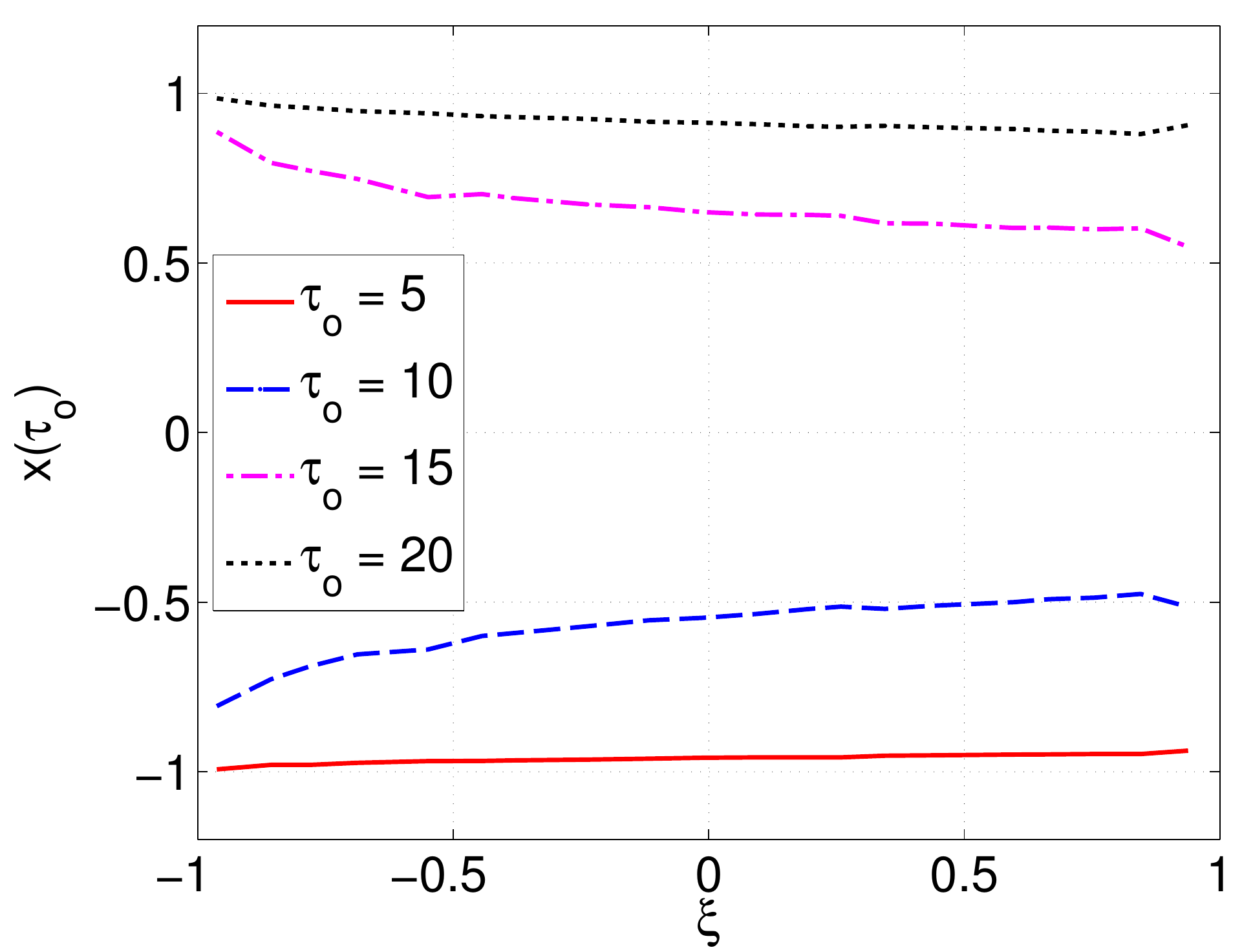}
		\label{fig4.1.5b}
		}
	\caption{Rigid body dynamics -- Different trajectories $x(\tau)$ in the warped time scale $\tau$ and $x(\tau)$ as a function of the random variable $\xi$.}
	\label{fig4.1.5}
\end{figure}

The time-warping is carried out using the estimated parameters and the responses are interpolated into the transformed time line $\tau$, leading to in-phase trajectories $x(\tau)$ (see \figref{fig4.1.5a}). As expected, $x(\tau)$ are smooth functions of $\xi$ at all instants, which allows the effective use of PCEs (\figref{fig4.1.5b}). 

Principal component analysis (PCA) is then conducted on the obtained
transformed trajectories. The first $18$ principal components are
retained in order to achieve a PCA truncation error $\epsilon_1 = {
  \sum\limits_{i= K' + 1}^{K} \lambda_i } / { \sum\limits_{i= 1}^{K}
  \lambda_i}$ smaller than $1 \times 10^{-3}$.  The first eight
principal components are plotted in \figref{fig:rigpcacmps} in the Supplement.
\figref{fig:rigpcaerror} depicts the PCA truncation error $\epsilon_1$
as a function of the number of retained principal components, the LOO
error $\epsilon_2$ of the PCE for the coefficient of each principal
component and the upper bound of the total error of the PCA-PCE model.
It shows that the PCA truncation error $\epsilon_1$ decreases
exponentially with the number of retained principal components.  Using
PCE to represent the first PCA coefficient, the obtained relative LOO
error is $7.7 \times 10^{-3}$. It is also clear that it is harder to
represent the higher mode PCA coefficients by PCEs, as was observed in
\cite{BlatmanIcossar2013}. However, it is worth noting that most of the
stochastic features of the response is captured by the first few
components.
%\begin{figure}[!ht]
%\centering
%\subfigure[Principal components 1-4]
%	{
%		\includegraphics[width=0.45\linewidth]{ex1pca_components_g1.pdf}
%	}
%\subfigure[Principal components 5-8]
%	{
%		\includegraphics[width=0.43\linewidth]{ex1pca_components_g2.pdf}
%	}
%		\caption{Rigid body dynamics -- The first eight principal components.}
%		\label{fig:rigpcacmps}
%\end{figure}
%%
\begin{figure}[!ht]
\centering
	{
		\includegraphics[width=0.45\linewidth]{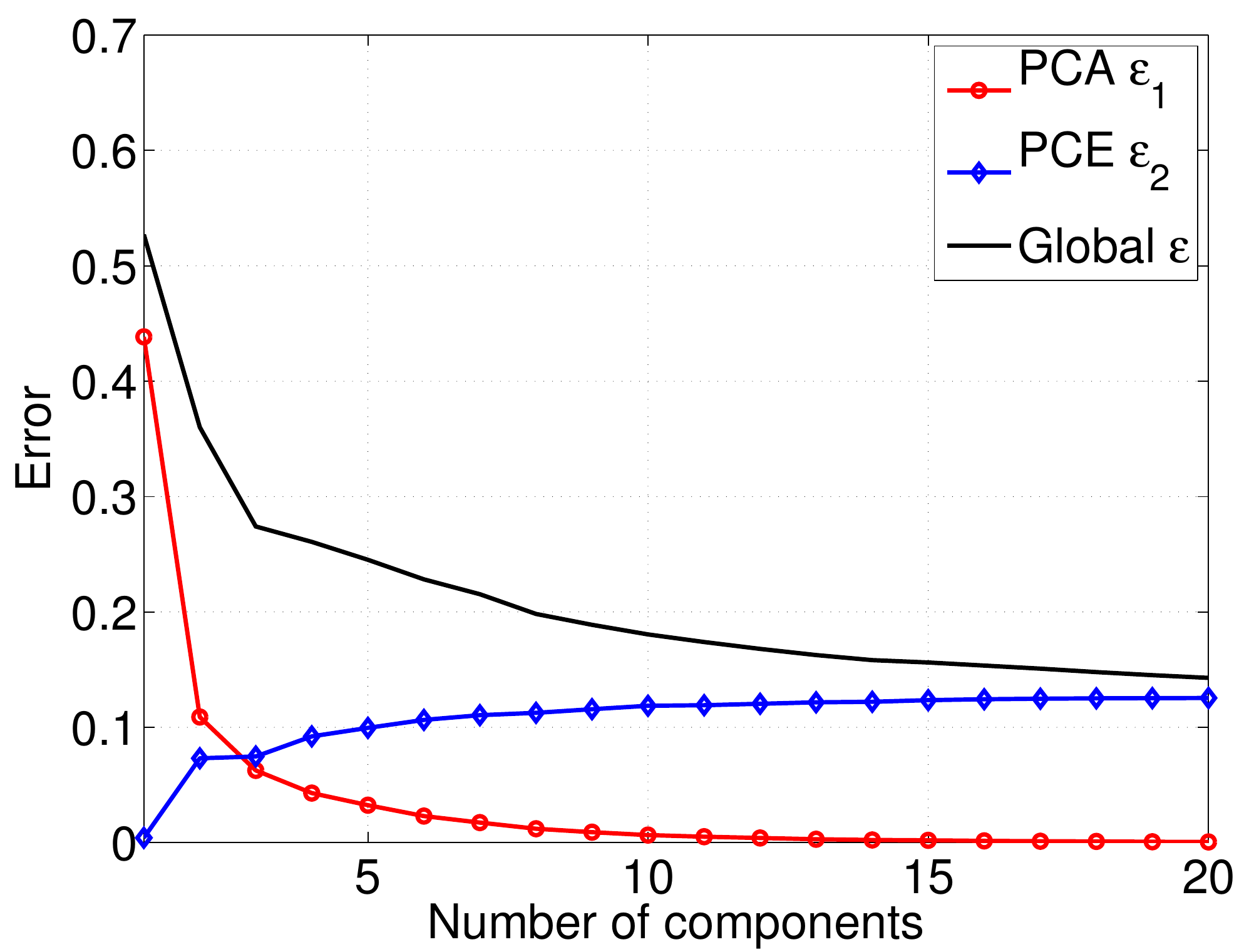}
	}
		\caption[Rigid body dynamics -- PCA and PCE truncation-induced error]{Rigid body dynamics -- PCA truncation-induced error $\epsilon_1$ and PCE approximation error $\epsilon_2$ normalized by $\text{trace}(\tilde{\matSigma}) $ with $\tilde{\matSigma}$ being the empirical covariance matrix of the sample set $\cy$ and the upper bound $\epsilon = \prt{ \sqrt{\epsilon_1} + \sqrt{\epsilon_2} }^2$ of the total error.}
		\label{fig:rigpcaerror}
\end{figure}

\figref{fig4.1.7} depicts two specific realizations of the angular velocity $x(t)$ predicted by time-warping PCEs, which are plotted together with the predictions by time-frozen sparse PCEs and the actual responses obtained by the numerical solver.
As mentioned previously, one observes that starting from $15~s$, the direct approach encounters instability, which results in inaccurate predictions. The time-warping approach allows one to improve notably the quality of the surrogate model. The predictions by time-warping PCEs are in excellent agreement with the actual responses. A relative error exceeding $0.1$ is recorded in only $79$ simulations among $10,000$ validations.

In \figref{fig4.1.8}, the time-dependent mean and standard deviation of the response are plotted. Time-frozen PCEs allow one to represent the mean trajectory with relatively small discrepancy compared to the trajectory obtained with the MCS. It can faithfully predict the standard deviation at the early instants $t<15~s$, however becomes suddenly unstable afterwards.
In contrast, time-warping PCEs provide estimates of the statistics that are almost indistinguishable from the MCS estimates. The relative errors between the reference and predicted mean and standard deviation are $7.31 \times 10^{-4}$ and $7.19 \times 10^{-4}$, respectively.

%%%%%%%%%%%%%%%%%%%%%%%%%%%%%%%%%%%%%%%%%%%%%%%%%%%%%%%%%%%%%%%%%%%%%%%%%%%%%%%%%%%%%
\subsection{Kraichnan-Orszag model}
Let us investigate dynamical systems with random initial conditions, \eg
the so-called Kraichnan-Orszag three-mode problem. It was introduced by Kraichnan
\cite{Kraichnan1960} to model a system of several interacting shear
waves and later was studied by Orszag \cite{Orszag1967} in the case of
Gaussian initial conditions. This model is described by the following
system of ODEs:
\begin{equation}
 \left\{
 \begin{array}{l}
    \dot{x}(t) = y(t) \, z(t) , \\
    \dot{y}(t) = z(t) \, x(t), \\
    \dot{z}(t)  = -2 \, x(t) \, y(t).
 \end{array}
 \right.	
% \label{eq5.3.1}
\end{equation}

\begin{figure}[ht!]
	\centering
	\subfigure[$x(t)$ in the original time scale $t$]
		{
		\includegraphics[width=0.45\linewidth]{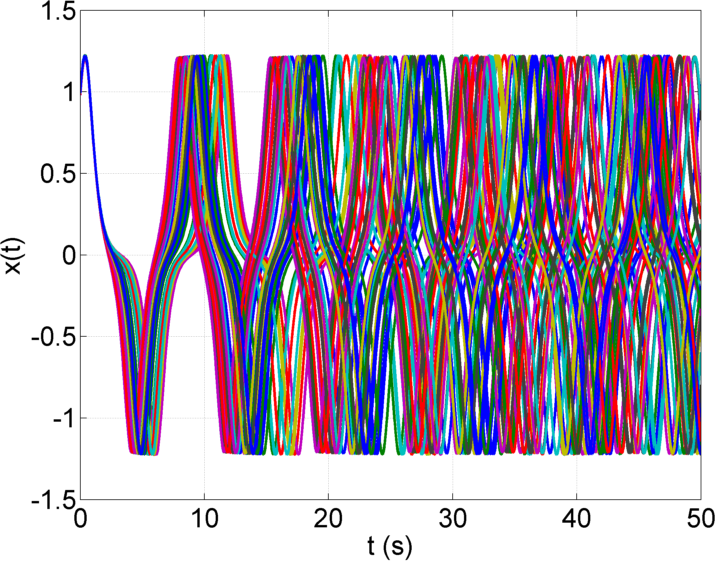}
		\label{fig:KO1Dxvst}
		}
	\subfigure[$x(\tau)$ in the warped time scale $\tau$]
		{
		\includegraphics[width=0.45\linewidth]{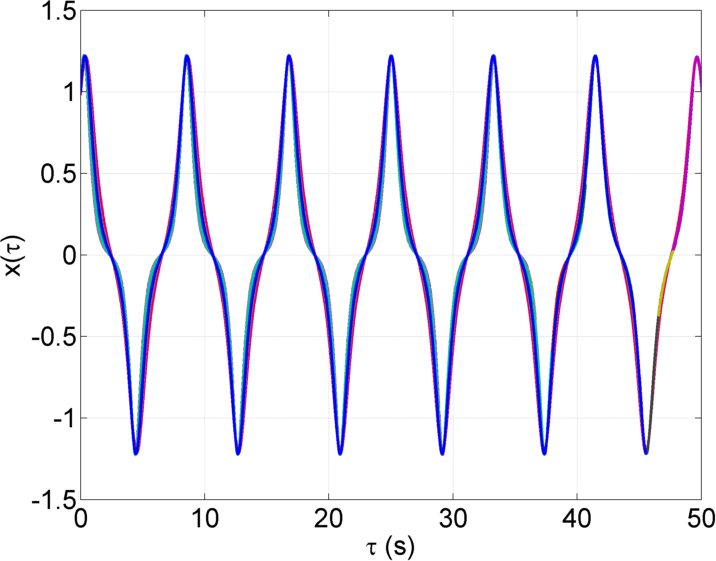}
		\label{fig:KO1Dxvstau}
		}
	\caption[Kraichnan-Orszag model -- Different trajectories in the original and warped time scales $t$]{Kraichnan-Orszag model -- $N=50$ different trajectories in the original and warped time scales. }
\end{figure}

The initial condition of $x(t)$ is considered stochastic, \ie $x(t=0) =
\alpha + 0.01 \, \xi$ with $\xi \sim \cu[-1,1]$ whereas $y(t=0) = 1.0$,
$z(t=0)= 1.0$.  Herein, we consider $\alpha = 0.99$ as investigated by Gerritsma et al.
\cite{Gerritsma2010} with the time-dependent PCEs.  Note that when
$\alpha$ is in the range $[0,0.9]$, the responses are insensitive to the
initial conditions. For $\alpha \in [0.9, 1]$, there is a strong
dependence of the responses on the initial state.  \figref{fig:KO1Dxvst}
depicts the large discrepancies between time-histories of $x(t)$ due to
a minor variability of the initial condition $x(t=0)$.

The surrogate model of the response $x(t)$ is computed with time-frozen and time-warping PCEs using an experimental design of size $N = 50$ (\figref{fig:KO1Dxvst}).
On the one hand, adaptive sparse PCEs with candidate bases up to total degree 20 are used for the time-frozen approach. On the other hand, a time-transform scheme $\tau = k \, t$ with one governing parameter is used for the time-warping scheme. The trajectories resulting from the time-warping process are depicted in \figref{fig:KO1Dxvstau}. The adaptive sparse PCE representing $k$ has the relative LOO error $2.2 \times 10^{-6}$. The first $13$ principal components are retained so that $99.9\%$ of the response's variance is explained.
%\figref{fig:KO1Dpcp} depicts the first four principal components.
The relative LOO errors of PCEs for the first two components are $9.4 \times 10^{-5}$ and $7 \times 10^{-3}$, respectively.

\begin{figure}[!ht]
\centering
\subfigure[$\xi=0.6294$]
	{
		\includegraphics[width=0.45\linewidth]{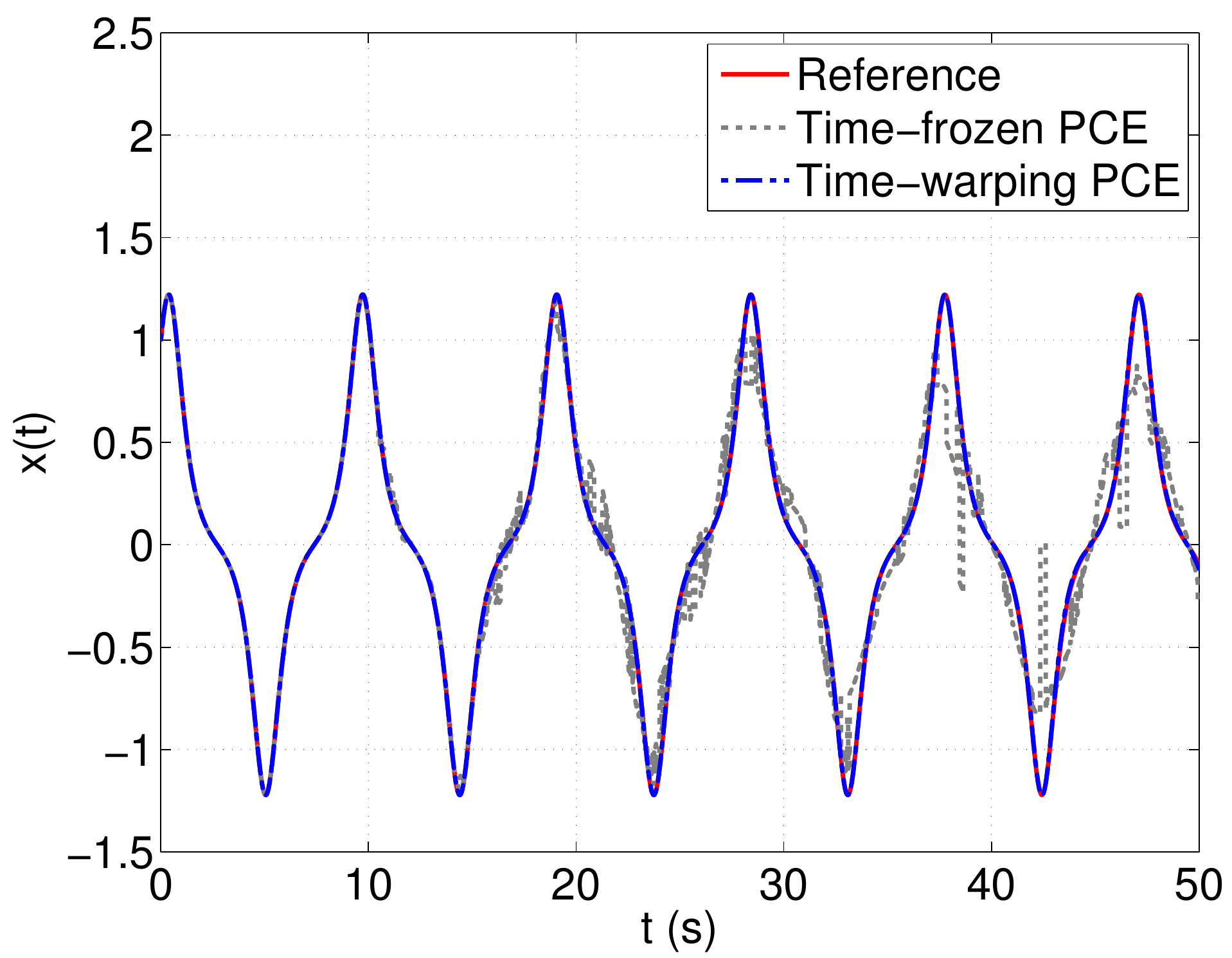}
	}
\subfigure[$\xi=-0.7460$]
	{
		\includegraphics[width=0.45\linewidth]{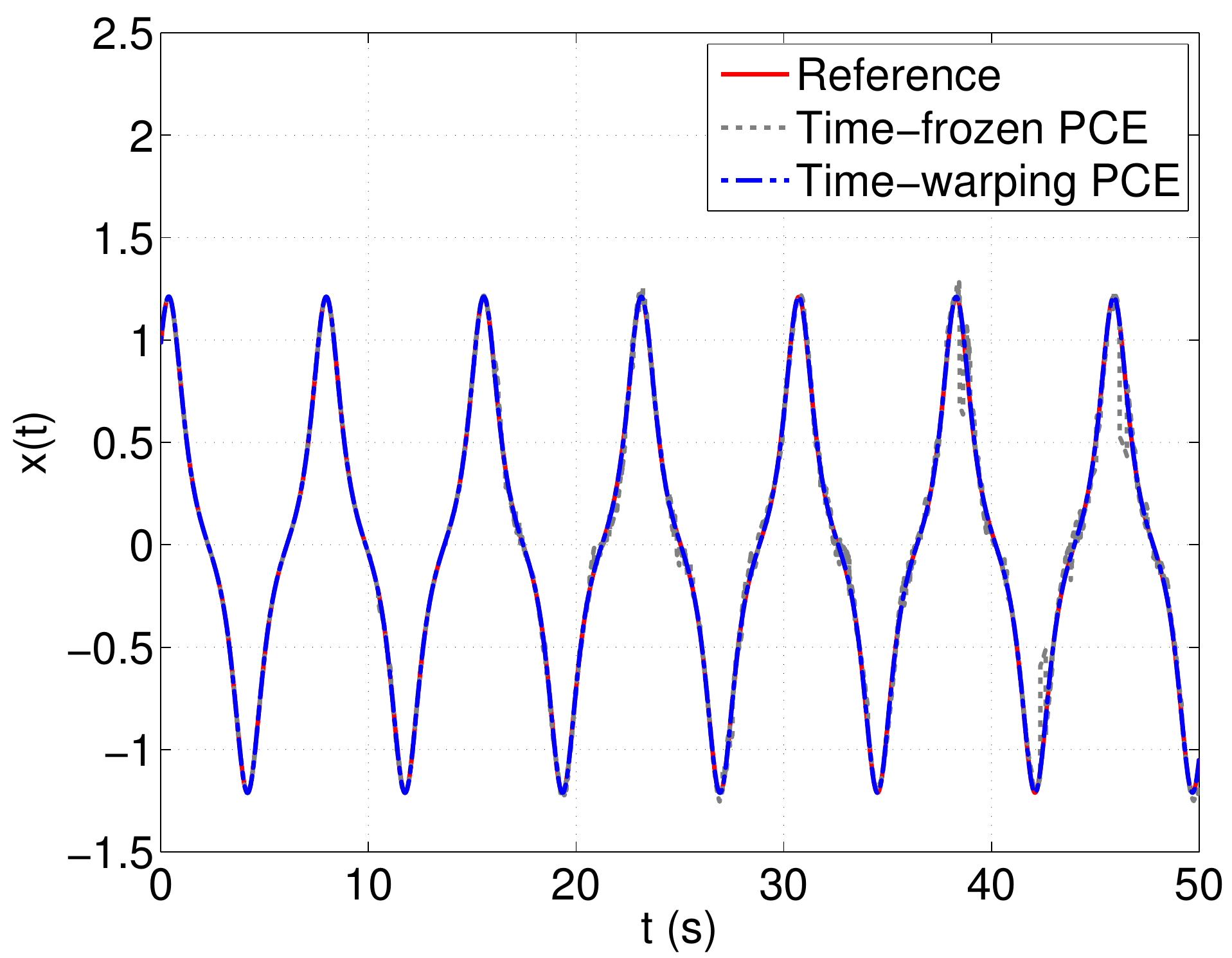}
	}
		\caption{Kraichnan-Orszag model -- Two particular trajectories and their predictions by time-warping PCEs.}
		\label{fig:KO1D2predict}
\end{figure}

\begin{figure}[!ht]
	\centering
	\subfigure[Mean trajectory]
	{
	\includegraphics[width=0.45\linewidth]{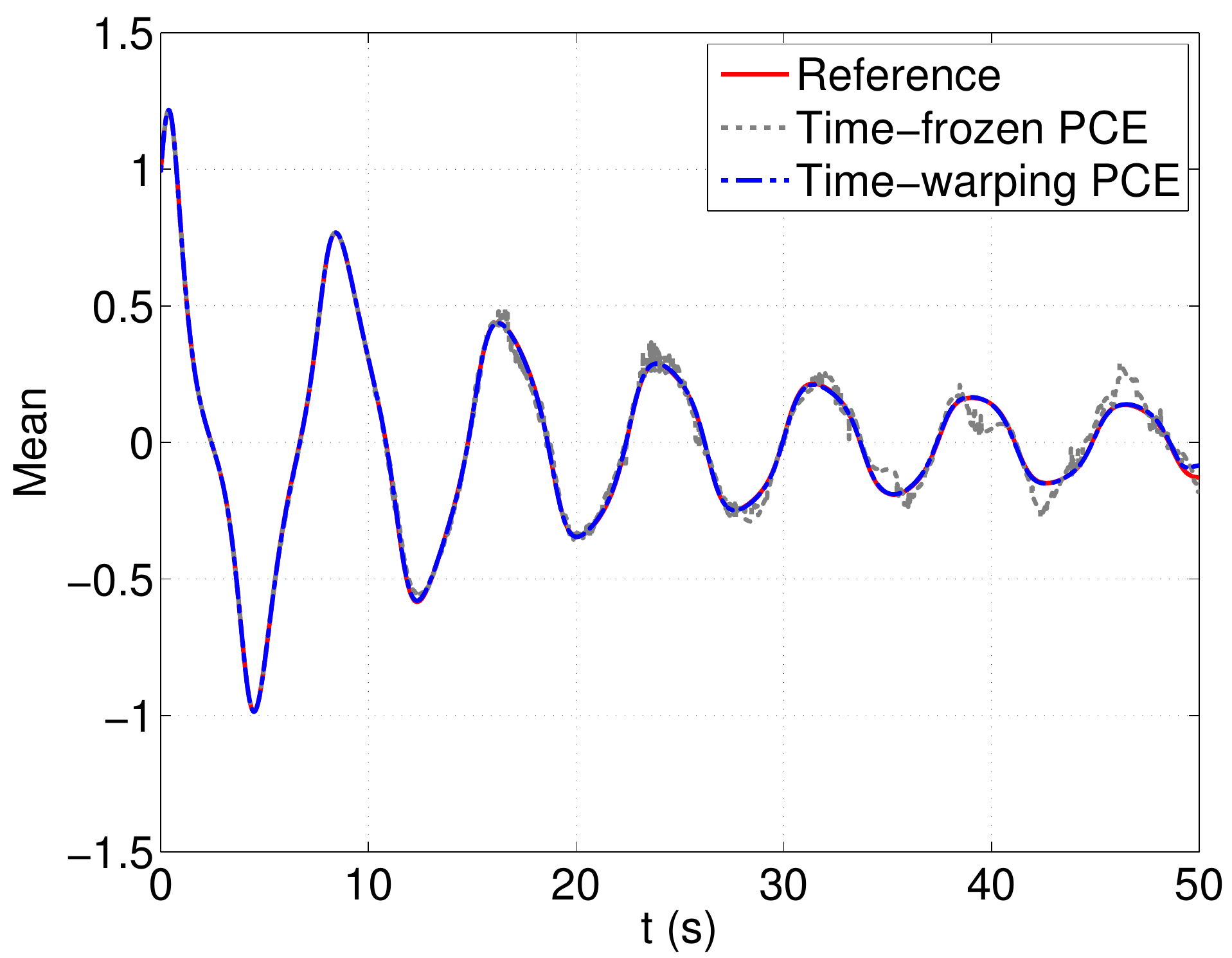}
	}
	\subfigure[Standard deviation trajectory]
	{
	\includegraphics[width=0.45\linewidth]{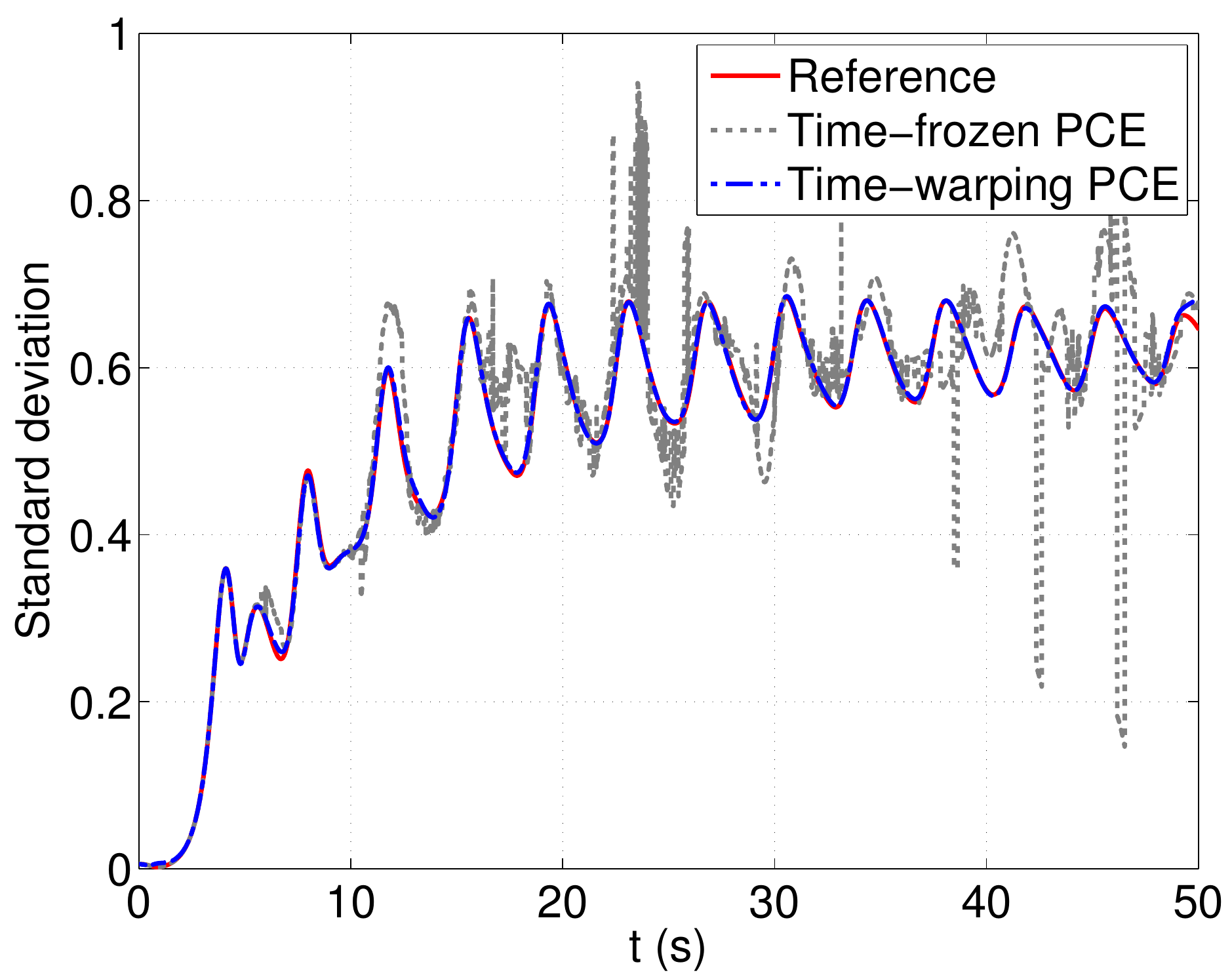}
	}
	\caption{Kraichnan-Orszag model -- Mean and standard deviation of the trajectories: comparison of the two approaches.}
	\label{fig:KO1Dmeanstd}
\end{figure}

The time-warping PCE model is then validated by accessing the accuracy of its predictions. \figref{fig:KO1D2predict} plots two specific predictions of the surrogate model which are graphically indistinguishable from the actual time-histories obtained with the original Matlab solver. 
Only $1.27\%$ of the $10,000$ predictions experiences a relative error larger than $0.1$.
Regarding the mean and standard deviation trajectories (\figref{fig:KO1Dmeanstd}), the time-warping approach leads to respective relative errors $2.1 \times 10^{-4}$ and $5.3 \times 10^{-4}$, which shows an excellent agreement between the predictions and the references.
These figures also show that the time-frozen sparse PCEs computed with the same experimental design of size $50$ lead to predictions which are not sufficiently accurate.

This numerical application illustrates the potential application of the
proposed time-warping approach to systems subject to uncertain initial
conditions. The excellent performance of the approach is even more
impressive given {the fact that} the
responses are strongly sensitive with respect to a minor variability of
the initial condition.
%
%\FloatBarrier

%%%%%%%%%%%%%%%%%%%%%%%%%%%%%%%%%%%%%%%%%%%%%%%%%%%%%%%%%%%%%%%%%%%%%%%%%%%%%%%%%%%%%
\subsection{Oregonator model}

We consider now the Oregonator model which describes the dynamics of a
well-stirred, homogeneous chemical system governed by a three species
coupled mechanism. Note that this benchmark problem was used by Le Ma\^{i}tre et al.
\cite{LeMaitre2009} to illustrate the intrusive
time-transform approach. This chemical system undergoes an oscillation
governed by the following system of ODEs:
\begin{equation}
 \left\{
 \begin{array}{l}
  \dot{x}(t)= k_1 \, y(t) - k_2 \, x(t) \, y(t) + k_3 \, x(t) - k_4 \, x(t)^2, \\
  \dot{y}(t)= -k_1 \, y(t) - k_2 \, x(t) \, y(t) + k_5 \, z(t), \\
  \dot{z}(t)= k_3 \, x(t) - k_5 \, z(t),
 \end{array}
 \right.
 \label{eq4.2.1}
\end{equation}
in which $(x,y,z)$ denotes the three species concentration and the
coefficients $k_i, \, i=1 \enum 5$ are the reaction parameters.
Hereafter, all the reaction parameters are considered independent random
variables with uniform and normal distributions (see Table \ref{tab:1}).
It is worth noting that Le Ma\^{i}tre et al. \cite{LeMaitre2009}
considered only $k_4$ and $k_5$ as uniform random variables while fixing
the remaining parameters (\ie $k_1 = 2,\, k_2 = 0.1, \, k_3 = 104$).
The initial condition is $(x_0, y_0, z_0) =(6,000; 6,000; 6,000)$, which
corresponds to a deterministic mixture.  We aim at building PCEs of the
concentration $x(t)$ as a function of the random parameters $\vexi =
(k_1, \, k_2, \, k_3, \, k_4, \, k_5)$.
\begin{table}[!ht]
\caption{Reaction parameters of the Oregonator model}
\centering
\begin{tabular}{|c|c|c|c|c|}
\hline
Parameters & Distribution & Mean & Standard deviation & Coefficient of variation  \\
\hline
$k_1$ & Uniform & $2$ & $0.2/\sqrt{3}$ & $0.0577$ \\\hline
$k_2$ & Uniform & $0.1$ & $0.005/\sqrt{3}$ & $0.0289$ \\\hline
$k_3$ & Gaussian & $104$ & $1.04$ & $0.01$ \\\hline
$k_4$ & Uniform & $0.008$ & $4 \times 10^{-4} /\sqrt{3}$ & $0.0289$  \\\hline
$k_5$ & Uniform & $26$ & $2.6/\sqrt{3}$ & $0.0577$ \\
\hline
\end{tabular}
\label{tab:1}
\end{table}

\begin{figure}[!ht]
	\centering
%	\subfigure[Original time scale $t$]
%	{
%	\includegraphics[width=0.45\linewidth]{ex4x_vs_t.pdf}
%	\label{fig4.3.1a}
%	}
%	\subfigure[Warped time scale $\tau$]
%	{
%	\includegraphics[width=0.45\linewidth]{ex4x_vs_tau.pdf}
%	\label{fig4.3.1b}
%	}
	\subfigure[Original time scale $t$ (zoom)]
		{
		\includegraphics[width=0.45\linewidth]{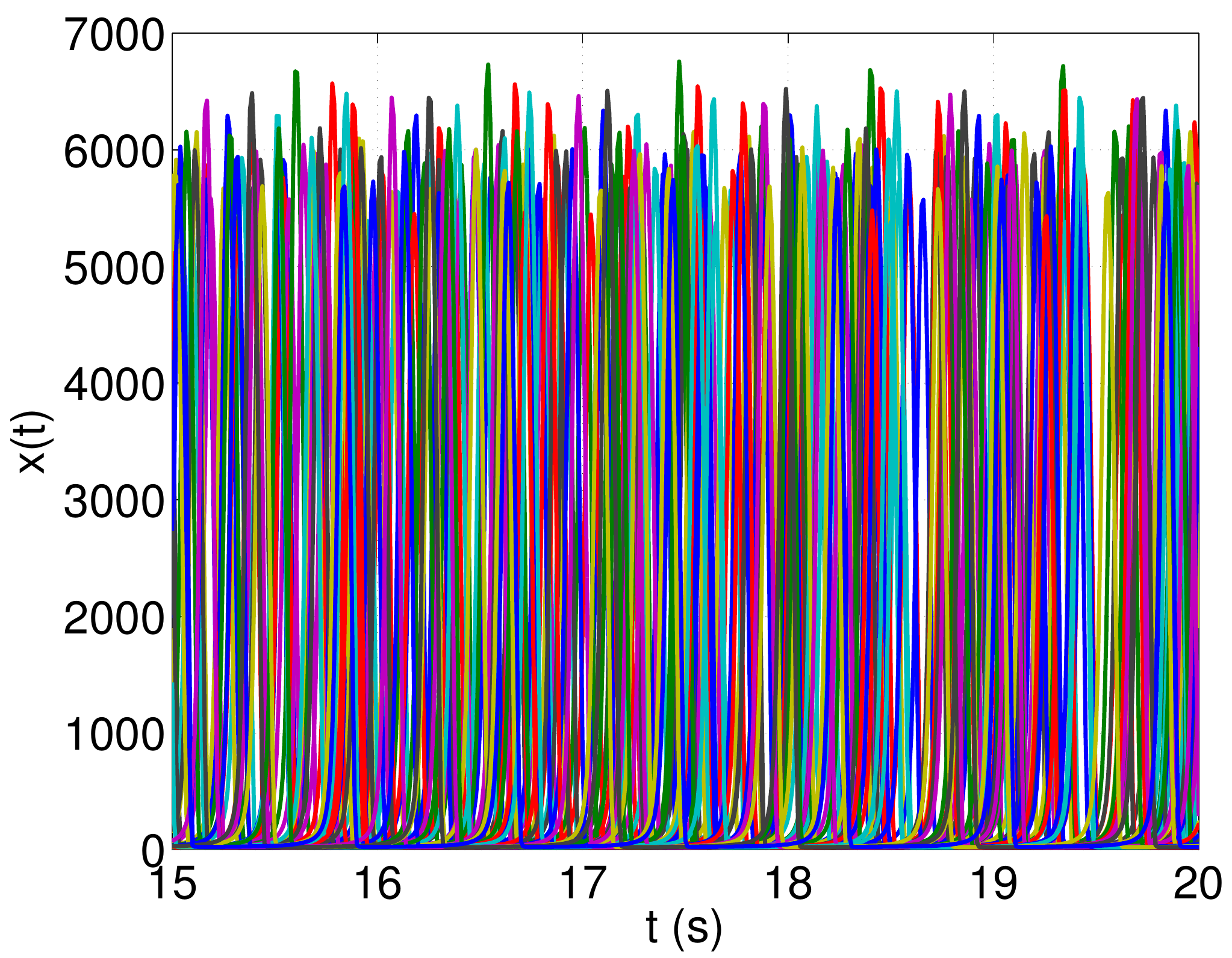}
		\label{fig4.3.1a}
		}
	\subfigure[Warped time scale $\tau$ (zoom)]
		{
		\includegraphics[width=0.45\linewidth]{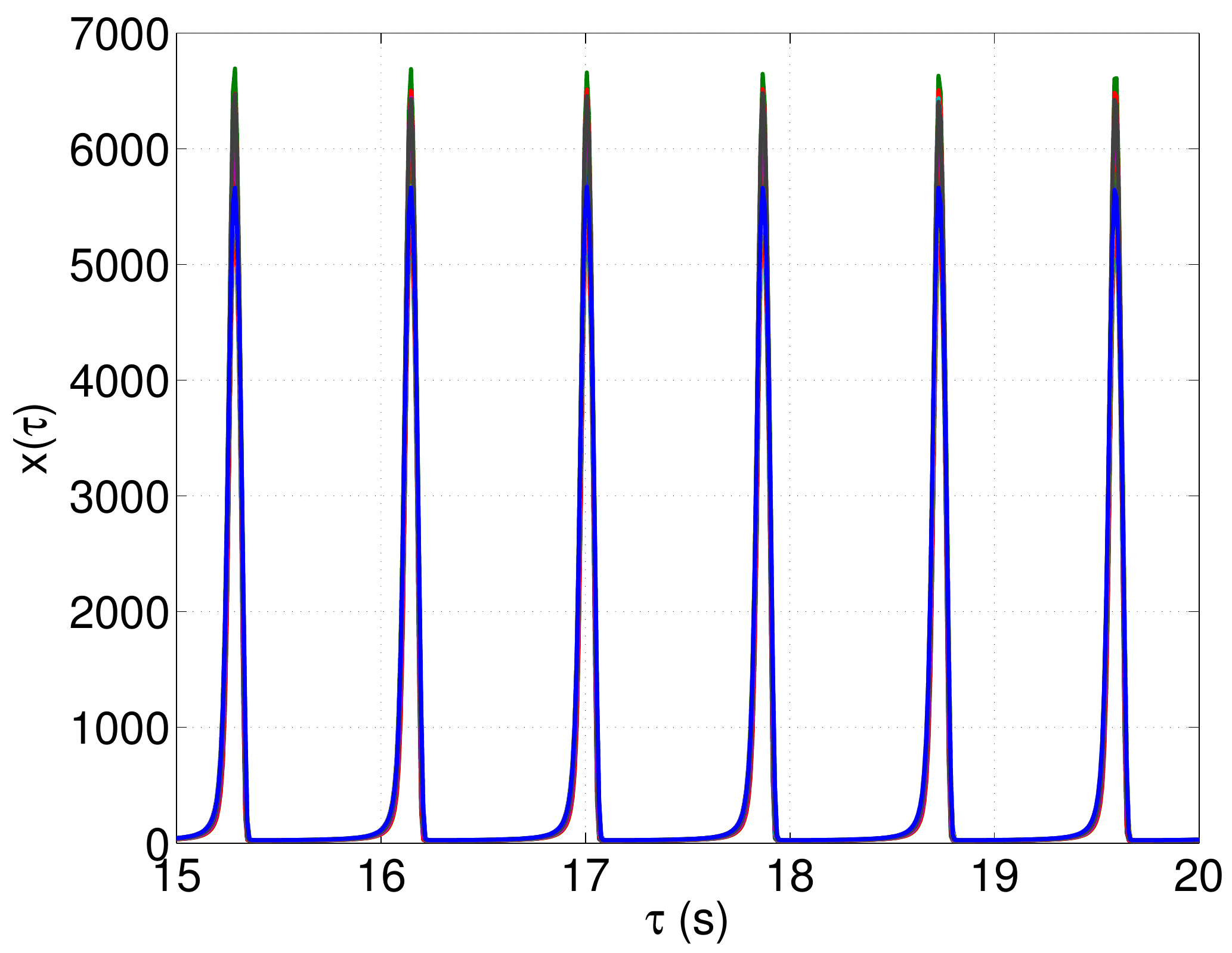}
		\label{fig4.3.1b}
		}
	\caption[Oregonator model -- Different trajectories of the response in the original and transformed time scales]{Oregonator model -- $N=50$ different trajectories of the response $x(t)$. For the sake of clarity, the figures are zoomed in the range $[15,20]$. \figref{fig4.3.1s} depicts the entire time horizon.}
	\label{fig4.3.1}
\end{figure}

\begin{figure}[!ht]
	\centering
%	\captionsetup[subfigure]{justification=centering}
%	\subfigure
%		[$\vexi=(1.8970,   0.1001,  104.3676,    0.0077,25.5417)$]
%		{
%			\includegraphics[width=0.45\linewidth]{ex4validation_1_long.pdf}
%			}	
%	\subfigure
%		[$\vexi = (1.9481,\,    0.0999,\,  102.7929,\,    0.008,\,   27.6482)$]
%			{
%			\includegraphics[width=0.45\linewidth]{ex4validation_2_long.pdf}
%			}
	\subfigure
	[$\vexi=(1.8970,   0.1001,  104.3676,    0.0077,25.542)$]
	{
		\includegraphics[width=0.45\linewidth]{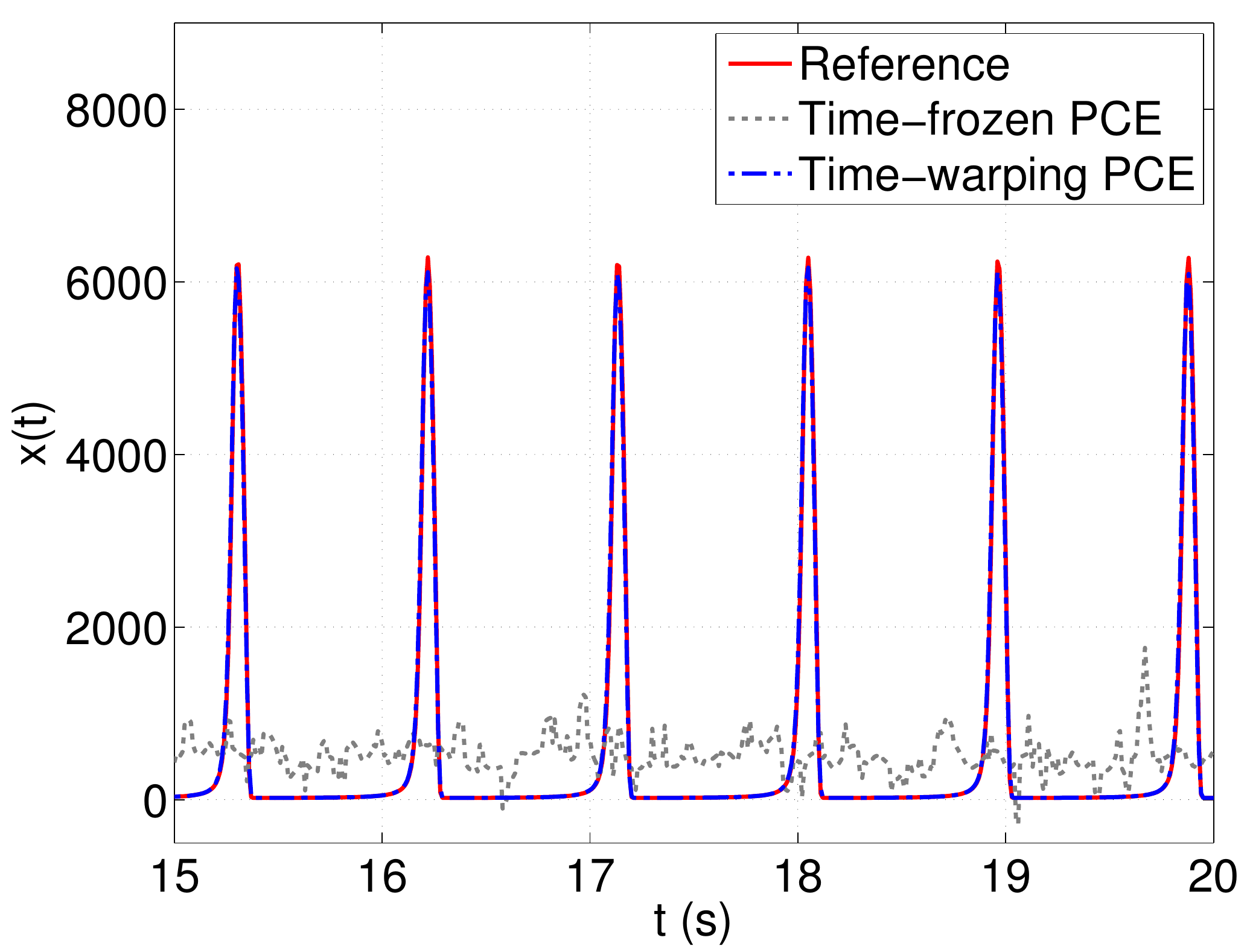}
		}	
	\subfigure
	[$\vexi = (1.9481,\,    0.0999,\,  102.7929,\,    0.008,\,   27.648)$]
		{
		\includegraphics[width=0.45\linewidth]{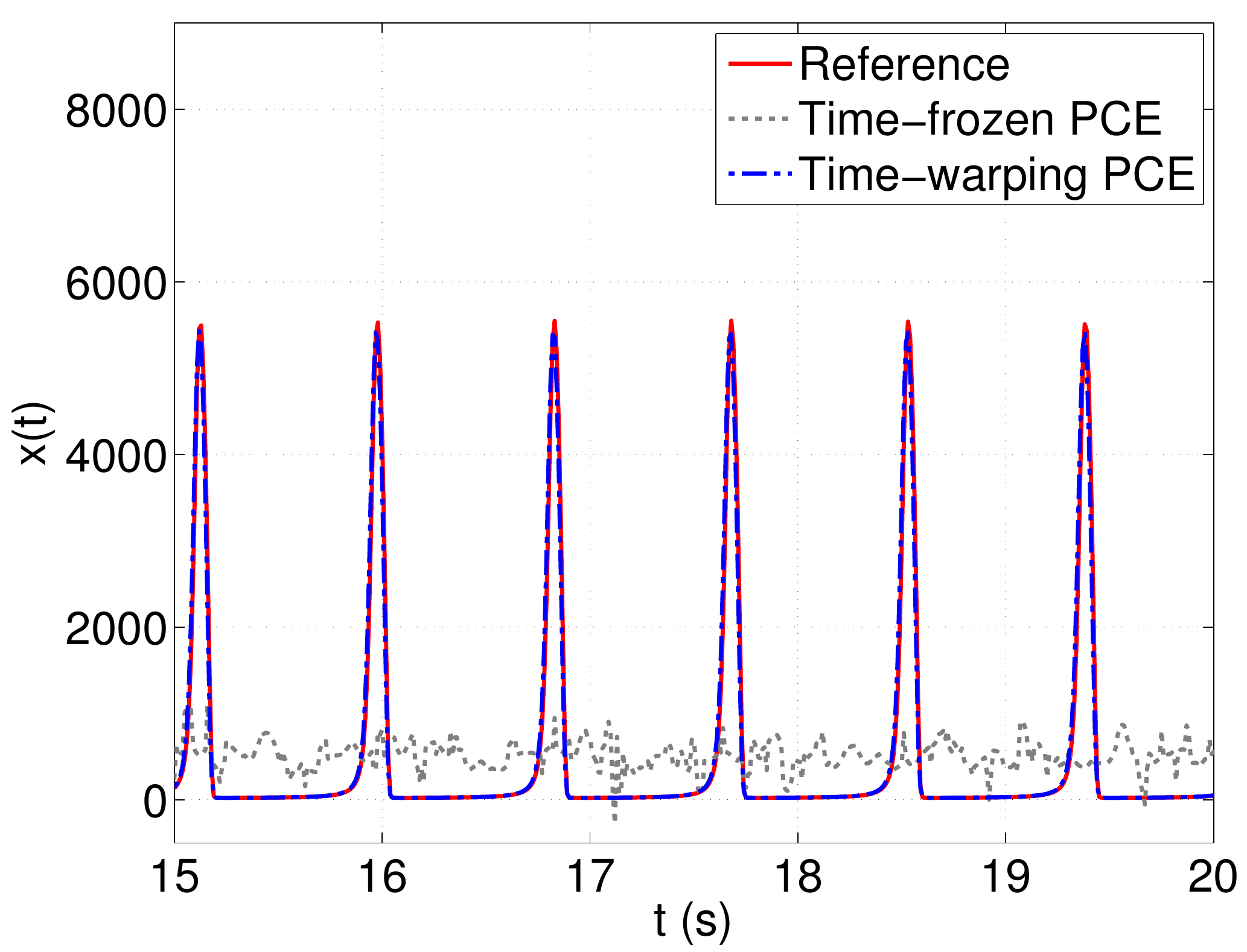}
		}
	\caption[Oregonator model -- Two particular trajectories and their predictions by time-frozen and time-warping PCEs.]{Oregonator model -- Two particular trajectories $x(t)$ and their predictions by time-frozen and time-warping PCEs. For the sake of clarity, the figures are zoomed in the range $[15,20]$. \figref{fig4.3.3s} depicts the entire time horizon.}
	\label{fig4.3.3}
\end{figure}

\figref{fig4.3.1a} depicts 50 trajectories among $500$ realizations of $x(t)$, which are used as the experimental design for fitting time-frozen PCEs. One notices that after 5 seconds, the different trajectories are completely out-of-phase. Time-frozen sparse PCEs with candidate polynomials up to total degree 20 are used. 
%The corresponding LOO error (\figref{fig4.3.2}) exhibits a relatively fast escalation. 
The PCE model actually starts degenerating at $t=3~s$. In particular, \figref{fig4.3.3} shows that when used for predicting the responses, time-frozen PCE provide negative values of the concentration at some instants, which is non physical for the considered problem.
%

% % % % % time-warping
We now apply the proposed non-intrusive time-warping approach to this problem. Note that only \emph{50 trajectories} of $x(t)$ are used as an experimental design for this approach. A linear time-transform $\tau = k\, t + \phi$ is again utilized. The parameters $k$ and $\phi$ are determined and sparse PCEs of $k$ and $\phi$ are then computed. 
The relative LOO errors of the PCE models for $k$ and $\phi$ are respectively $4.42 \times 10^{-5}$ and $4.8 \times 10^{-2}$, which indicate a high accuracy. The response trajectories are interpolated into the transformed time line $\tau$ (\figref{fig4.3.1b}) and adaptive sparse PCEs with candidate polynomials up to total degree 20 combined with PCA are then used. 
The first $18$ components are retained in PCA to obtain a truncation error $\epsilon_1$ smaller than $1 \times 10^{-2}$. 
%The first eight components are depicted in \figref{fig:oregpcacmps}. 
%\figref{fig:oregpcaerror} shows the convergence of the errors with respect to the number of principal components. 
The PCEs for the first two coefficients have relative errors $7.57 \times 10^{-4}$ and $1.5 \times 10^{-3}$, respectively.

\begin{figure}[!ht]
	\centering
	\subfigure[Mean trajectory]
	{
	\includegraphics[width=0.45\linewidth]{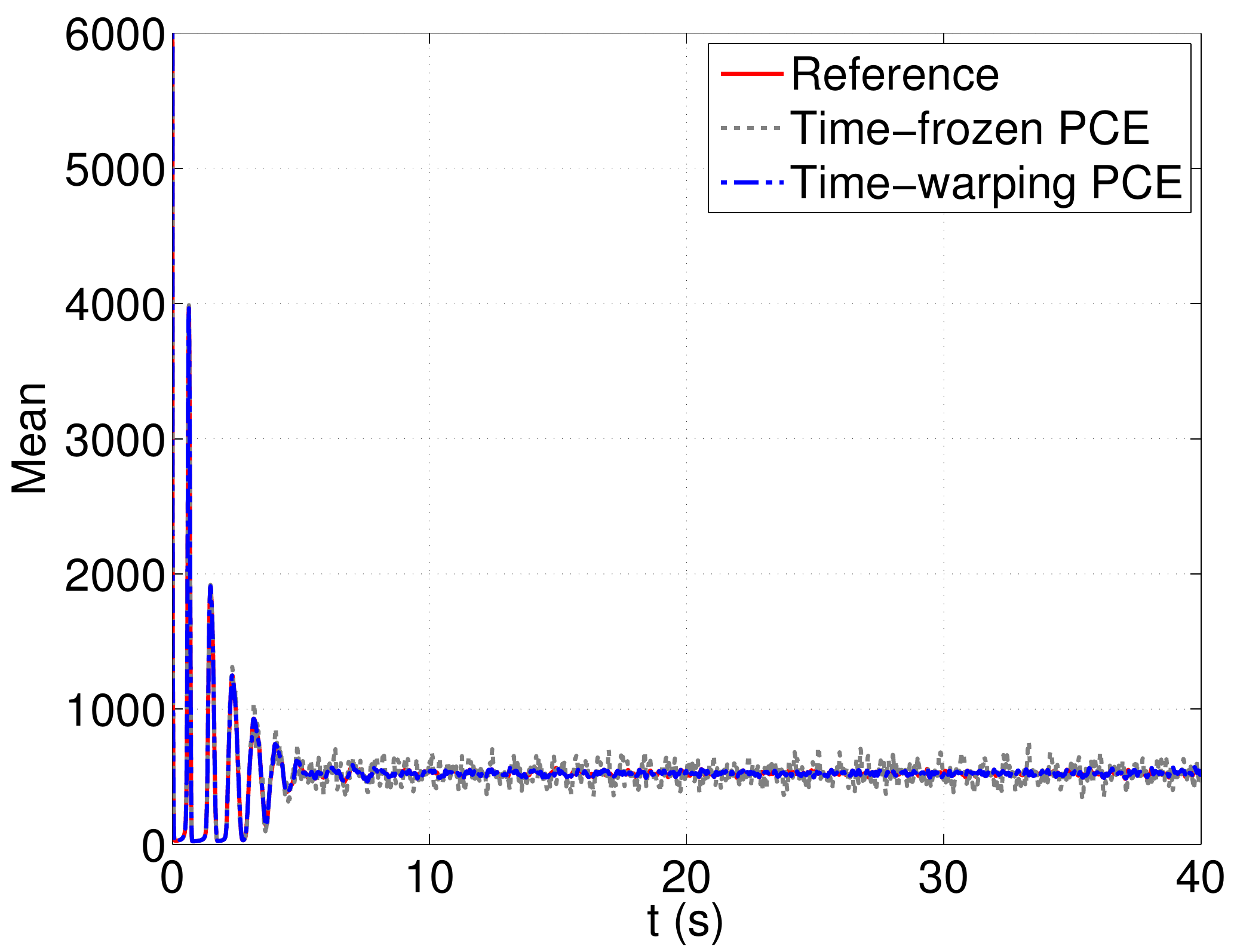}
	}
	\subfigure[Standard deviation trajectory]
	{
	\includegraphics[width=0.45\linewidth]{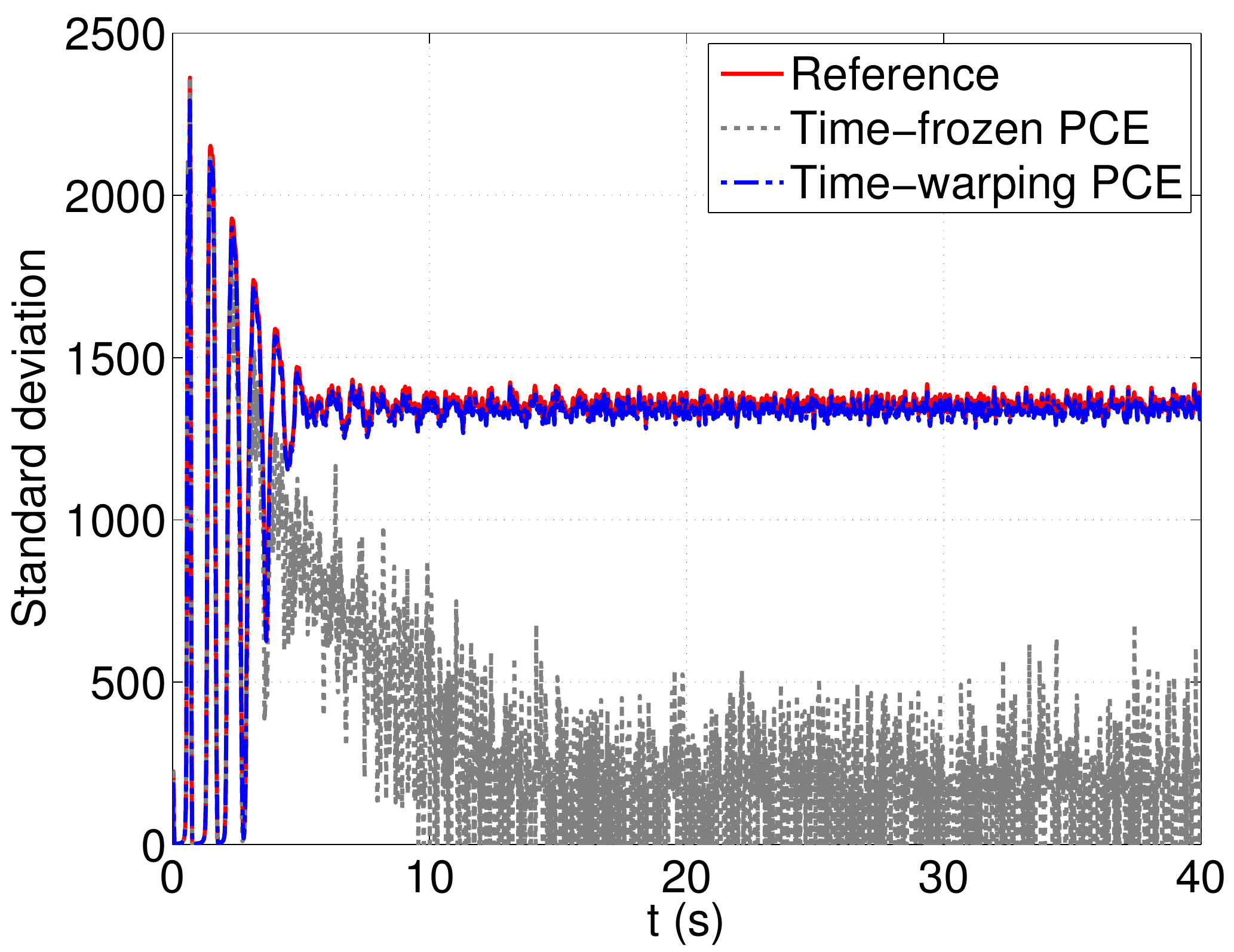}
	}
	\caption{Oregonator model -- Mean and standard deviation of $x(t)$: comparison of the two approaches.}
	\label{fig4.3.4}
\end{figure}

A validation set of $10,000$ trajectories is used to get reference trajectories of the concentration $x(t)$.
\figref{fig4.3.3} depicts two particular realizations computed by the numerical solver (Matlab ordinary differential equation solver \texttt{ode45}, using a time step $\Delta_t = 0.01$ for the total duration $T=40$~s) and predictions by PCEs with and without time-warping. It is shown that without time-warping, PCEs fail to capture the oscillatory behavior of the response. 
In contrast, the use of time-warping allows PCEs to predict the response with great accuracy. Only $1.24\%$ of the predictions (among $10,000$ samples) has a relative error larger than $0.1$.
\figref{fig4.3.4} depicts the statistics of $x(t)$ predicted by time-frozen and time-warping PCEs in comparison with MCS-based trajectories. Without time-warping, the estimates by PCEs differ significantly from the reference trajectories already from $3$~s. The discrepancies then quickly increase in time. For instance, PCEs without time-warping estimate a decreasing trend in time for the standard deviation, whereas the latter actually oscillates around a constant value (around 1400) with high frequency. By introducing the time-warping pre-processing, one can use sparse PCEs to capture the complex behavior of the time-dependent statistics of the response all along the trajectories.
The relative error for the mean and standard deviation trajectories are $3.11 \times 10^{-4}$ and $3.6 \times 10^{-3}$, respectively.

\begin{figure}[!ht]
	\centering
	\subfigure[$\vexi=(1.8970,   0.1001,  104.3676,    0.0077,25.542)$]
		{
		\includegraphics[width=0.45\linewidth]{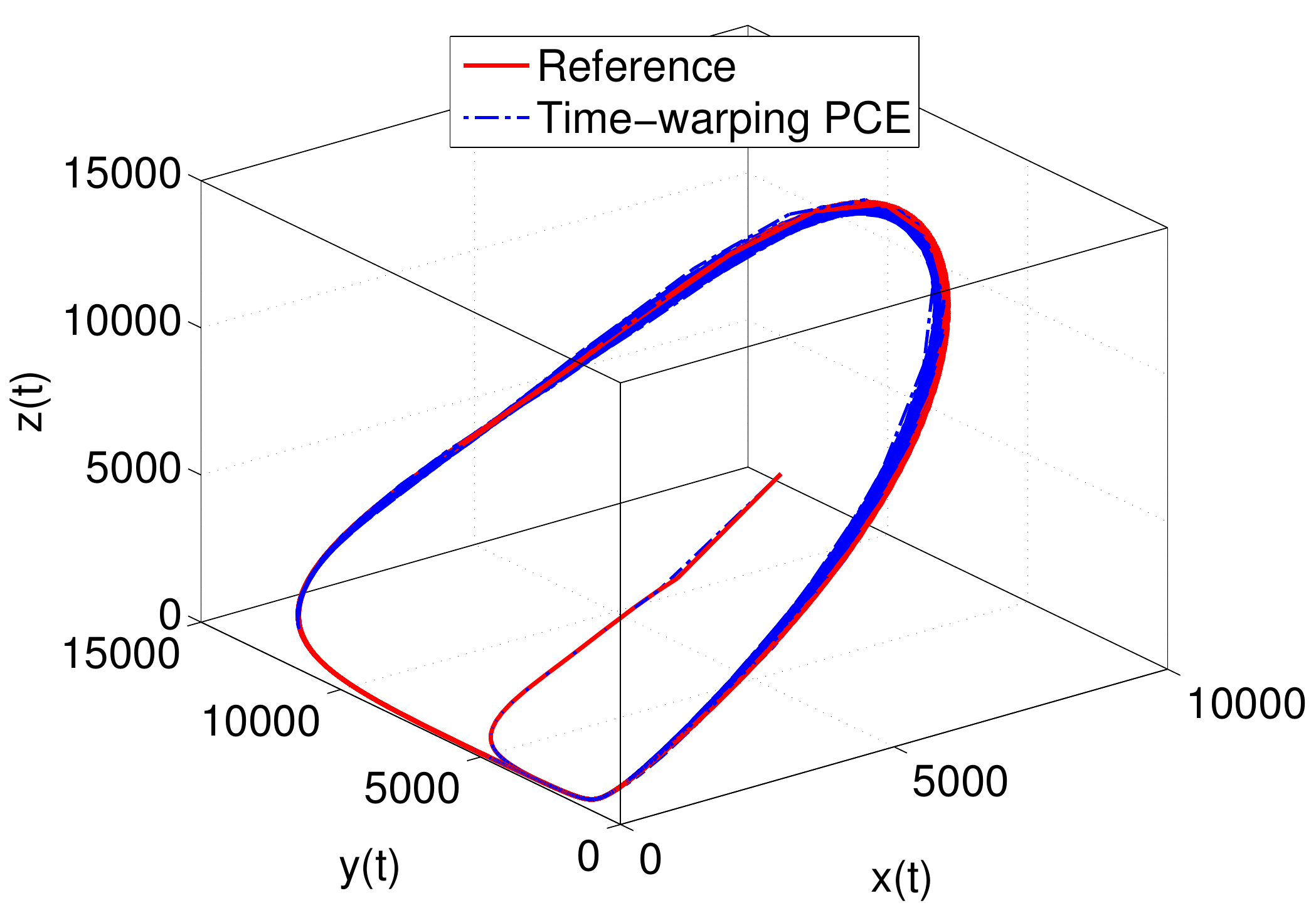}
		}
	\subfigure[$\vexi = (1.9481,\,    0.0999,\,  102.7929,\,    0.008,\,   27.648)$]
		{
		\includegraphics[width=0.45\linewidth]{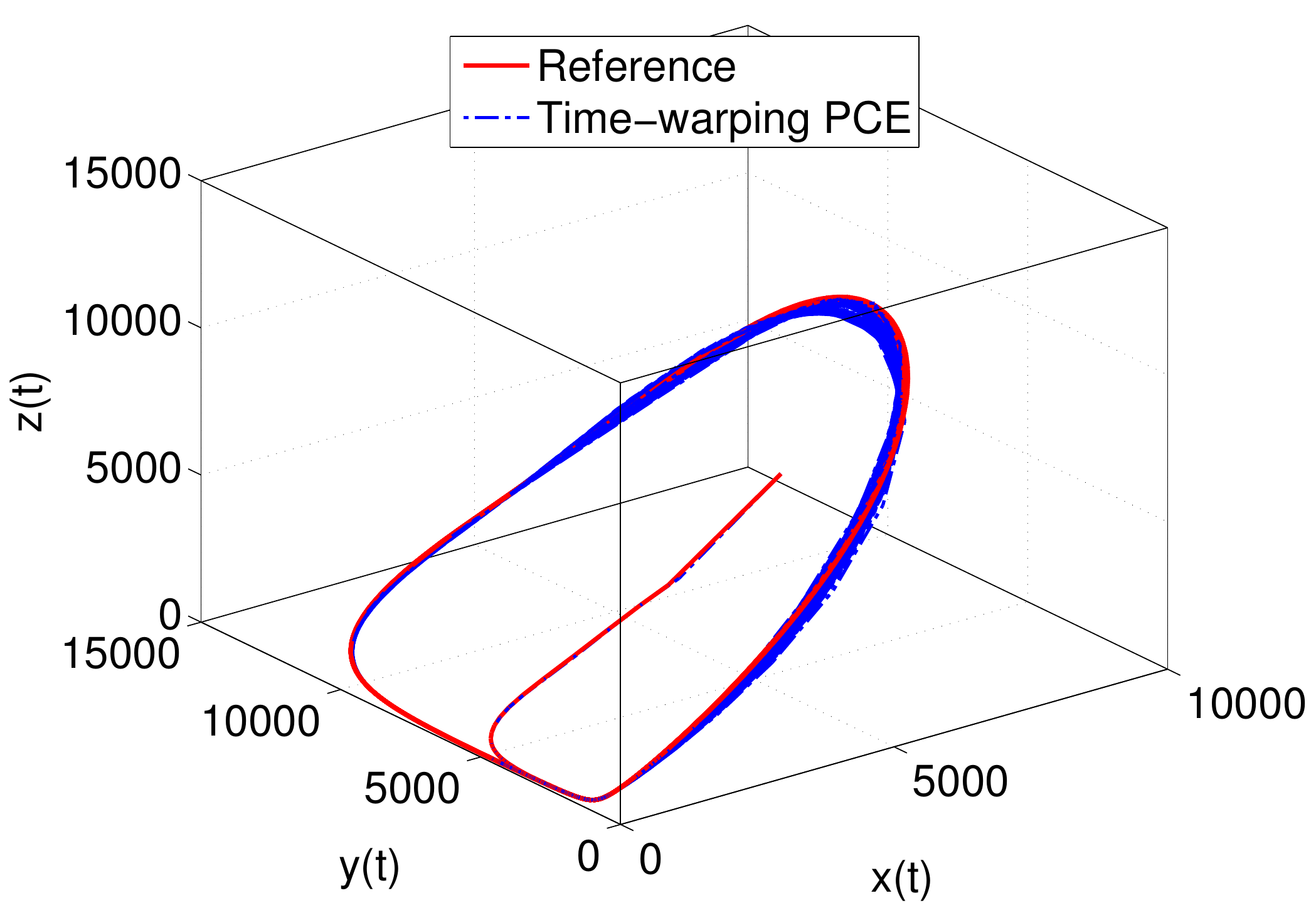}
		}
	\subfigure[Mean trajectory]
	{
	\includegraphics[width=0.45\linewidth]{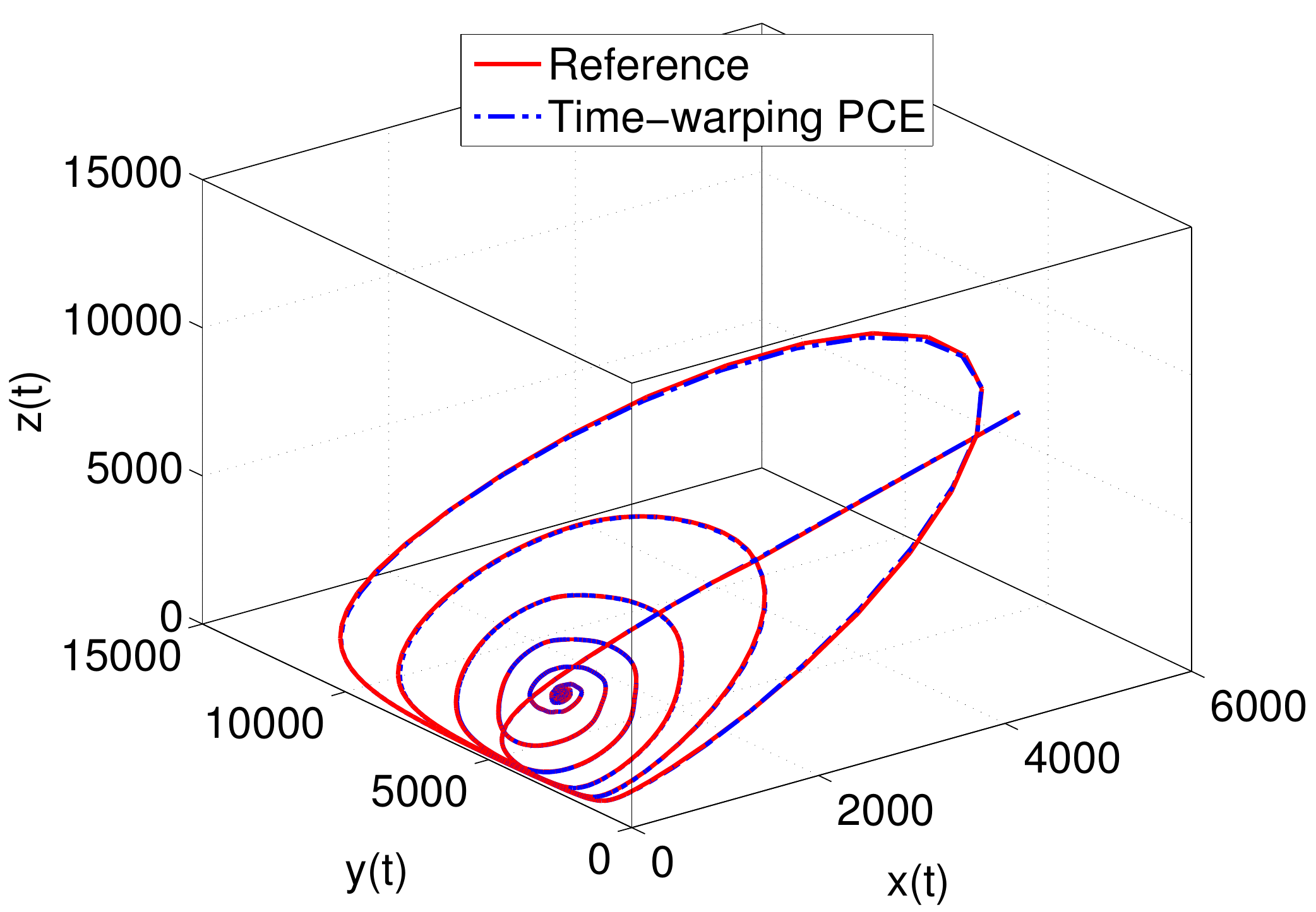}
	}
	\subfigure[Standard deviation trajectory]
	{
	\includegraphics[width=0.45\linewidth]{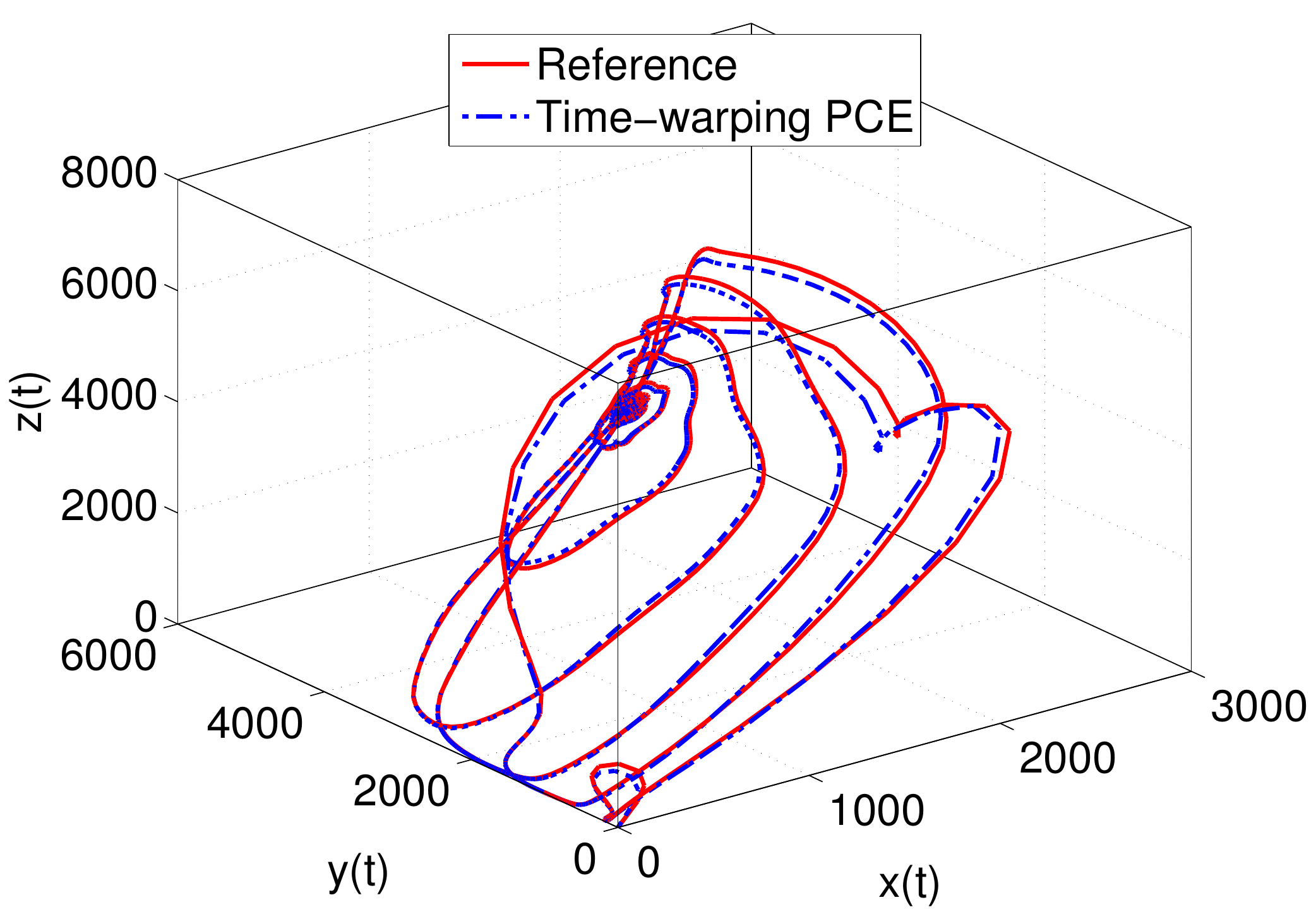}
	}
	\caption{Oregonator model --  Trajectories of $(x(t),y(t),z(t))$ predicted by time-warping PCEs \emph{vs.} the reference trajectories.}
	\label{fig4.3.5}
\end{figure}

Finally, the time-warping PCE scheme is applied to surrogate the responses $y(t)$ and $z(t)$ of the system using the same experimental design of size $50$ and the same procedure. \figref{fig4.3.5} shows a great agreement between two specific trajectories, the mean and standard deviation of $(x,y,z)$ in the state-space predicted by time-warping PCEs and the reference functions.
%

%%%%%%%%%%%%%%%%%%%%%%%%%%%%%%%%%%%%%%%%%%%%%%%%%%%%%%%%%%%%%%%%%%%%%%%%%%%%%%%%%%%%%
\subsection{Forced vibration of a Bouc-Wen oscillator}
In the previous case studies, self-oscillating systems were considered.
In this example, we show that the proposed approach is also applicable
to forced-vibration systems.  Let us now consider the SDOF Bouc-Wen
oscillator \cite{Kafali2007} subject to a stochastic excitation. The
equation of motion of the oscillator reads:
\begin{equation}
 \left\{
 \begin{array}{l}
    \ddot{y}(t) + 2 \, \zeta \, \omega \, \dot{y}(t) + \omega^2 (\rho \, y(t) + (1-\rho) \, z(t) ) =  - x(t) \, , \\
    \dot{z}(t) = \gamma \dot{y}(t)  - \alpha \, \abs{\dot{y}(t)} \, \abs{z(t)}^{n-1} z(t) - \beta \, \dot{y}(t) \, \abs{z(t)}^n \, .
 \end{array}
 \right.	
% \label{eq5.3.1}
\end{equation}
in which $\zeta$ is the damping ratio, $\omega$ is the fundamental frequency, $\rho$ is the post- to pre-yield stiffness ratio, $\gamma$, $\alpha$, $\beta$, $n$ are parameters governing the hysteretic loops and the excitation $x(t)$ is a sinusoidal function given by $x(t) = A \, \sin (\omega_x \, t)$.

Deterministic values are used for the following parameters of the Bouc-Wen model: $\rho =0$, $\gamma=1$, $n=1$, $\beta=0$. The remaining parameters $\vexi = \prt{\zeta,\,\omega,\, \alpha,\, A,\, \omega_x}$ are considered independent random variables with associated distributions given in Table~\ref{tab:boucparam}.
\begin{table}[!ht]
\caption{Uncertain parameters of the Bouc-Wen model}
\centering
\begin{tabular}{|c|c|c|c|c|}
\hline
Parameters & Distribution & Mean & Standard deviation & Coefficient of variation \\
\hline
$\zeta$ & Uniform & 0.02 & 0.002 & $0.1$ \\\hline
$\omega$ & Uniform & $2 \, \pi$ & $0.2 \, \pi$ & $0.1$\\\hline
$\alpha$ & Uniform & $50$ & $5$ & $0.1$ \\\hline
$A$ & Uniform & $1$ & $0.1$ & $0.1$ \\\hline
$\omega_x$ & Uniform & $\pi$ & $0.1 \, \pi$ & $0.1$ \\
\hline
\end{tabular}
\label{tab:boucparam}
\end{table}

\begin{figure}[!ht]
	\centering
	\subfigure[Original time scale $t$]
	{
	\includegraphics[width=0.45\linewidth]{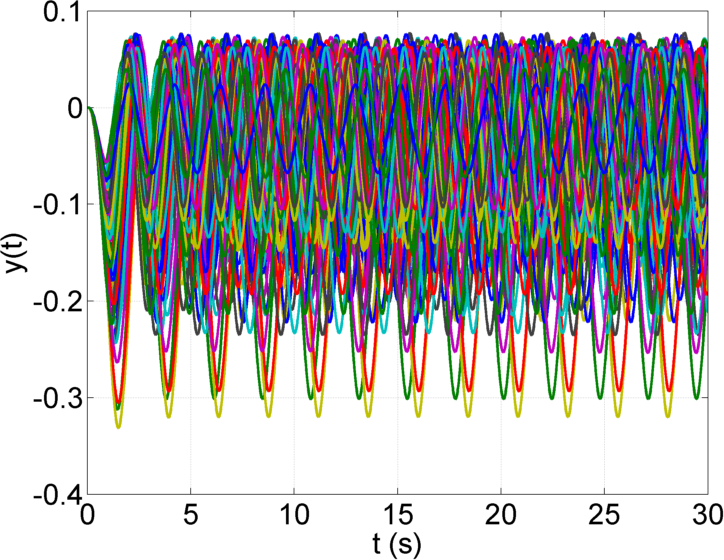}%{ex5y_vs_t.pdf}
	\label{fig:boucyta}
	}
	\subfigure[Warped time scale $\tau$]
	{
	\includegraphics[width=0.45\linewidth]{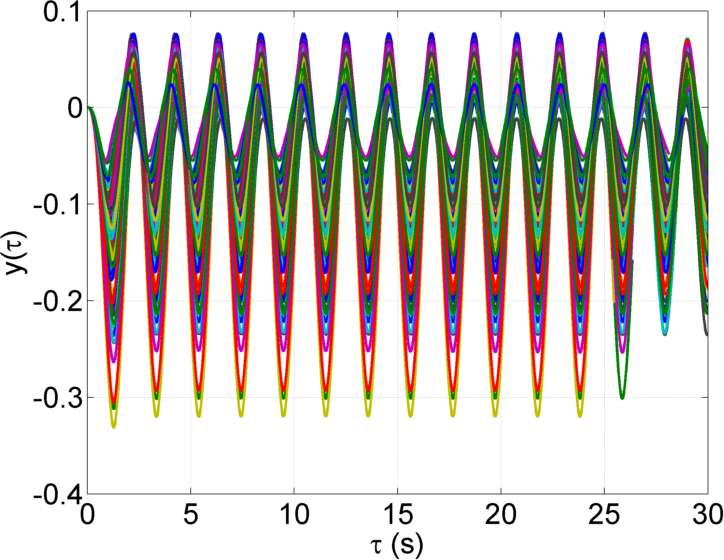}%{ex5y_vs_tau.pdf}
	\label{fig:boucytb}
	}
	\caption[Bouc-Wen oscillator -- Different trajectories of the solution in the original and warped time scales.]{Bouc-Wen oscillator -- $N=100$ different trajectories of the solution in the original time scale $t$ and in the transformed time line $\tau$.}
	\label{fig:boucyt}
\end{figure}

One aims at representing the oscillator displacement $y(t)$ as a function of the uncertain input parameters using time-frozen and time-warping PCEs. To this end, $100$ simulations of the oscillator are carried out using the Matlab solver \texttt{ode45} with time increment $\Delta_t = 0.005$~s for the total duration $T = 30$~s and initial condition $y(t=0) = 0$, $\dot{y}(t=0) = 0$. The displacement trajectories are depicted in \figref{fig:boucyta}.

First, the time-frozen sparse PCEs are computed with candidate polynomials up to total degree $20$.
For this case study, a time-warping scheme $\tau = k \, t$ with only one parameter is used. After the time-warping process, the trajectories become in-phase as depicted in \figref{fig:boucytb}. Adaptive sparse PCE representing $k$ has the relative LOO error $5 \times 10^{-5}$. In order to achieve a truncation error $\epsilon_1$ smaller than $1 \times 10^{-3}$, $13$ first principal components are retained in PCA. 
%\figref{fig:boucpcacmps} depicts the first eight principal components.
The relative LOO errors of PCEs for the first two components are $6 \times 10^{-3}$ and $6.21 \times 10^{-2}$, respectively.

\begin{figure}[!ht]
	\centering
	\subfigure
	[$\vexi = \prt{0.0191, 5.6208, 57.3581, 0.9401, 2.8577}$]
	{
	\includegraphics[width=0.45\linewidth]{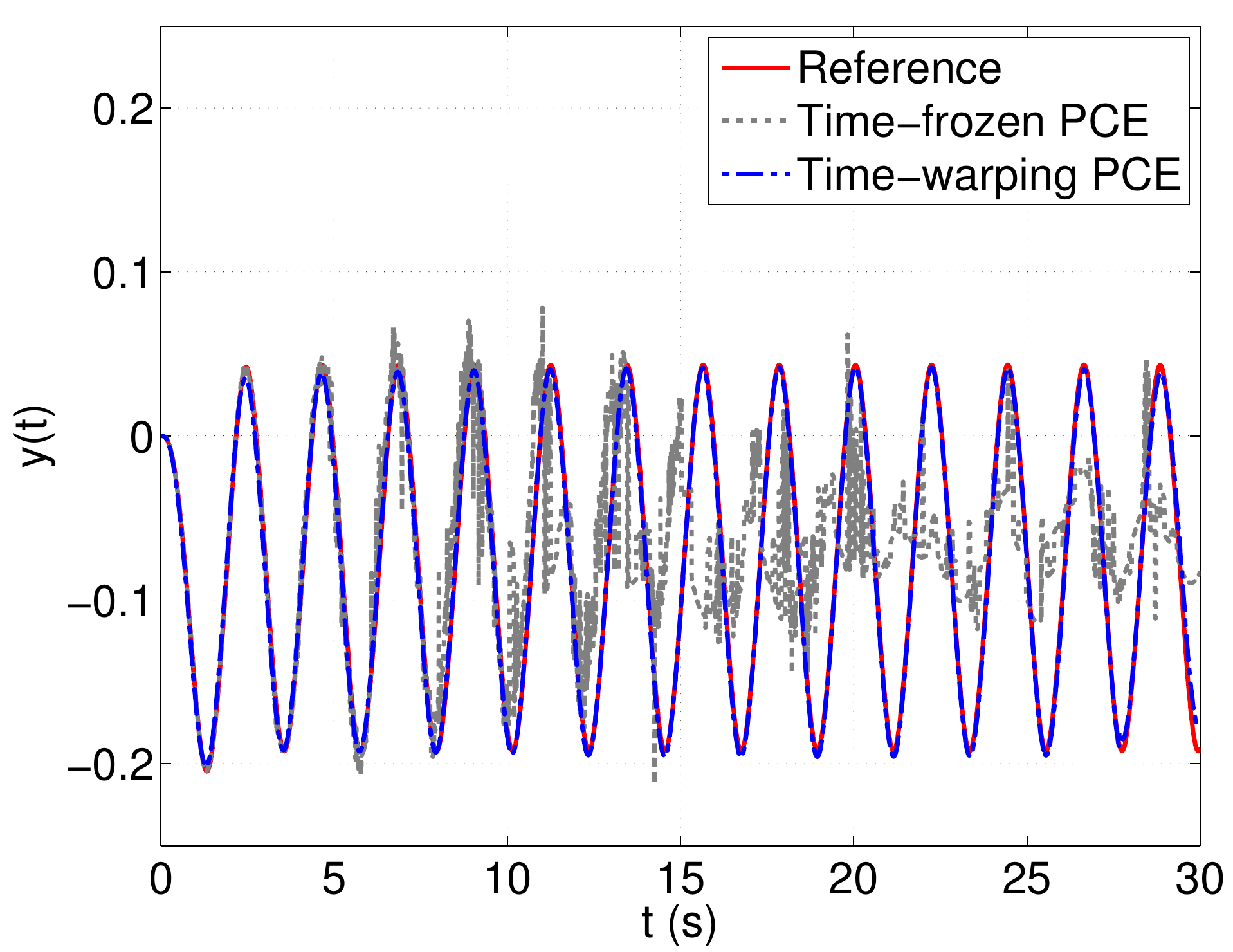}
	}	
	\subfigure
	[$\vexi = \prt{0.0196, 6.1226, 46.9916, 1.0291, 3.4542}$]
	{
	\includegraphics[width=0.45\linewidth]{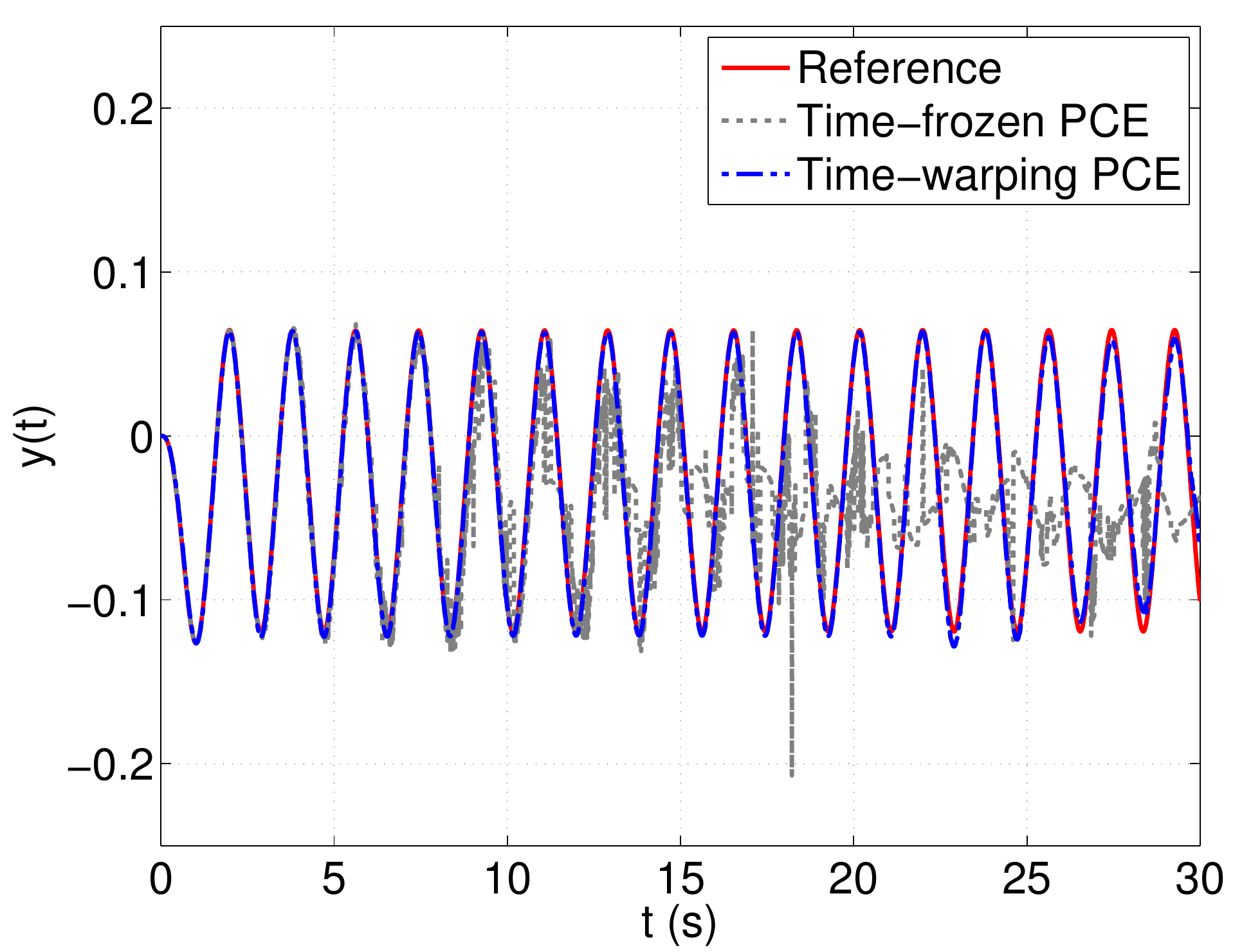}
	}
	\caption{Bouc-Wen oscillator -- Two particular trajectories and their predictions by the two approaches.}
	\label{fig:bouc2predict}
\end{figure}

\begin{figure}[!ht]
	\centering
	\subfigure[Mean trajectory]
	{
	\includegraphics[width=0.45\linewidth]{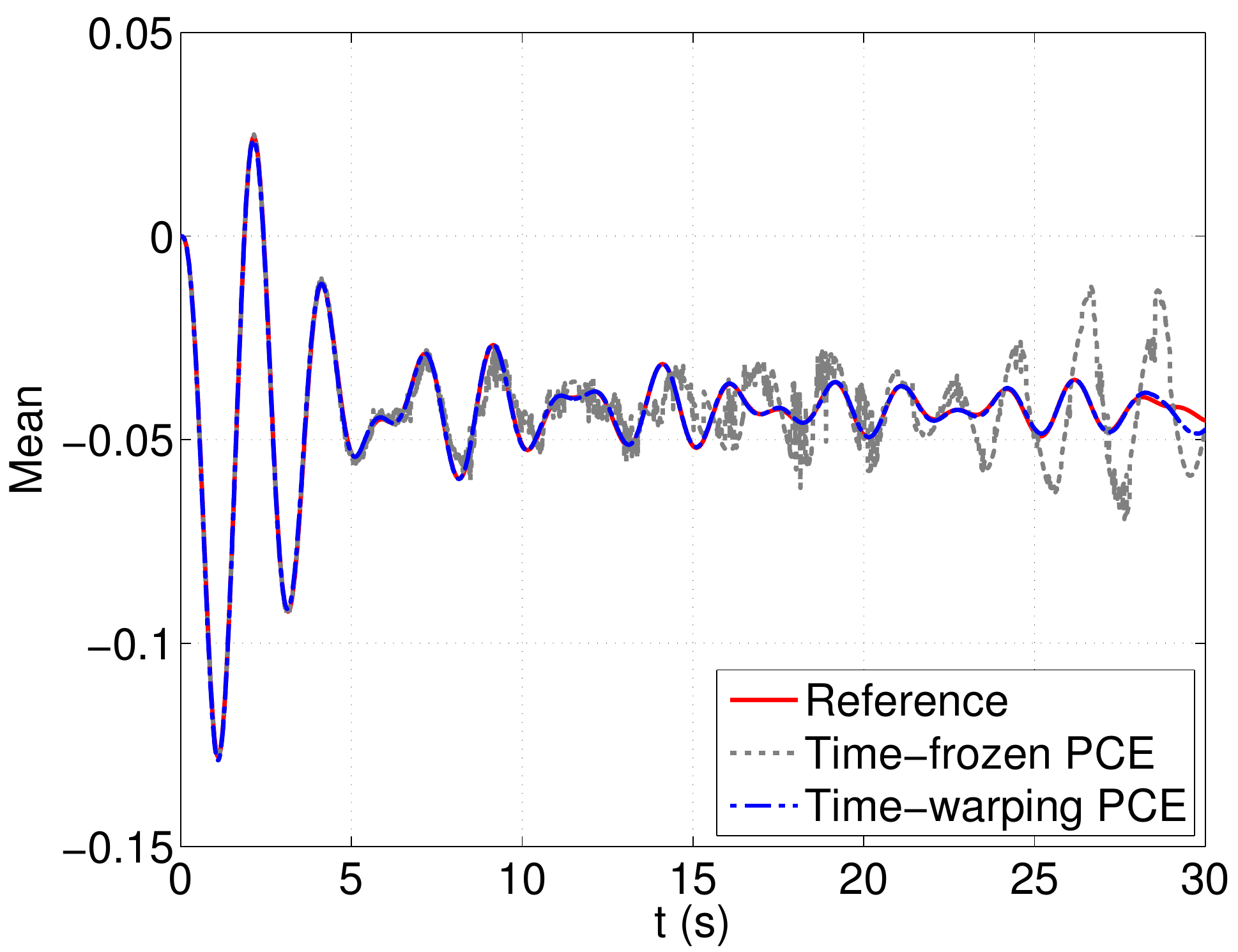}
	}
	\subfigure[Standard deviation trajectory]
	{
	\includegraphics[width=0.45\linewidth]{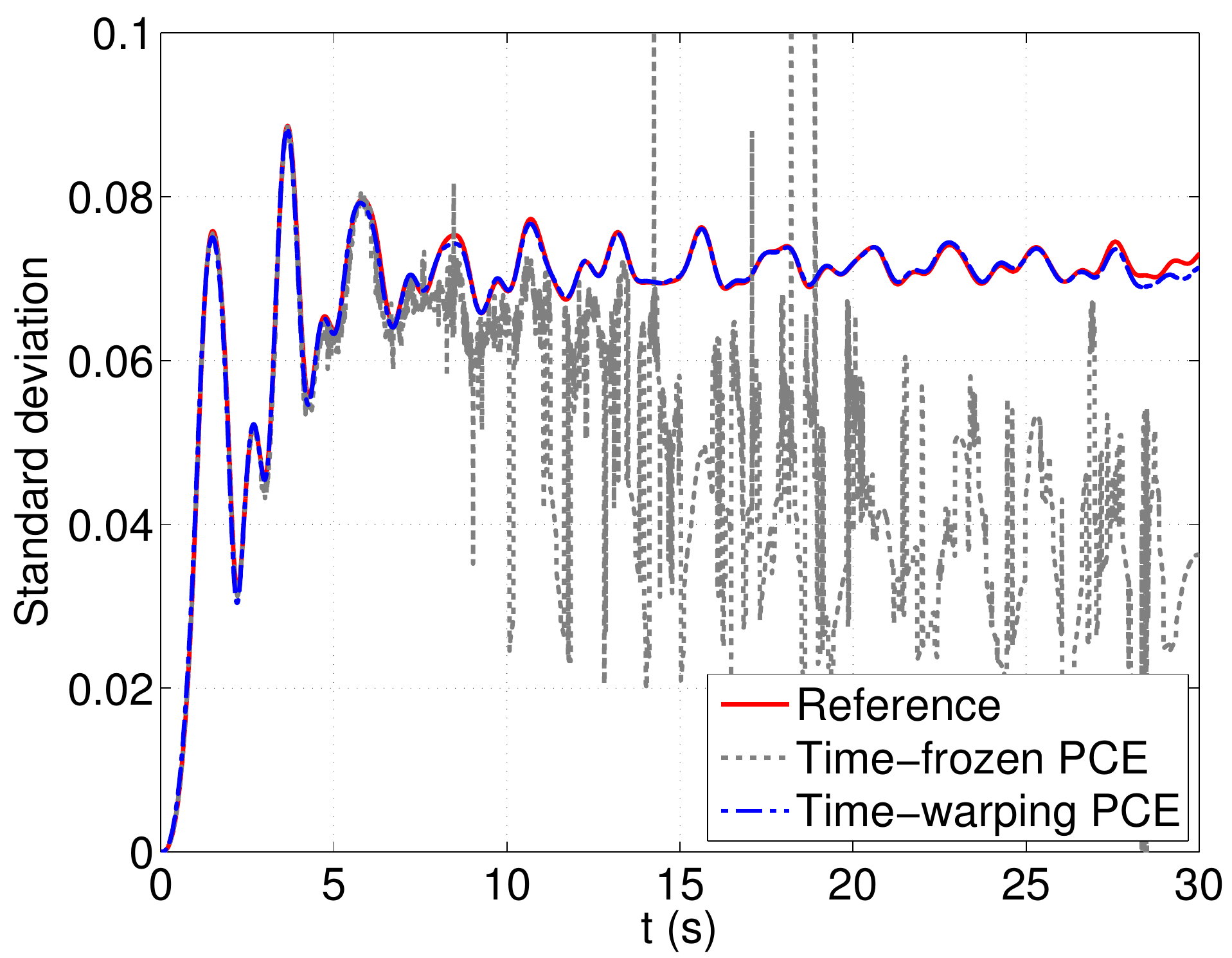}
	}
	\caption{Bouc-Wen oscillator -- Mean and standard deviation of the trajectories: comparison of the two approaches.}
	\label{fig:boucmeanstd}
\end{figure}

Let us validate the accuracy of the time-warping PCE model. In \figref{fig:bouc2predict}, two specific predictions of the PCE model are plotted against the actual responses obtained with the original Matlab solver. A remarkable agreement can be observed. 
Among $10,000$ validations, only $4.87\%$ has a relative error larger than $0.1$.
Regarding the time-dependent mean and standard deviation of the oscillator, time-warping PCE-based estimates outstandingly match the reference trajectories (\figref{fig:boucmeanstd}). Only a minor discrepancy can be observed at the end of the considered time duration $T=30$~s, which is due to the modest number of simulations used as the experimental design. The corresponding relative errors are both $2.4 \times 10^{-3}$. On the contrary, time-frozen PCEs exhibit a low level of accuracy after $5$ seconds.

It is worth noting that in the current case study, we considered both uncertainties from the mechanical properties and the excitations. In particular, complicated hysteretic behavior was investigated. To the best of the authors' knowledge, this is the first time that such a system is considered in the literature of uncertainty quantification for the purpose of deriving time-dependent surrogate models.
%%%%%%%%%%%%%%%%%%%%%%%%%%%%%%%%%%%%%%%%%%%%%%%%%%%%%%%%%%%%%%%%%%%%%%%%%%%%%%%%%%%%%
%%%%%%%%%%%%%%%%%%%%%%%%%%%%%%%%%%%%%%%%%%%%%%%%%%%%%%%%%%%%%%%%%%%%%%%%%%%%%%%%%%%%%
\section{Discussion}
The various numerical applications in chemical and mechanical engineering have proved the effectiveness of the time-warping PCE approach, which may be shortly explained as follows.
It was observed that when represented in the space of the temporal variable $t$, the system's responses are increasingly non-linear functions of the uncertain parameters.
When projecting the responses onto a \emph{suitable} space, in this case the transformed time line $\tau$, the resulting trajectories become smooth functions of the uncertain input parameters, whose complexity does hardly increase with time. Therefore, PCEs can be applied effectively to the projected responses and represent well the solutions at late instants.
In this paper, a measure of similarity was proposed to define a suitable space for projecting the responses, which exploits the periodicity of the trajectories.
Further investigations are required to clearly determine such a suitable space in a more general case.

In the proposed approach, the virtual time $\tau$ is a function of the uncertain parameters $\vexi$.
In other words, the basis $\tau$ onto which the responses are projected is not deterministic. {This is a feature shared by the approach based on multiscale stochastic preconditionners \cite{Alexanderian2012, Alexanderian2014}.}
This differs significantly from approaches commonly used in the literature, in which the response trajectories are first projected onto a set of \emph{deterministic} reduced basis determined a priori using a set of numerical simulations of the system.
This is usually done with a simple \emph{linear} transform, for instance data compression techniques such as principal component analysis or wavelet decomposition.

When analyzing further, one discovers a particular feature which constitutes a major difference between the classical time-frozen PCE approach and the proposed time-warping method.
The PC coefficients $y_{\veb}(\tau)$ in the time-warping representation (\eqrefe{eq:timewarpPCE}) are functions of $\tau$, therefore being dependent on $\vexi$. This contradicts the representation of time-frozen PCEs (\eqrefe{eq:timefroPCE}), in which $t$ and $\vexi$ intervene in the solution in a separated manner.

From a more general perspective, the effectiveness of the approach can be explained by analyzing the functionalities of the time-warping process and PCEs. The most important feature of an oscillatory trajectory consists in its spectral content, which is characterized by the vibration periodicity. The other feature is the temporal content characterized by the vibration amplitude.
The pre-processing step handles partially the dynamics of the system by dealing with the frequency content. Using the time-warping process, the resulting trajectories have similar frequencies and phases. In other words, in terms of frequencies, the transformed trajectories exhibit a similar dynamical behavior, which is close to that of the reference trajectory. The other aspect of the dynamics, \ie the random temporal amplitude of the trajectories, is handled with sparse PCEs. As a summary, the dynamics is captured by the time-warping process, whereas the uncertainties are represented by PCEs.

As explained, sparse PCEs alone are not capable of dealing with the
dynamics. The proposed approach illustrates a novel way to solve
stochastic dynamical problems, in which a specialized technique might be
used to capture the dynamical aspect whereas sparse PCEs are used to
propagate uncertainties.  From this perspective, Yaghoubi et al. \cite{Yaghoubi2016}
have recently applied the warping-based approach in the frequency domain
to surrogate the frequency response function of mechanical systems.
This principle is further developed by Mai et al. \cite{Mai2016IJ4UQ2} to tackle
more complex problems in which non-linear uncertain structures subject
to stochastic motions are of interest and where the response
trajectories are non-stationary, \ie they do not show pseudo-periodic
oscillations.  The projection of the responses onto a special basis made
of auto-regressive functions will allow us to represent the non-linear
dynamical behavior of the systems.

In addition, it is worthwhile mentioning that the proposed methodology is fully non-intrusive, \ie the surrogate models of the systems' response trajectories are obtained by using a pre-computed set of trajectories related to an experimental design. In this respect, the methodology is readily applicable to any other problems featuring randomized limit cycle oscillations.

{Finally, it is noteworthy that the current approach exhibits some limitations. First of all, a linear time transform was used for all the considered numerical applications. More generalized transforms involving a non-linear dependence of the transformed time on the physical temporal variable, see \eg \cite{Alexanderian2012,Alexanderian2014}, might be considered in future researches. A multi-linear stochastic time transform similar to the approach introduced in \cite{Yaghoubi2016} in the frequency domain should be investigated to handle
the responses of uncertain dynamical systems in the transient and stationary phases or
address the complex random polychromatic responses.}
%%%%%%%%%%%%%%%%%%%%%%%%%%%%%%%%%%%%%%%%%%%%%%%%%%%%%%%%%%%%%%%%%%%%%%%%%%%%%%%%%%%%%
%%%%%%%%%%%%%%%%%%%%%%%%%%%%%%%%%%%%%%%%%%%%%%%%%%%%%%%%%%%%%%%%%%%%%%%%%%%%%%%%%%%%%
\section{Conclusions and perspectives}
Polynomial chaos expansions (PCEs) represent an effective metamodeling technique which has been efficiently used in several practical problems in a wide variety of domains. It is, however, well known that PCEs fail when modeling the stochastic responses at late instants of dynamical systems. In this paper, we pointed out the cause of the failure, which is mainly associated with the large dissimilarities between distinct responses introduced by the variability of the uncertain parameters. 

To address the above issue, we suggested an approach which consists in representing the responses into a virtual time line where the similarities between different response trajectories are maximized. The virtual time line is obtained by warping, \ie scaling and shifting, the original time grid.
The parameters governing the trajectory-dependent time warping are determined by means of a global optimization problem using an objective function herein introduced to quantify the similarity between distinct trajectories.
The proposed approach allows one to effectively solve complex benchmark problems from mechanics and chemistry using only low-order PCEs.
This approach also suggests that when representing the original response quantities onto a suitable transformed space, the complexity of the responses may reduce significantly, thus allowing more effective application of PCEs. In general, pre-processing the experimental design before applying PCEs is a promising approach that needs further investigation.

\appendix
\section*{APPENDIX}

%%%%%%%%%%%%%%%%%%%%%%%%%%%%%%%%%%%%%%%%%%%%%%%%%%%%%%%%%%%%%%%%%%%%%%%%%%%%%%%%%%%%%
\section{Rigid body dynamics}
\begin{figure}[h]
\centering
\subfigure[Principal components 1-4]
{
  \includegraphics[width=0.45\linewidth]{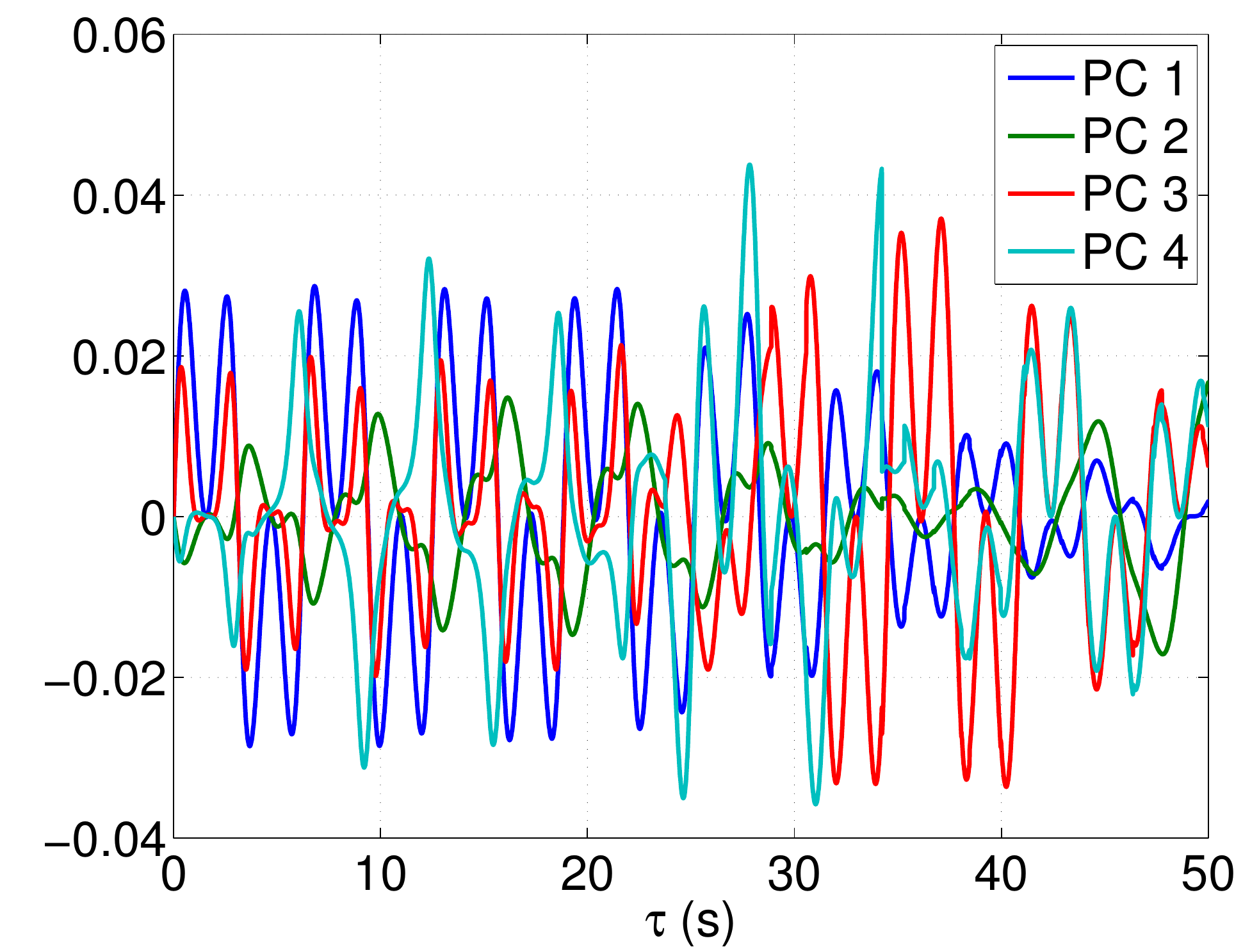}
}
\subfigure[Principal components 5-8]
{
  \includegraphics[width=0.43\linewidth]{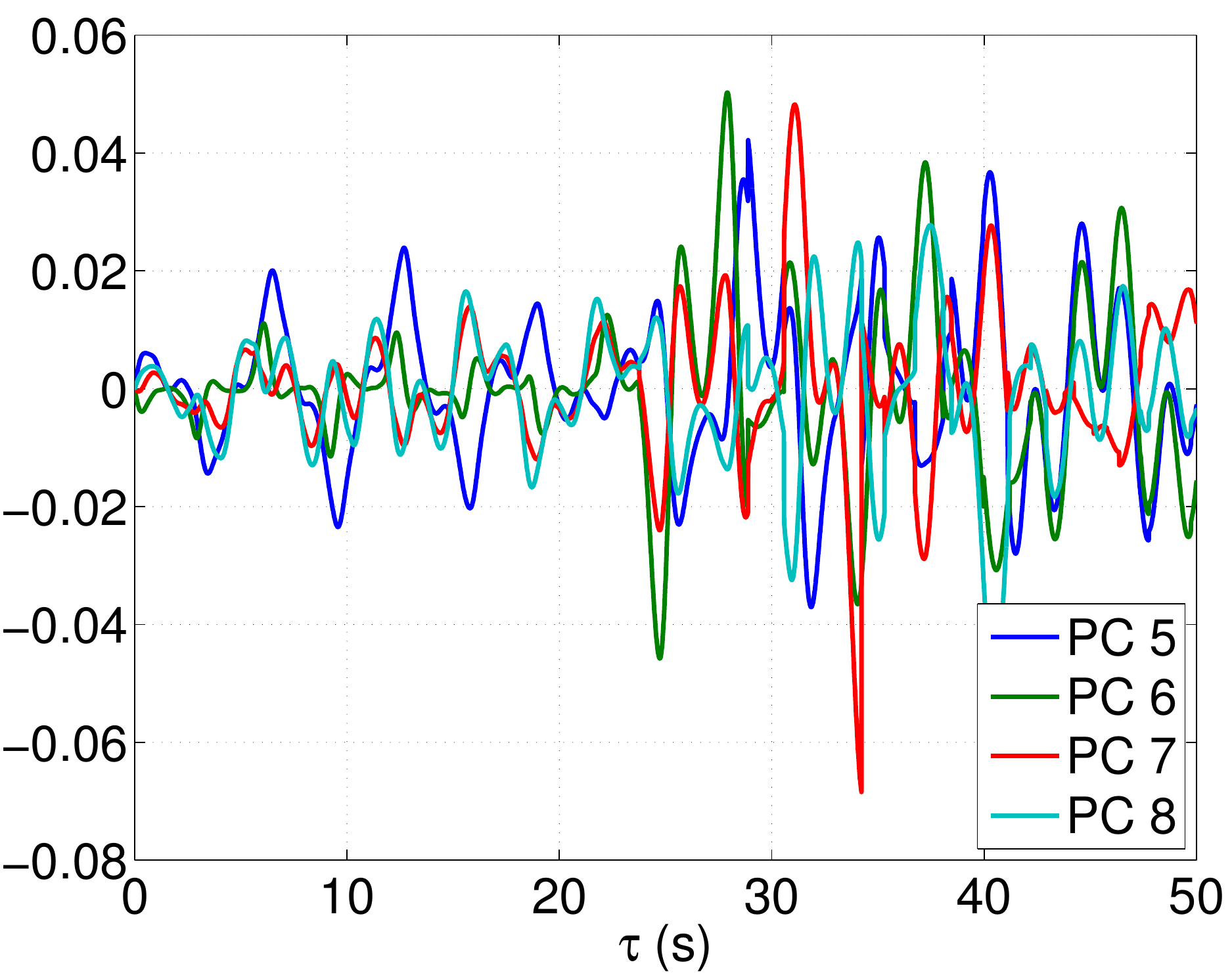}
}
\caption{Rigid body dynamics -- The first eight principal components.}
\label{fig:rigpcacmps}
\end{figure}

This section presents supplementary results of the investigation on the
rigid body system. Figure~\ref{fig:rigpcacmps} presents the eight first
components obtained from the principal component analysis of the
trajectories in the time-warped scale. 

%

%%%%%%%%%%%%%%%%%%%%%%%%%%%%%%%%%%%%%%%%%%%%%%%%%%%%%%%%%%%%%%%%%%%%%%%%%%%%%%%%%%%%%
\section{Oregonator model}
\begin{figure}[!ht]
  \centering
  \subfigure[Original time scale $t$]
  {
    \includegraphics[width=0.45\linewidth]{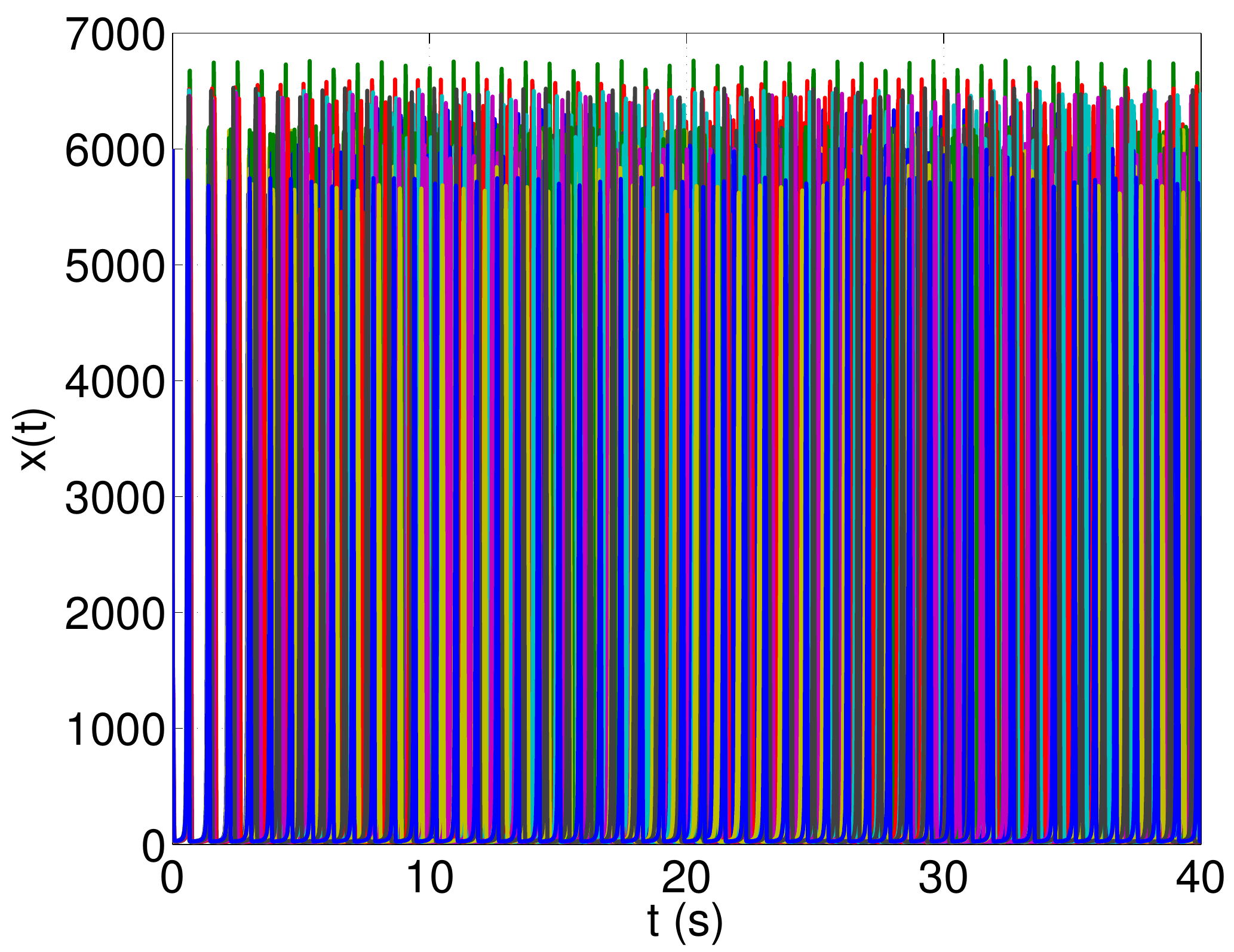}
    % \label{fig4.3.1a}
  }
  \subfigure[Warped time scale $\tau$]
  {
    \includegraphics[width=0.45\linewidth]{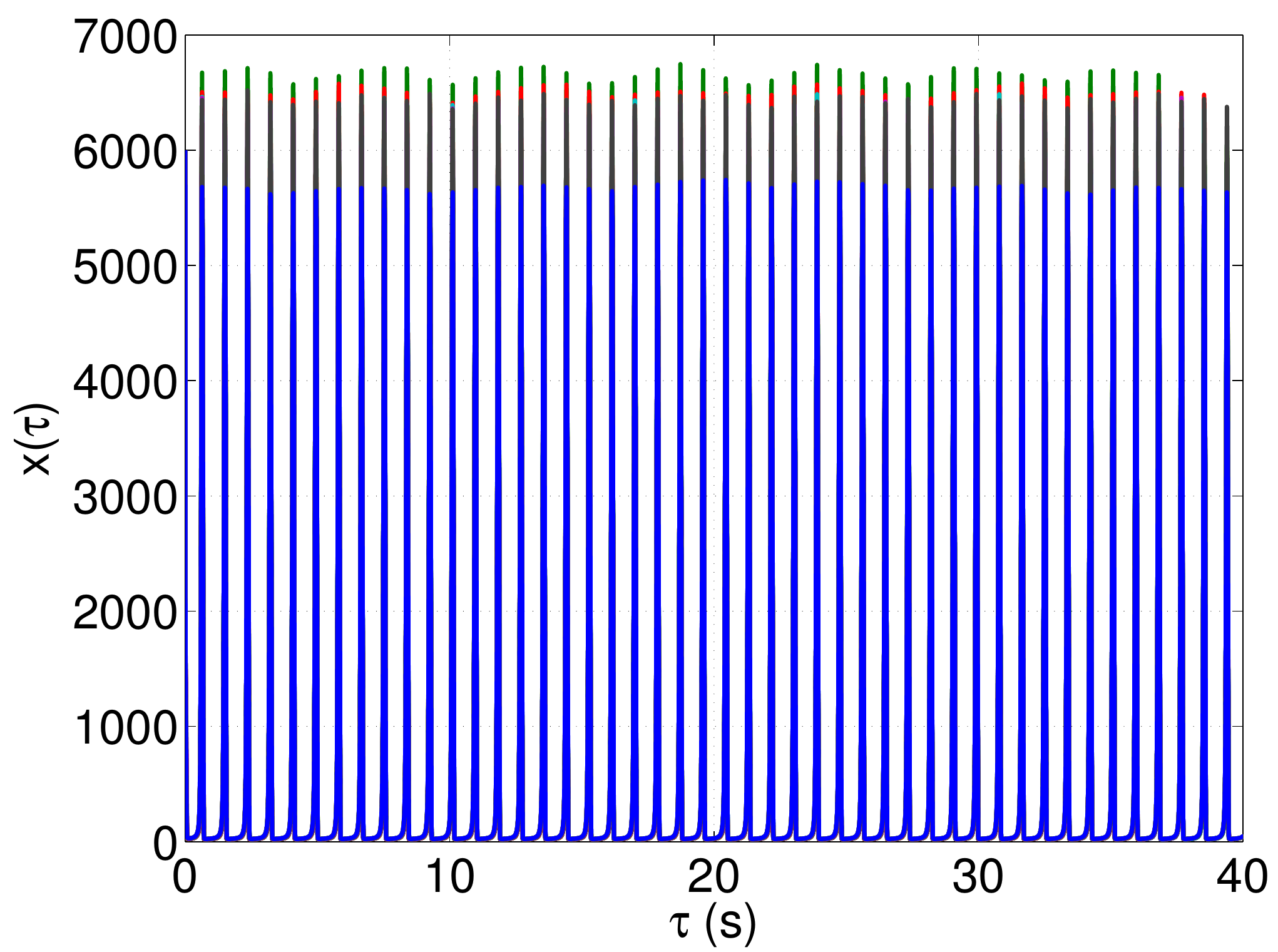}
    % \label{fig4.3.1b}
  }
  \caption[Oregonator model -- Different trajectories of the response in the original and transformed time scales]{Oregonator model -- $N=50$ different trajectories of the response $x(t)$.}
  \label{fig4.3.1s}
\end{figure}

This section presents supplementary results of the investigation on the
Oregonator model. Figure~\ref{fig4.3.1s}  presents 50 trajectories
plotted in the original time scale (Fig. (A)) and after time warping
(Fig.~(B)). It is visually obvious that the time-warping pre-processing
aligns well these trajectories with each other.

\begin{figure}[h]
  \centering
  \subfigure
  [$\vexi=(1.8970,   0.1001,  104.3676,    0.0077,25.5417)$]
  {
    \includegraphics[width=0.45\linewidth]{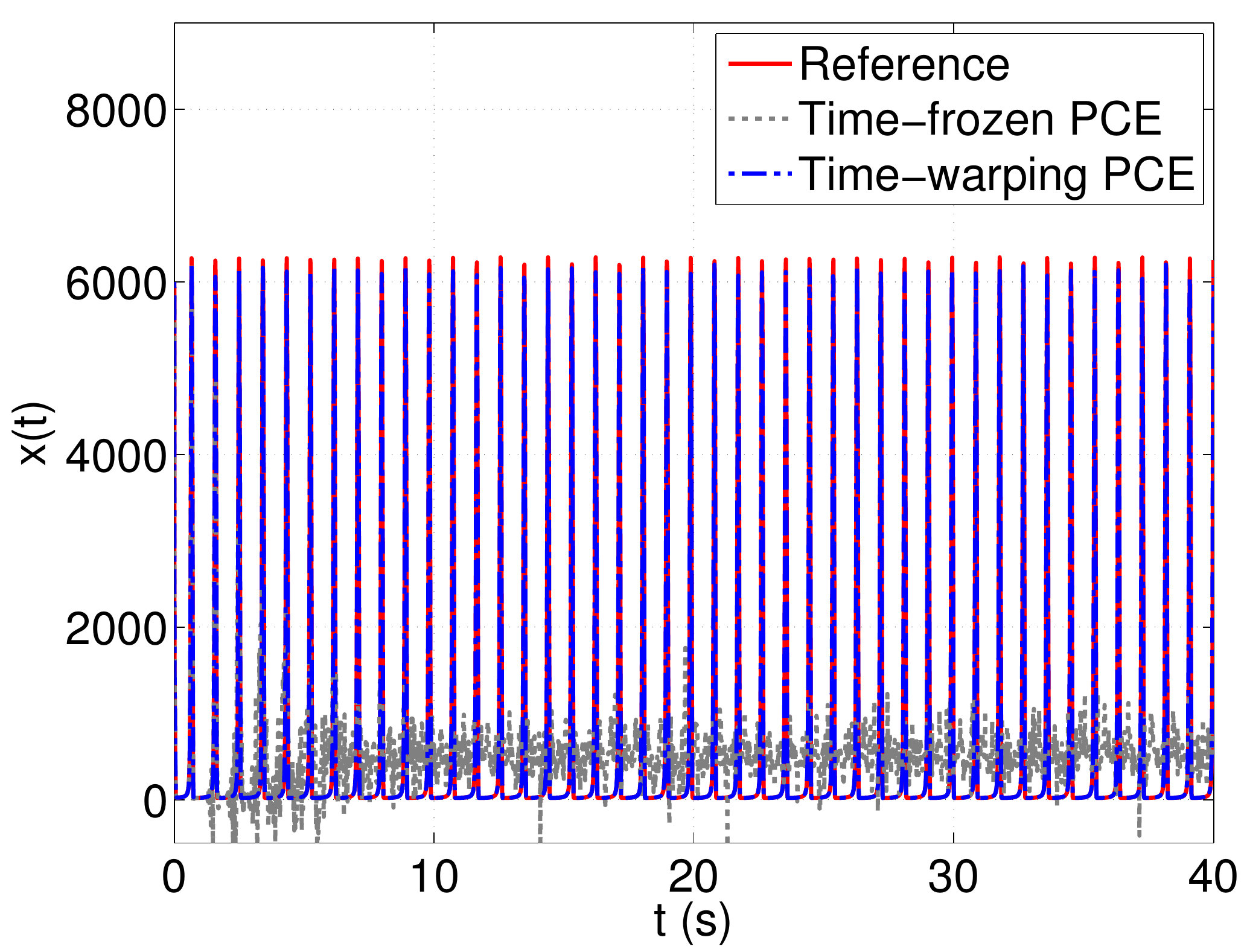}
  }	
  \subfigure
  [$\vexi = (1.9481,\,    0.0999,\,  102.7929,\,    0.008,\,   27.6482)$]
  {
    \includegraphics[width=0.45\linewidth]{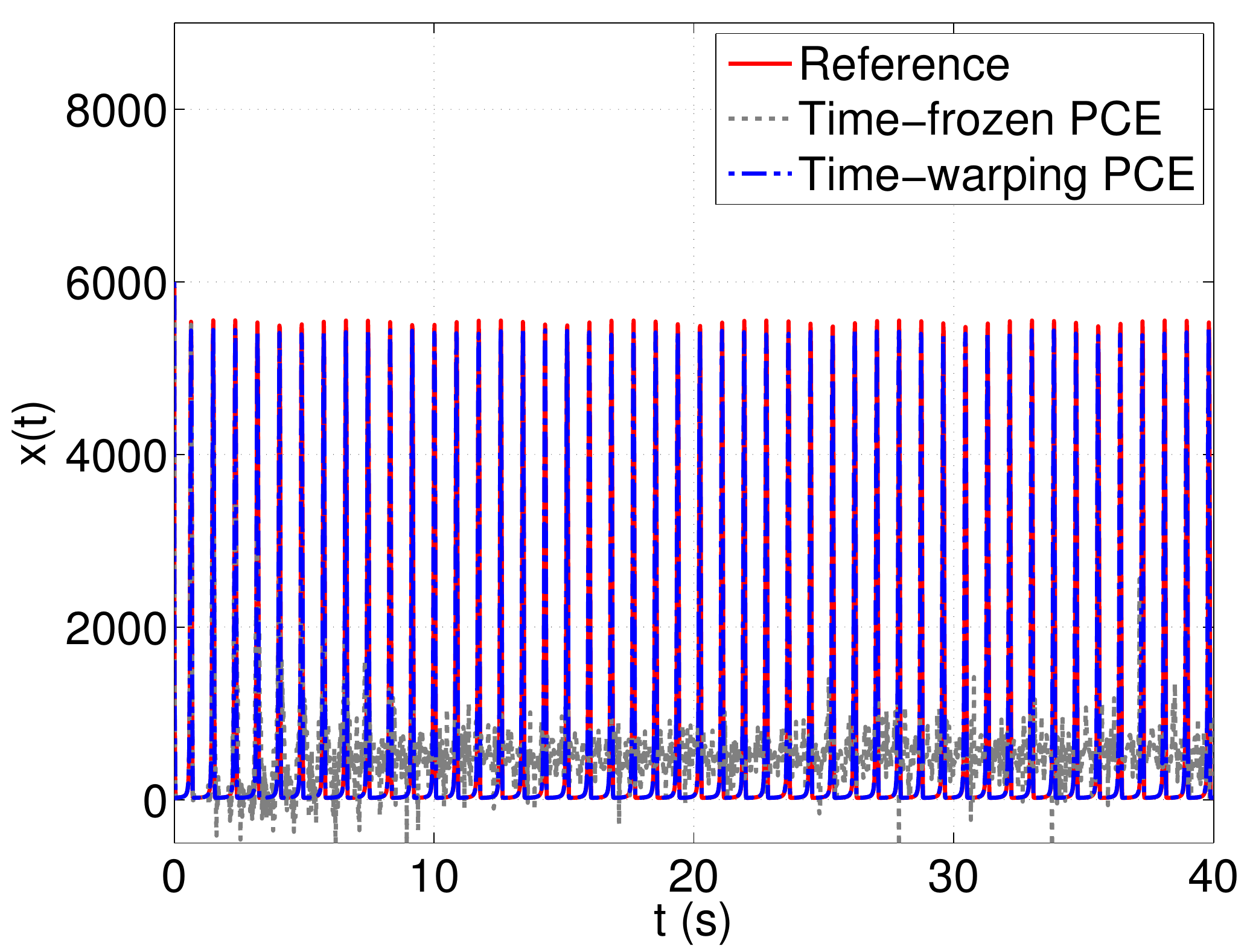}
  }
  \caption[Oregonator model -- Two particular trajectories and their predictions by time-frozen and time-warping PCEs.]{Oregonator model -- Two particular trajectories $x(t)$ and their predictions by time-frozen and time-warping PCEs.}
  \label{fig4.3.3s}
\end{figure}

Figure~\ref{fig4.3.3s} shows two particular trajectories obtained from
the original Oregonator model, as well as their prediction using
time-frozen and time-warping PCE. Time-frozen PCE essentially generates
numerical noise after a few seconds, whereas the prediction by
time-warping PCE is accurate until the latest time instants.

%%%%%%%%%%%%%%%%%%%%%%%%%%%%%%%%%%%%%%%%%%%%%%%%%%%%%%%%%%%%%%%%%%%%%%%%%%%%%%%%%%%%%
\section{Duffing oscillator}
Let us consider a non-linear damped single-degree-of-freedom (SDOF)
Duffing oscillator under free vibration, which is described by the
following equation of motion:
\begin{equation}
 \ddot{y}(t) + 2 \, \omega \, \zeta \, \dot{y}(t) + {\omega}^2 \,(y(t) + \epsilon \, y^3(t) ) =  0.
 \label{eq4.3.1}
\end{equation}
The oscillator is driven by uncertain parameters $\vexi = \prt{\zeta,\,
  \omega,\, \epsilon}$ described in Table~\ref{tab:2}. The initial
conditions are considered deterministic with $y(t=0)=1$ and
$\dot{y}(t=0)=0$.  Note that a simplified form of this equation which
represents an undamped linear oscillator was used in other publications
for illustrating the time-dependent generalized polynomial chaos
\cite{Gerritsma2010}, the intrusive time-transform approach
\cite{LeMaitre2009} and the flow map composition PCEs
\cite{Luchtenburg2014}.
\begin{table}[!ht]
\caption{Duffing oscillator -- Probabilistic model of the uncertain parameters}
\centering
\begin{tabular}{|c|c|c|c|c|}
\hline
Parameters & Distribution & Mean & Standard deviation & Coefficient of variation  \\
\hline
$\zeta$ & Uniform & $0.03$ & $0.015/\sqrt{3}$ & $0.2887$ \\\hline
$\omega$ & Uniform & $ 2\, \pi$  & $\pi/\sqrt{3}$ & $0.2887$ \\\hline
$\epsilon$ & Uniform & $-0.5$ & $0.25/\sqrt{3}$ & $0.2887$\\
\hline
\end{tabular}
\label{tab:2}
\end{table}

\begin{figure}[!ht]
	\centering
	\subfigure[Original time scale $t$]
	{
	\includegraphics[width=0.45\linewidth]{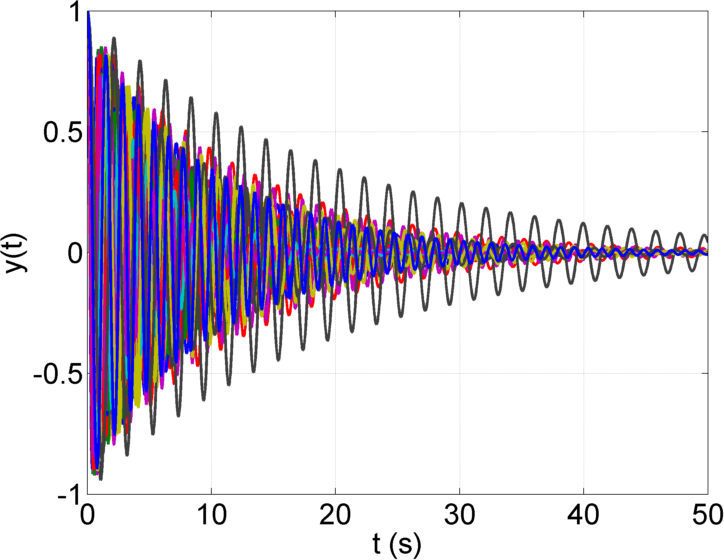}
	\label{fig:dufyt}
	}
	\subfigure[Warped time scale $\tau$]
	{
	\includegraphics[width=0.45\linewidth]{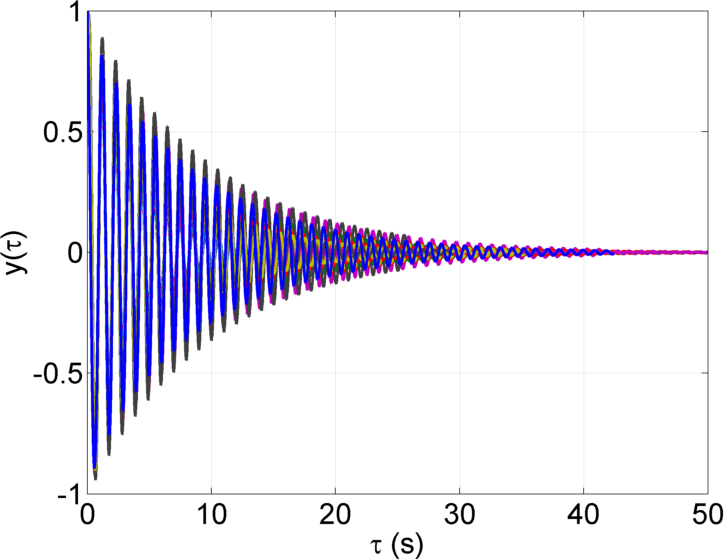}
	\label{fig:dufyvstau}
	}
	\caption[Duffing oscillator -- Different trajectories of the response in the original and warped time scales.]{Duffing oscillator -- $N=50$ different trajectories of the response in the original and warped time scales.}
	\label{fig:dufytau}
\end{figure}

Hereafter, we aim at building PCEs of the displacement $y(t)$ as a function of the random variables $(\zeta, \, \omega, \, \epsilon)$.
First, we use $200$ trajectories of $y(t)$ as experimental design to compute time-frozen sparse PCEs of adaptive degree up to 20. 
Next, we use the time-warping approach, which requires only 50 trajectories $y(t)$ as experimental design.
The $50$ trajectories in the original time scale are plotted in \figref{fig:dufyt}. The same trajectories after time-warping are plotted in \figref{fig:dufyvstau}.
A linear time-warping with two parameters, \ie $\tau = k \, t  + \phi$, is used for each trajectory.
Using sparse PCEs of degree up to 20, the metamodels of $k$ and $\phi$ are obtained with relative LOO errors $1.87\times 10^{-5}$ and $2.08 \times 10^{-4}$ respectively, which indicates a high level of accuracy.
PCA is then applied to retrieve eight principal components that results in the PCA truncation error smaller than $1 \times 10^{-3}$. 
The relative LOO errors of PCE models for the first two components are $8 \times 10^{-4}$ and $4 \times 10^{-3}$, respectively.

\begin{figure}[!ht]
	\centering
	\subfigure
	[$\vexi = (0.0403,    5.0455,   -0.7186)$]
	{
	\includegraphics[width=0.45\linewidth]{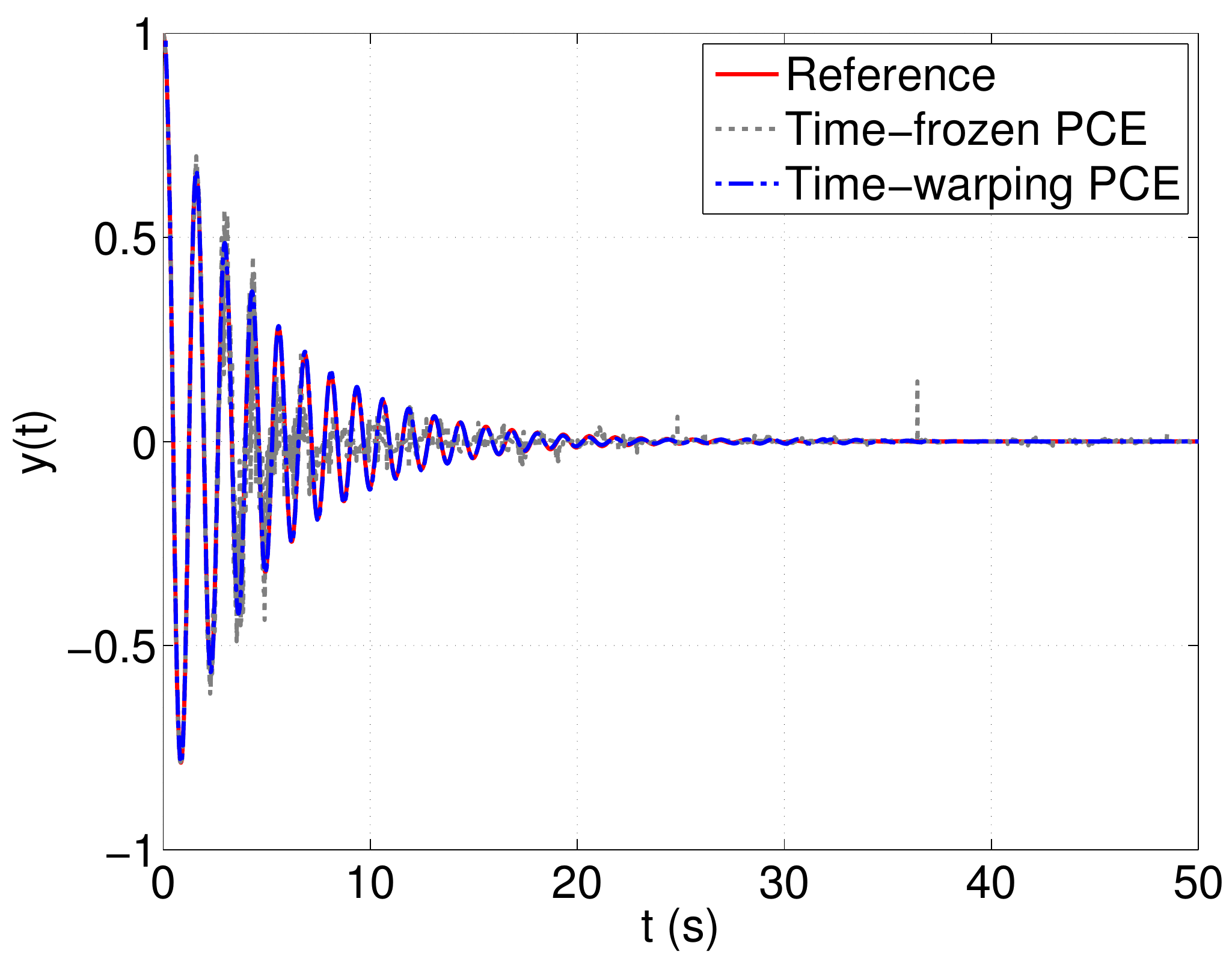}
	}	
	\subfigure
	[$\vexi = (0.0222,    4.9974,   -0.5007)$]
	{
	\includegraphics[width=0.45\linewidth]{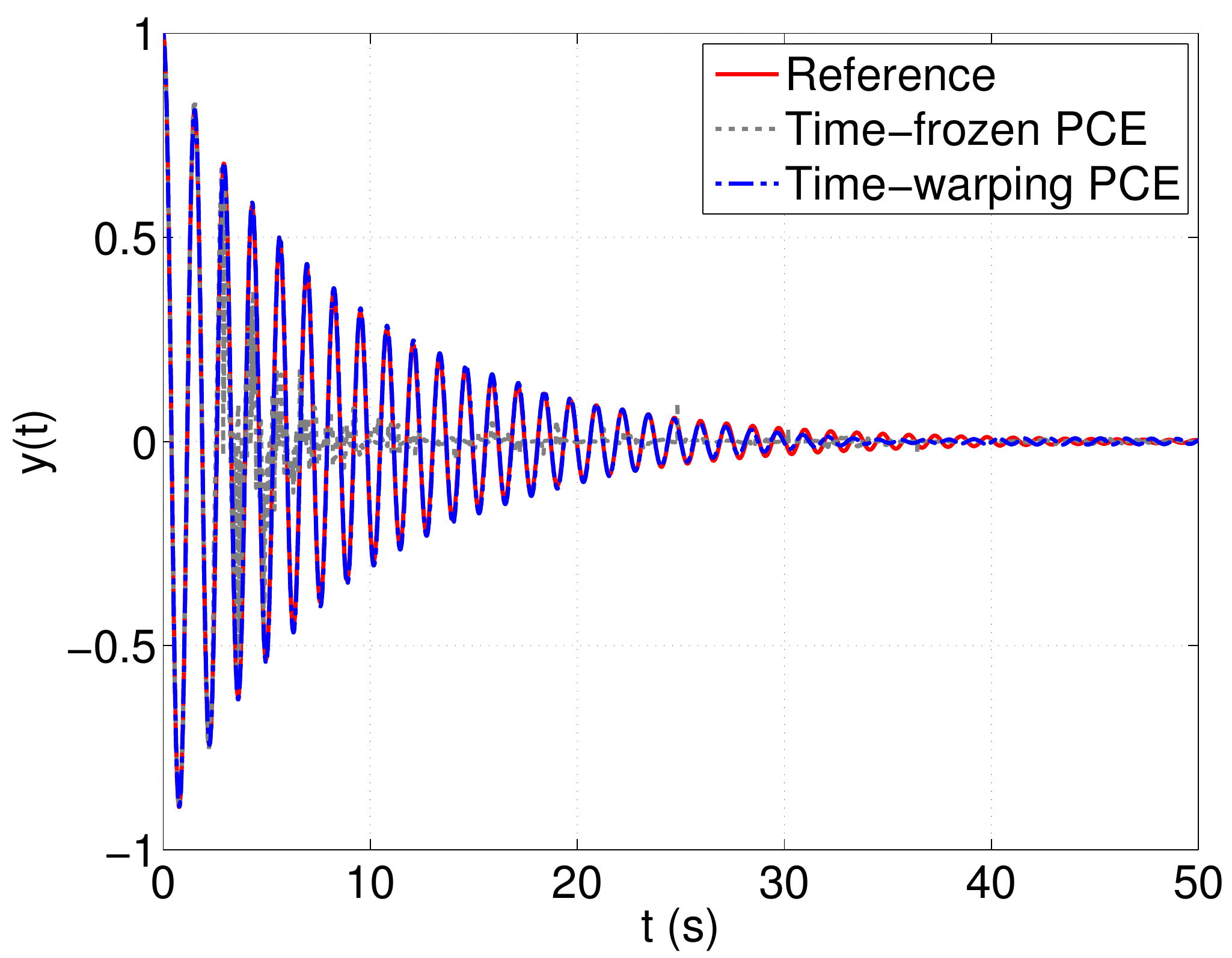}
	}
	\caption{Duffing oscillator -- Two particular trajectories and their predictions by time-frozen and time-warping PCEs.}
	\label{fig4.2.3}
\end{figure}

An independent validation set of $10,000$ runs is used to judge the accuracy of the PCE models. \figref{fig4.2.3} presents two specific realizations of the displacement $y(t)$ obtained with two distinct sets of parameters $(\zeta,\omega,\epsilon)$. 
Without time-warping, PCEs are capable of predicting the response at the early time instants ($t<3~s$), then their accuracies degenerate with time, resulting in incorrect predictions. By introducing the time-warping of the trajectories, PCEs can faithfully capture the damped oscillatory behaviour. 
Only $0.18\%$ of $10,000$ predictions exhibits a relative error exceeding $0.1$.
Note that an experimental design of size 200 is used for time-frozen PCEs, whereas only 50 trajectories are used for computing time-warping PCEs.
This emphasizes the fact that the time-warping pre-processing of the response allows one to build accurate PCEs at an extremely small computational cost.

\begin{figure}[!ht]
	\centering
	\subfigure[Mean trajectory]
	{
	\includegraphics[width=0.45\linewidth]{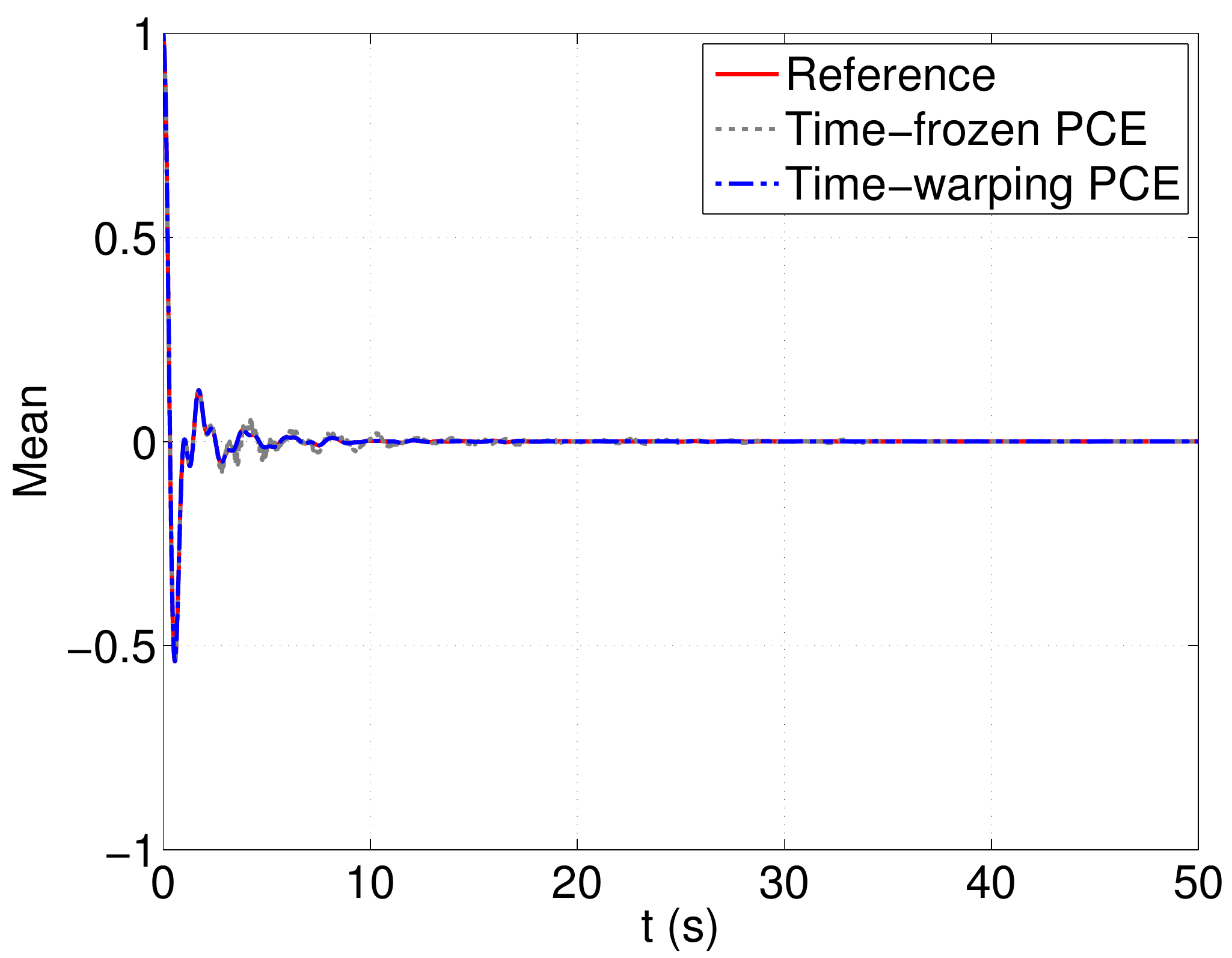}
	}
	\subfigure[Standard deviation trajectory]
	{
	\includegraphics[width=0.45\linewidth]{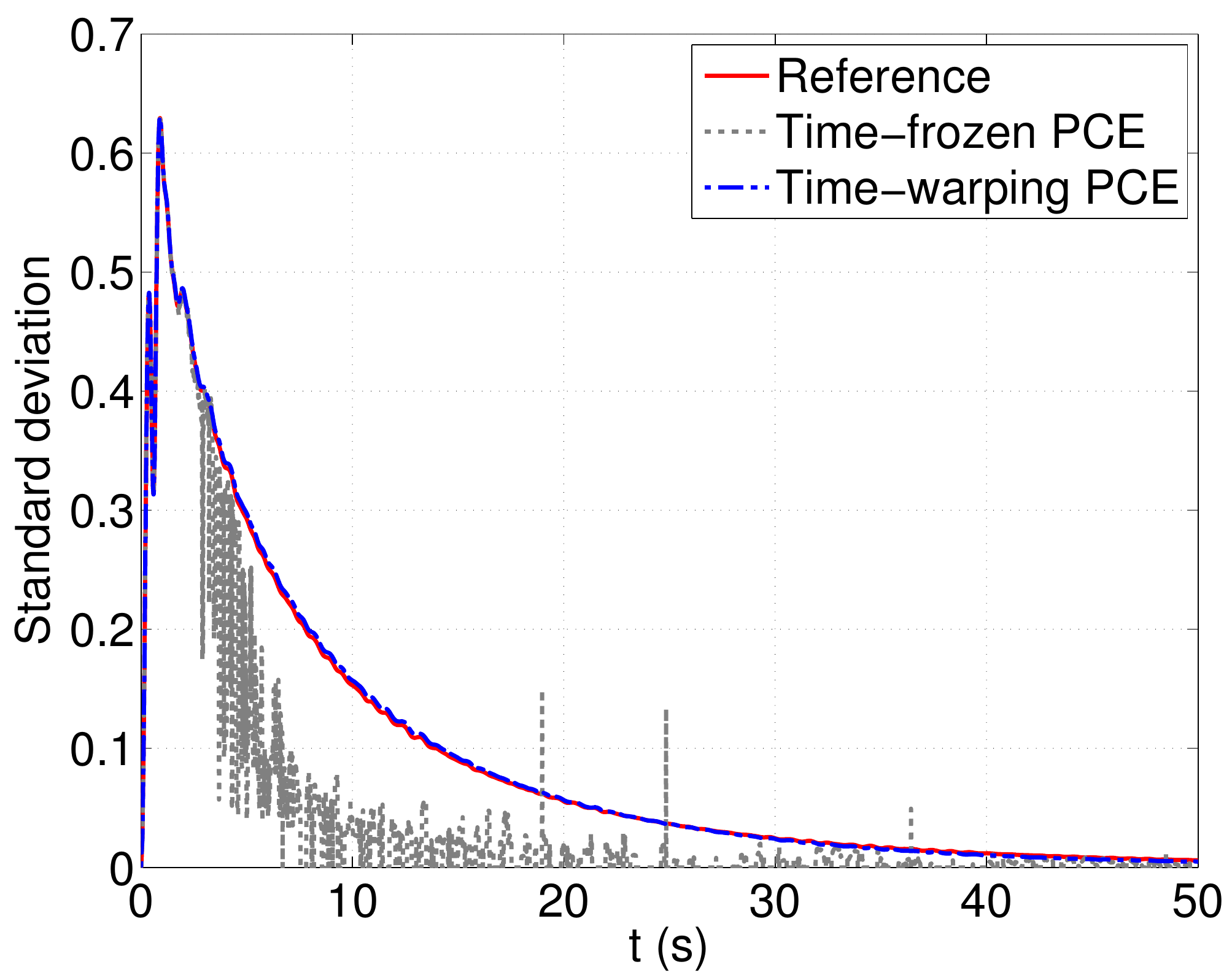}
	}
	\caption{Duffing oscillator -- Mean and standard deviation of the trajectories: comparison of the two approaches.}
	\label{fig4.2.4}
\end{figure}

In terms of time-dependent statistics (\figref{fig4.2.4}), time-frozen PCEs can predict rather well the mean trajectory, however fail to represent the standard deviation after early instants ($t>3~s)$. In contrast, the time-warping approach provides excellent accuracy on the mean and standard deviation time histories. The relative discrepancies between mean and standard deviation time histories predicted by time-warping PCEs with the reference trajectories are $3.27 \times 10^{-5}$ and $3.47 \times 10^{-4}$, respectively.

%\bibliography{biblioRSUQ,biblioSCHOEBI}
\bibliographystyle{chicago}
\bibliography{bib}

\end{document}